\documentclass[twocolumn,showpacs,superscriptaddress,nofootinbib,prd]{revtex4}

\usepackage{shortcuts}

\usepackage{graphicx}
\usepackage{color}
\usepackage{amsmath}
\usepackage{amssymb}

\begin{document}

\title{Faster computation of adiabatic EMRIs using resonances} 

\author{Rebecca Grossman} 
\email{becky@phys.columbia.edu}
\affiliation{Physics Department, Columbia University, New York, NY
  10027} 
\author{Janna Levin} 
\email{janna@astro.columbia.edu}
\affiliation{Department of Physics and Astronomy, Barnard College of
  Columbia University, 3009 Broadway, New York, NY 10027}
\affiliation{Institute for Strings, Cosmology and Astroparticle
  Physics, Columbia University, New York, NY 10027} 
\author{Gabe
  Perez-Giz} 
\email{gabe@phys.columbia.edu} 
\affiliation{Physics
  Department, Columbia University, New York, NY 10027}

\begin{abstract}
\label{sec:abs}

Motivated by the prohibitive computational cost of producing adiabatic
extreme mass ratio inspirals, we explain how a judicious use of
resonant orbits can dramatically expedite both that calculation and
the generation of snapshot gravitational waves from geodesic sources.
In the course of our argument, we clarify the resolution of a
lingering debate on the appropriate adiabatic averaging prescription
in favor of torus averaging over time averaging.

\end{abstract}


\maketitle

\section{Introduction}
\label{sec:intro}

Stellar mass compact objects inspiraling into supermassive black holes
(SMBHs) will be important astrophysical sources of gravitational waves
(GWs) for future space-based detectors.  Accurate GW templates for
such extreme mass-ratio inspirals (EMRIs) require detailed knowledge
of the motion of the source, so there has been a community effort to
calculate EMRI trajectories.  If we neglect the gravitational
self-force of the small object, its orbit is a Kerr geodesic that, up
to parameters specifying the initial position, is characterized by
three constant orbital parameters: an energy $E$, an azimuthal angular
momentum $L_{z}$, and the Carter constant $Q$.  Determining the
inspiral is tantamount to calculating how the self-force causes both
the positional parameters and the orbital parameters to evolve in
time.

Despite ongoing efforts, direct evaluation of the self-force in the
Kerr case is still not possible.  Accordingly, there have been
parallel efforts to approximate its effects.  The focus of this paper
is the adiabatic approximation, which captures the slow secular
evolution of $\elqall$ by solving a system of ordinary differential
equations (ODEs) of the form
\begin{subequations}
\label{eq:ELzQ_adiab_ODEs}
\begin{alignat}{1}
  \D{t}{E} &= \mathcal{F}_{E} \elqset
  \label{eq:E_adiab_ODE}
  \\
  \D{t}{L_{z}} &= \mathcal{F}_{L_{z}} \elqset
  \label{eq:Lz_adiab_ODE}
  \\
  \D{t}{Q} &= \mathcal{F}_{Q} \elqset
  \label{eq:Q_adiab_ODE}
  \quad .
\end{alignat}
\end{subequations}
For now, it suffices to know that the righthand sides (RHSs) of
equations (\ref{eq:ELzQ_adiab_ODEs}) are so costly to evaluate that
these equations will have to be integrated using a numerical grid.
More specifically, the $\elq$ velocity field will be pre-computed only
on a dense mesh of points in $\elq$-space.  Real-time integration of
(\ref{eq:ELzQ_adiab_ODEs}) will then rely on derivative values
interpolated off of that grid.

\begin{figure}[t]
  \centering
  \includegraphics[width=55mm]{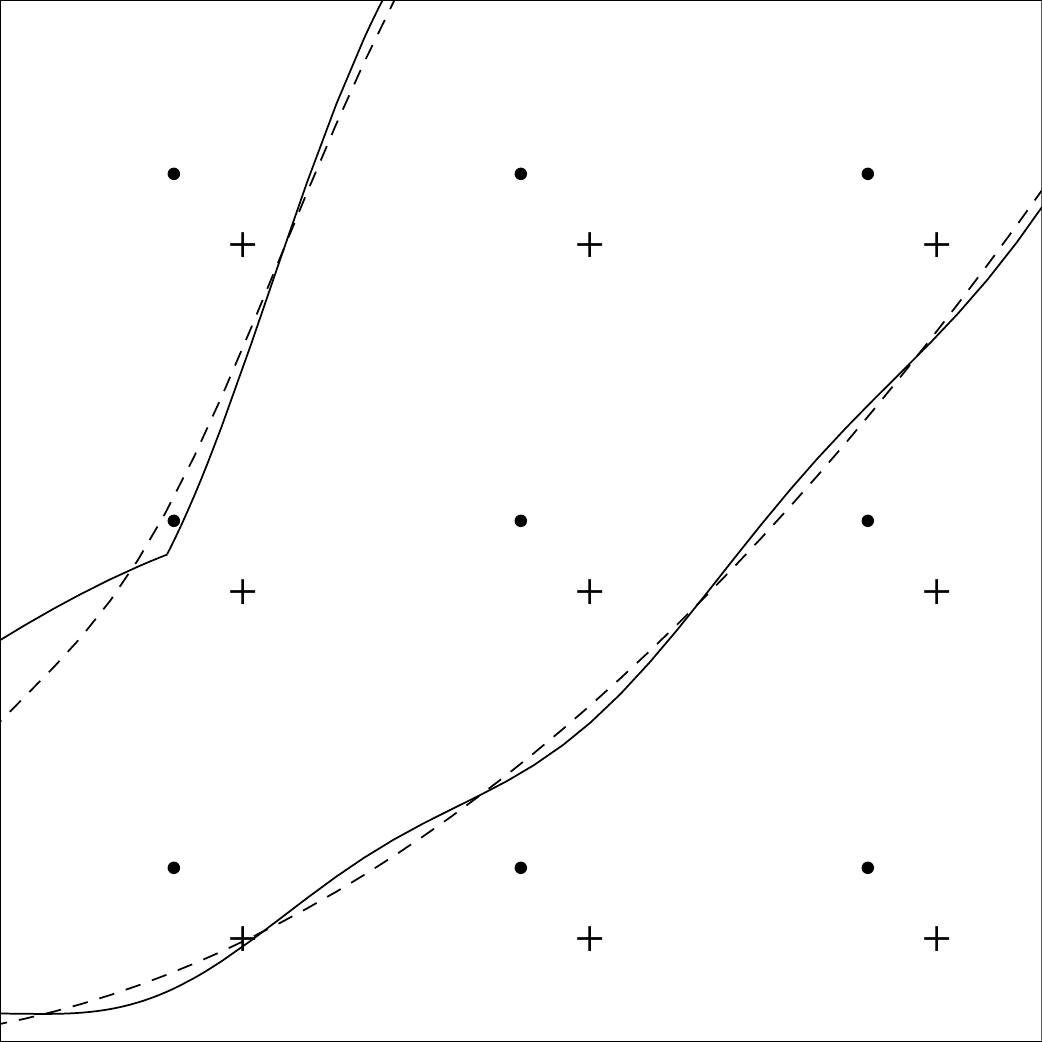}
  \hfill
  \caption{Above is a heuristic depiction of two possible numerical
    grids that could be used to generate adiabatic approximations
    (dashed curves) to true inspirals (solid curves) in the orbital
    parameter space.  The dots represent a set of resonant grid points
    and the plus signs a set of non-resonant grid points.  The
    resulting adiabatic curves are the same in either case but
    significantly less costly to produce with the resonant grid.  A
    true inspiral may evolve in a way that is not well-approximated
    adiabatically as it approaches a low-order resonance, as on the
    left.  That divergence, if it occurs, happens regardless of
    whether the resonant point is used as part of the numerical grid.}
  \label{fig:grid}
\end{figure}

In this paper, we advocate building such grids using only points
corresponding to resonant geodesics, for which the frequencies of the
radial and polar motions are rationally related.  As we will see,
intermediate calculations that comprise the bulk of the computational
expense can be recycled among several Fourier modes on resonant grid
points but must be recomputed for every mode in the non-resonant case.
We estimate that, compared to using non-resonant grid points, our
prescription could reduce the computational cost of an EMRI grid by an
order of magnitude or more.  The resonant-grid prescription will also
facilitate faster computation of GW snapshots from geodesic sources.

We represent our proposal schematically in Figure \ref{fig:grid}.
First, the RHSs of equations (\ref{eq:ELzQ_adiab_ODEs}) are evaluated
directly on a grid of either resonant points (dots) or non-resonant
points (plus signs).  At any other point, the RHS values can be
interpolated from the values at the grid points.  The adiabatic
eqautions (\ref{eq:ELzQ_adiab_ODEs}) are continuous and smooth, so
regardless of which grid is used, integrating them produces the same
adiabatic solutions (dashed curves).  The only difference is that the
resulting adiabatic curves cost significantly less to generate with
the resonant grid.

Ref.\ \cite{mino2005} and more recently, Ref.\ \cite{flanagan2010}
have noted that such adiabatic approximations may fail to capture
important features of the true inspiral (solid curves) near
low-order\footnote{A resonance is low-order if the numerator and
  denominator of the rational frequency ratio are both small
  integers.} resonances.  Heuristically speaking, those authors argue
that while an adiabatic solution may remain fairly faithful to an
inspiral that steers clear of resonant points (lower right of the
figure), those approximations may fare much worse for an inspiral that
transits near a resonant point (middle left of the figure).  To
pre-empt possible confusion, we remark that there is no inconsistency
between this observation and our proposal.  The decision to include
any particular $\elq$ point, resonant or not, in the numerical grid is
unrelated to whether the resulting adiabatic curves will faithfully
reflect EMRI motion near that same point.  The ironic coincidence is
that the points where the adiabatic approximation is most likely to
fare poorly\footnote{To balance the argument, Ref.\ \cite{mino2005}
  also offers plausible reasons why the adiabatic approximation may
  still be valid near resonances.}  are also the optimal grid points
for generating adiabatic curves.

The rest of this paper is organized as follows.  In Section
\ref{sec:background}, we review some relevant features of resonant
Kerr orbits in both physical space and phase space.  In Section
\ref{sec:adiab-insp-thru-flux-balance}, we summarize how one arrives
at the adiabatic equations of motion and clarify why an averaging
perscription required to derive those equations must be a torus
average rather than a time average, an issue that has raised some
debate in the literature
\cite{mino2005,drasco2005,pound2005,pound2007:2,pound2007:1,hinderer2008}.
Partly to help make the averaging argument and partly because we will
focus on frequency-domain approaches to solving the adiabatic
equations, Section \ref{sec:adiab-insp-thru-flux-balance} also
provides some necessary mathematical background on Fourier analysis in
both the non-resonant and resonant cases.  With a clear view of the
adiabatic program now in hand, Section \ref{sec:definite} presents the
main result of the paper, namely a concrete prescription for
computational savings that frequency-domain EMRI codes can leverage by
using resonances.  Finally, Section \ref{sec:spec} speculates about
how a more unorthodox use of resonances could offer additional
efficiencies provided it can be practically implemented.

\section{Resonant Kerr Orbits}
\label{sec:background}

The paramount role of resonant orbits was the central theme of an
earlier series of papers \cite{levin2008,grossman2011,levin2008:2}.
(We use the terms ``resonant'', ``closed'', and ``periodic''
interchangeably.)  A spectrum of closed orbits, which manifests as a
spectrum of multi-leaf clovers, entirely structures black hole
dynamics.  Although completely closed orbits must return to their
initial values\footnote{$\varphi$ need only return to its initial
  position mod $2\pi$.} of $(r,\theta,\varphi)$ simultaneously, only
the $r$-$\theta$ periodicity detailed in Ref.\ \cite{grossman2011} is
relevant to the present work.  The rational number associated with the
$r$-$\theta$ frequencies determines the multi-leaf clover
geometry. What's more, that rational obediently stacks in energy
monotonically: lower rationals correspond to lower energies than do
higher rationals.  Such orbits constitute a measure zero set but are
nonetheless dense in the phase space.  Every non-resonant orbit is
arbitrarily close to some resonant orbit.

We first consider resonant orbits in {\it physical space} and then
again in {\it phase space}.  Since we will be concerned with functions
that do not depend explicitly on either the azimuthal angle or on
coordinate time, it will suffice for us to restrict attention to
geodesic motion in two coordinates $(r,\theta)$ in physical space and
to the projection of the motion into a $4D$ submanifold of the phase
space spanned by $(r,\theta)$ and their conjugate momenta.

\subsection{Resonant Orbits in Physical Space}
\label{sec:r_theta}

The black hole is completely characterized by its mass $M$ and spin
$a$. The geodesic of the lighter companion is characterized by four
dimensionless constants $\mu, E,L_z,Q$.  In Boyer-Lindquist
coordinates and dimensionless units, which is equivalent to setting
$M=\mu=1$, the radial and polar Kerr equations of motion can be
written as
\begin{eqnarray} 
\label{eq:r_eq_motion}
\dot{r} &=& \pm\sqrt{R\left(r\right)}
\\
\label{eq:th_eq_motion}
\dot{\theta} &=& \pm\sqrt{\Theta\left(\theta\right)}
\quad ,
\end{eqnarray}
where
\begin{eqnarray} 
\label{eq:R}
R\left(r\right) &\equiv& -\left(1-E^{2}\right)r^{4} + 2r^{3} 
\\
\nonumber
& &-\left[a^{2}\left(1-E^{2}\right)+L_{z}^{2}\right]r^{2}+2\left(aE-L_{z}\right)^{2}r-Q\Delta
\\
\label{eq:TH}
\Theta\left(\theta\right) &\equiv& Q-\cos^{2}\left(\theta\right)\left\{a^{2}\left(1-E^{2}\right)+\frac{L_{z}^{2}}{\sin^{2}{\theta}}\right\}
\end{eqnarray}
and
\begin{eqnarray}
\label{eq:Delta_defn}
\Delta &\equiv& r^{2} - 2r + a^{2}
\quad .
\end{eqnarray}

An overdot denotes differentiation with respect to Mino time
\cite{mino2003}, $\lambda$, related to proper time $\tau$ by
\begin{alignat}{1}
\label{eq:Sigma_defn}
\frac{d\tau}{d\lambda} &=\Sigma 
\equiv r^{2}+a^{2}\cos^2{\theta}
\quad .
\end{alignat}
The advantage of using Mino time is that the $r$ and $\theta$
equations of motion decouple and that each is a function only of one
coordinate.  To make connections with observations, we will often care
about how certain quantities evolve with respect to coordinate time
$t$.  However, coordinate time turns out to be mathematically
cumbersome, so throughout this paper, we perform all intermediate
calculations related to such quantities by first changing variables to
Mino time.

Solving equations (\ref{eq:R}) and (\ref{eq:TH}) for the radial and
polar turning points, we find that the radial coordinate varies
between a periastron $r_{p}$ and an apastron $r_{a}$ and that the
polar coordinate similarly varies between some minimum value
$\theta_{min}$ and maximum value $\theta_{max}={\pi}-\theta_{min}$.
All turning points depend only on the constants $\elqall$.  We
introduce the simplifying notation
\begin{alignat}{1}
\label{eq:defn_angle_variable_vector}
  \elqvec &\equiv \elqset
\end{alignat}
for those 3 orbital parameters and reserve the symbol $\elqC$ to refer
to any one of $\elqall$ individually.

The radial and polar coordinates are each periodic with respective
Mino periods
\begin{subequations}
\label{eq:N_theta_r}
\begin{alignat}{1}
\Lambda_r &= 2\int_{r_p}^{r_a} \frac{dr}{\sqrt{R(r)}}
\label{eq:L_r}
\\
\Lambda_{\theta} &= 4\int_{\theta_{min}}^{\pi/2}\frac{d\theta}{\sqrt{\Theta(\theta)}}
\label{eq:L_theta}
\end{alignat}
\end{subequations}
and corresponding frequencies
\begin{subequations}
\label{eq:freqs}
\begin{alignat}{1}
\label{eq:r_freq}
\omr &= \frac{2\pi}{\Lambda_{r}}
\\
\omth &= \frac{2\pi}{\Lambda_{\theta}}
\label{eq:theta_freq}
\quad .
\end{alignat}
\end{subequations}
The radial and polar velocities are also periodic with the same
corresponding periods and frequencies.  If the frequency ratio
\begin{alignat}{1}
\label{eq:q_r_theta}
1 + q_{r\theta} &\equiv \frac{\omth}{\omr}
\quad 
\end{alignat}
is a rational number, an $r$-$\theta$ projection of the resulting
orbit closes after a finite time.

Note from equations (\ref{eq:R}),~(\ref{eq:TH}) and
(\ref{eq:N_theta_r}) that the frequencies and $q_{r\theta}$ depend
only on the constants $\elqvec$.  $q_{r\theta}$ is also a topological
invariant and thus coordinate independent.  A $q_{r \theta}$ for which
the relatively prime numerator and denominator are both low-valued
integers will be referred to as ``low-order''. We arbitrarily call
low-order resonant orbits those for which the numerator and
denominator of $q_{r\theta}$ are each less than 10.

\begin{figure}
  \centering
  \includegraphics[width=85mm]{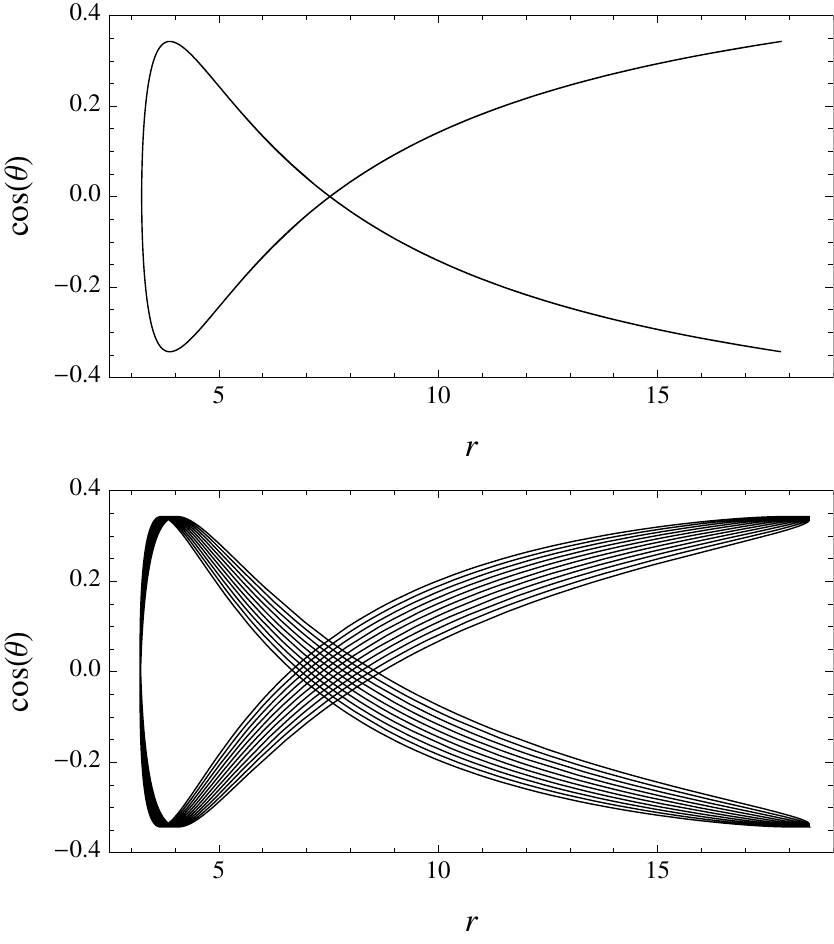}
  \hfill
  \caption{Top: A low-order periodic orbit with $q_{r\theta} =
    \frac{1}{2}$, $a=0.9$, $E=0.954788$, $L_{z}=2.65115$ and
    $Q=0.944969$, projected on the $r$-$\theta$ plane (we plot
    $r$-$\cos{\theta}$ to make the figure more viewable) with initial
    conditions $r_{0}=r_{a}=17.8148$ and
    $\theta_{0}=\theta_{min}=1.22079$. Bottom: A non-resonant orbit
    with $q_{r\theta}\approx\frac{125,857}{250,000}$, $a=0.9$,
    $E=0.956$, $L_{z}=2.65115$ and $Q=0.944969$, with initial
    conditions $r_{0}=r_{a}=18.4568$ and
    $\theta_{0}=\theta_{min}=1.22076$.}
  \label{fig:periodic_rcos}
\end{figure}

Projections of periodic orbits into the $r$-$\theta$ plane produce
Lissajous figures.  The top panel of Figure \ref{fig:periodic_rcos}
shows the Lissajous figure of a periodic orbit with a low-order $q_{r
  \theta}$, while the bottom panel shows the analogous projection of a
neighboring orbit with an irrational $q_{r \theta}$.

\begin{figure}
  \centering
  \includegraphics[width=75mm]{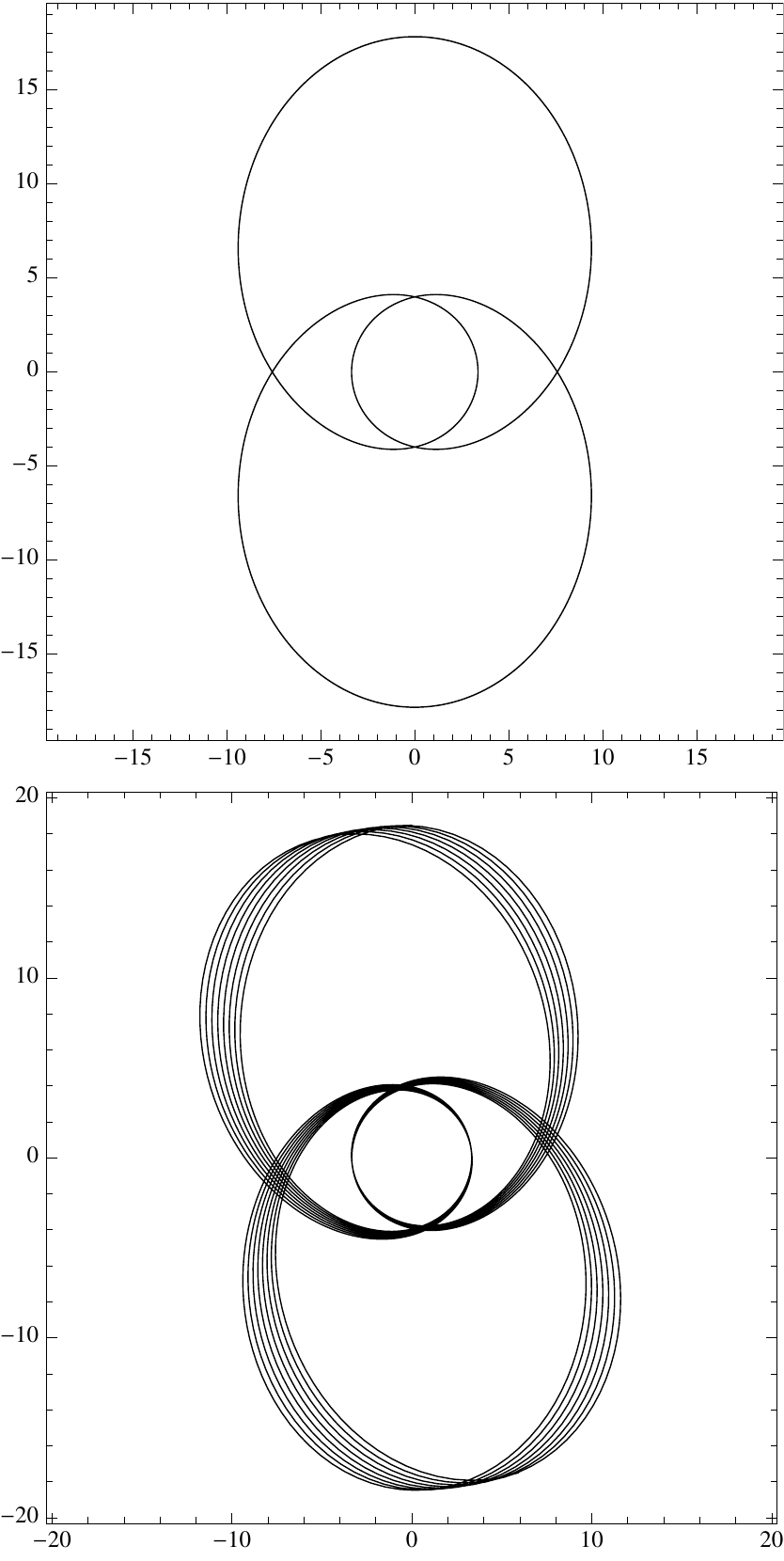}
  \hfill
  \caption{Top: The same orbits from Figure \ref{fig:periodic_rcos},
    projected into the orbital plane.}
  \label{fig:periodic_orb}
\end{figure}

The figures produced by projecting into the $r$-$\theta$ plane are
less topologically insightful than the figures in an orbital plane,
loosely defined in the Kerr system as the plane perpendicular to the
orbital angular momentum \cite{grossman2011}. In the orbital plane,
the rational $q_{r\theta}$ has powerful topological information and
can be interpreted as $q_{r\theta}=w+v/z$, where the integer $w$
represents the number of nearly circular whirls near periastron, the
integer $z$ is the number of elliptical leaves in the multi-leaf
clover pattern, and the integer $v$ is the order in which the leaves
are hit \cite{levin2008,grossman2011}.  To illustrate, the same two
orbits of Figure \ref{fig:periodic_rcos} are plotted in the orbital
plane in Figure \ref{fig:periodic_orb}.  The orbit in the top panel of
Figure \ref{fig:periodic_orb} has $q_{r\theta}=1/2$ and therefore
corresponds to a $2$-leaf clover, as is now evident. The bottom panel
non-resonant orbit is close to a resonant orbit with
$q_{r\theta}\approx\frac{125,857}{250,000}$ which would correspond to
a $250,000$-leaf clover that skips $125,857$ leaves in the pattern
each time. Notice that $\frac{125,857}{250,000} =
\frac{1}{2}+\frac{857}{250,000}$ so that the orbit is really a tight
precession of the $2$-leaf clover through an angle of
$\left(\frac{857}{250,000}\right)2\pi\approx 0.02154$ radians per
radial cycle.

\begin{figure*}
  \includegraphics[width=150mm]{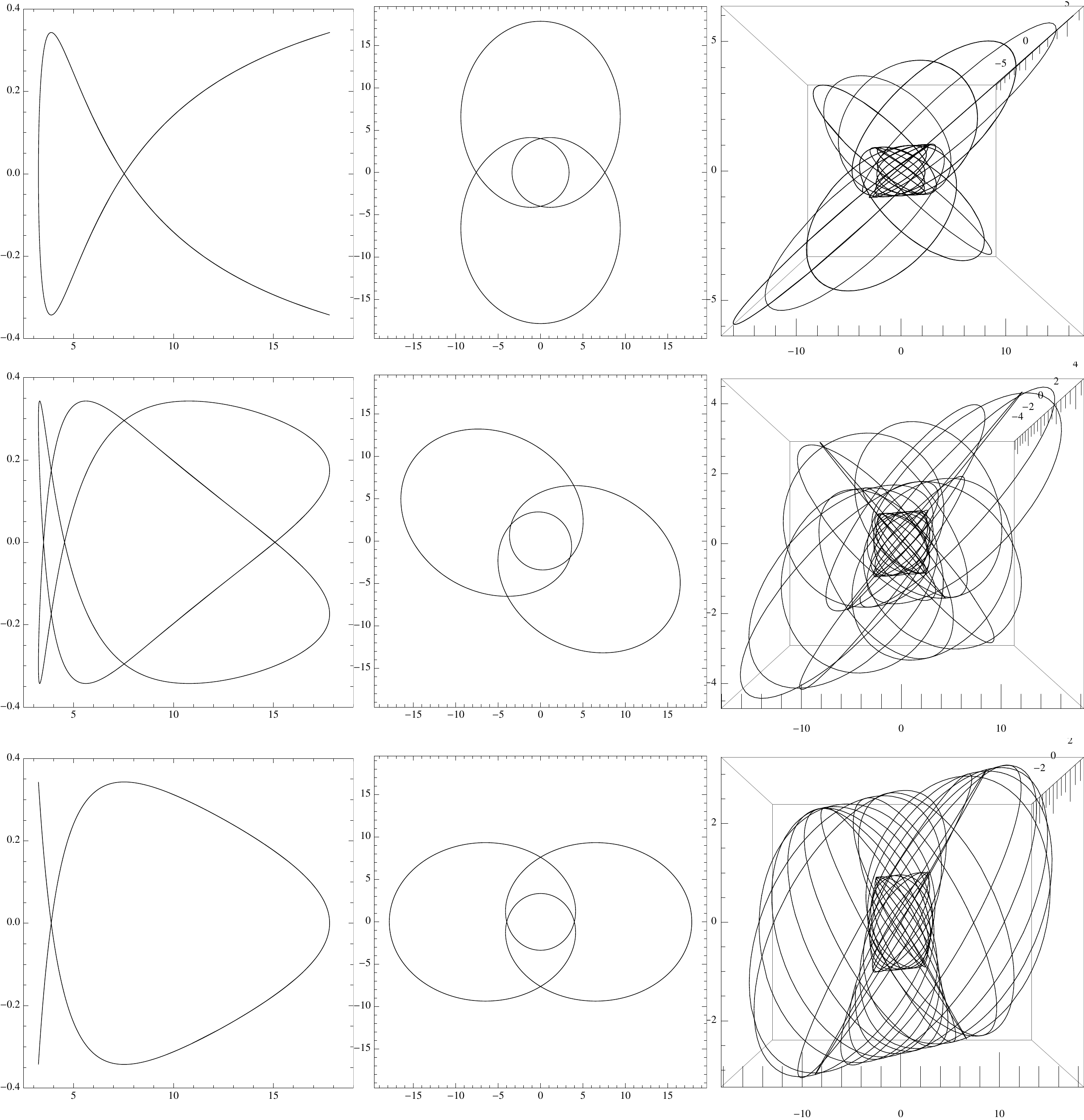}
  \hfill
  \caption{The above figures are all $q_{r\theta}=\frac{1}{2}$ orbits
    with $a=0.9$ and orbital parameters $L_{z}=2.65115$, $Q=0.944969$
    and $E=0.954788$.  The three figures in each row have the same
    initial coordinates.  The column on the left shows an
    $r$-$\cos{\theta}$ projection of the orbit, the middle column is a
    projection in the orbital plane, and the right column is the $3D$
    orbit.  All three rows have $r_{0}=r_{a}= 17.8148$.  The first row
    has $\theta_{0} = \theta_{min}=1.22079$ and is the same orbit
    shown in the top panel of Figs.\ \ref{fig:periodic_rcos} and
    \ref{fig:periodic_orb}, the middle row has $\theta_{0}=1.39579$,
    and the bottom row has $\theta_{0}=\frac{\pi}{2}$. }
  \label{fig:phasing}
\end{figure*}

While they do fix the turning points, the constants $\elqvec$ do not
uniquely fix the orbit \cite{levin2009}.  An orbit that hits apastron
at $\theta_{min}$ is not identical to the orbit that hits apastron at
a different value of $\theta$, as shown in Fig.\ \ref{fig:phasing}.
Since $q_{r\theta}$ depends only on constants, a $q_{r\theta}=1/2$
orbit is always a 2-leaf clover in the orbital plane
\cite{grossman2011}. However, orbits with different $r$-$\theta$
initial conditions $(r_{0},\theta_{0})$ are rotated relative to each
other in the orbital plane.

As Fig.\ \ref{fig:phasing} shows, the resulting orbits are genuinely
distinct in 3D.  Presumably, they could have distinct gravitational
wave emissions.  Interestingly, though, perihelion precession happens
on a faster time scale than plane precession. It is therefore
reasonable to suspect that all orbits with the same $q_{r\theta}$
generate similar waveforms and that the different plane precessions
induce modest differences in the modulations of the amplitude
\cite{2010arXiv1009.2533L}. We remain agnostic on the relative
importance of $r$-$\theta$ initial values on the waveform generated
and instead focus on efficient calculation of adiabatic inspirals.

\subsection{Resonant Orbits in Phase Space -- Phase Space Tori}
\label{sec:phase-space-tori}

In the 4D space spanned by $(r,p_r,\theta,p_\theta)$, all orbits
(resonant or non-resonant) lie on 2D tori that can be constructed as
the Cartesian product of two closed curves.  We obtain one of those
closed curves if we project an orbit into the $r$-$p_{r}$ plane. The
area of the curve is the familiar action $J_{r}$ used in action-angle
coordinates.  Analogously, the projection of the same orbit into the
$\theta$-$p_{\theta}$ plane yields another closed curve with area
$J_{\theta}$.  We now consider that pair of curves as a locus of
points on a 2D surface with the topology of the 2-torus
$\mathbb{S}^{1} \times \mathbb{S}^{1} \equiv \mathbb{T}^{2}$.  Every
set of orbital parameters $\elqvec$ defines one such torus that we
denote $\torelq$.

\begin{figure}
  \centering
  \includegraphics[width=90mm]{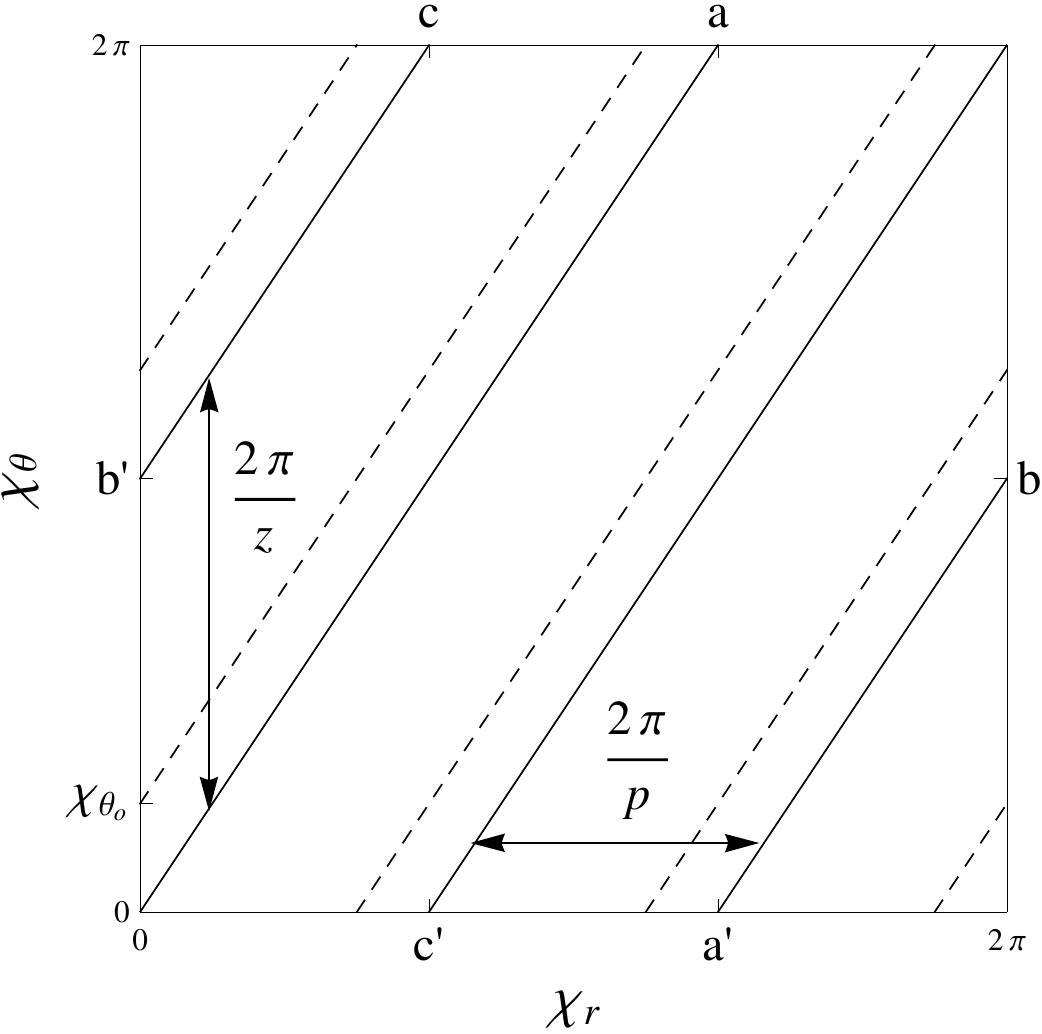}
  \hfill
  \caption{The above picture shows a resonant torus mapped to a square
    with the path of two resonant orbits traced out.  The solid line
    shows the path of a resonant orbit with
    $\chi_{r_0}=\chi_{\theta_0}=0$ and the orbit traced out by the
    dotted line has $\chi_{r_0}=0$ and $\chi_{\theta_0}=0.7894$.  The
    resonant torus and both resonant orbits have
    $\frac{\omth}{\omr}=1+q_{r\theta}=\frac{p}{z}
    =\frac{3}{2}$.}
  \label{fig:res_tor}
\end{figure}

The use of Mino time as an evolution parameter furnishes one (but
certainly not the only) coordinate system for $\torelq$, according to
the following construction.  As already mentioned, the motions
$r(\lambda)$ and $\theta(\lambda)$ are each individually periodic in
Mino time, with periods $\Tr$ and $\Tth$ (and frequencies $\omr$ and
$\omth$), respectively.  Scaling the evolution parameter $\lambda$ on
each of the $r$-$p_{r}$ and $\theta$-$p_{\theta}$ curves by $\omr$ and
$\omth$, respectively, leads to a natural definition of angle
variables $\angr \equiv \omr \lambda$ and $\angth \equiv \omth
\lambda$.  We choose a specific trajectory $\left( r(\lambda), p_{r}
(\lambda), \theta (\lambda), p_{\theta} (\lambda) \right)$ in order to
assign $\angr$ and $\angth$ values, respectively, along the
$r$-$p_{r}$ and $\theta$-$p_{\theta}$ curves, but the trajectory is
only a device that we discard once the torus coordinate system is in
place.  The points at $0$ and $2\pi$ in each of $\angr,\angth$ are
identified, so the torus can be represented as a $2\pi$-by-$2\pi$
square with opposite sides identified as in Fig.\ \ref{fig:res_tor}.
We will make a simplifying choice that $(r_a,\theta_{\text{min}})$
corresponds to the origin\footnote{Many references, including
  \cite{mino2003,mino2005,drasco2006,drasco2005}, instead tacitly
  choose the point $(r = r_{p}, p_{r} = 0, \theta =
  \theta_{\text{min}}, p_{\theta} = 0)$ as the origin of the torus
  coordinates.  We say ``tacitly'' because they refer to the
  individual orbit with those initial conditions as a fiducial
  geodesic to use in their analyses.  Another interpretation of that
  choice is that they are working not with one geodesic but with one
  \emph{torus} and that they have instead chosen a fiducial origin for
  a $\angr$-$\angth$ coordinate system on that torus.} of the torus.
Then, a reflection in the line $\angr = \pi$ corresponds to keeping
$r$ fixed and reversing the sign of $p_{r}$, and analagously for
reflections in $\angth = \pi$.  Note that each quadrant of the
toroidal square therefore contains the same $(r,\theta)$ pairs but
with all possible sign combinations for the momenta ($++, +-, -+,
--$).

Note that each $\angr$ corresponds to an ordered pair\footnote{Some
  references describe the mapping of functions of the form $F(r,
  \theta)$ to corresponding functions $F(\angr, \angth)$.  In fact, no
  function that enters an adiabatic EMRI calculation depends on $r$
  and $\theta$ alone. The notation $F(r, \theta)$ in those references
  is used because, once restricted to a torus, the value of each
  coordinate determines its conjugate momentum up to a sign.  Still,
  the values of those signs affect the value of the function.  We
  believe a notation such as $F(r, p_{r}, \theta, p_{\theta})$ for
  these pre-torus phase space functions is more appropriate.}
$(r,p_r)$ and each $\angth$ corresponds to an ordered pair
$(\theta,p_\theta)$.  We discuss alternative coordinate systems for
$\torelq$ in Appendix \ref{subsec:which_time} and elsewhere in this
article but will use only the $\angset$ coordinates for calculations.

On the compact $(\angr,\angth)$ square defined above, geodesic
trajectories are lines of slope $\omth/\omr = 1 + q_{r\theta}$.  With respect to Mino time,
those orbits are given parametrically by
\begin{subequations}
\label{eq:chi_of_lambda}
\begin{alignat}{1}
  \angr(\lambda) &= \omr \lambda + \offr
  \label{eq:chi_r_of_lambda}
  \\
  \angth(\lambda) &= \omth \lambda + \offth
  \label{eq:chi_theta_of_lambda}
  \quad .
\end{alignat}
\end{subequations}
Two different initial positions $\offvec$
and $\vec{\angsym}'_{o}$ produce distinct orbits unless
there exist real
numbers $x$ and $y$ that simultaneously satisfy
\begin{equation}
\label{eq:eqn_for_time_trans_offsets}
  \frac{y - \offth}
       {x - \offr}
  =
  \frac{\omth}{\omr}
  \quad ,
\end{equation}
and
\begin{eqnarray}
\label{eq:eqn_for_time_trans_torus_offsets}
  \offr' &=& x\text{ mod }2\pi
\\
\nonumber
\offth'  &=& y\text{ mod }2\pi
  \quad .
\end{eqnarray}
If these conditions are met, then the two different initial positions
produce time-translated versions of the same orbit.

When $\omth / \omr$ is irrational, we will call both
the torus and any orbits on that torus \termdef{non-resonant}.  Orbits
on non-resonant tori never close and instead sample the entire torus
ergodically: an orbit starting from any initial condition will pass
arbitrarily close to every point in the torus after some finite (but
possibly very long) time.  Therefore, non-resonant orbits with
different $\offvec\equiv\offset$ are arbitrarily close to time
translations of every other non-resonant orbit with the same
$\offvec$.  We will alternately refer to such orbits as aperiodic or
biperiodic.

When the frequency ratio $\omth / \omr$ is a rational
number $\frac{p}{z}$, we will call both the underlying torus and
orbits on that torus \termdef{resonant}.  Orbits that live on resonant
tori inherit the rational frequency ratio and thus always trace out
closed curves.  Since no single resonant orbit ergodically fills the
torus, even after infinite time, two resonant orbits with the same
$\elqvec$ but different $\left(\chi_{r_0},\chi_{\theta_0}\right)$ are
not necessarily time translations of each other.  {\it The set of all
  resonant orbits with the same} $\elqvec$ does fill the entire torus.
Because they return to their initial conditions after a finite time,
we will alternately refer to these orbits as periodic or singly
periodic.

Figure \ref{fig:res_tor} shows two resonant orbits on the resonant
torus defined by $E=0.954788$, $L_{z}=2.65115$, $Q=0.944969$.  Each is
thus a $q_{r\theta}=\frac{1}{2}$ orbit, or 2-leaf clover in the
orbital plane.  These are the same two orbits illustrated in physical
space in the top two rows of Figure \ref{fig:phasing}.  The two orbits
are distinguished by their initial position $\offvec$ on the torus.
The solid line orbit, which starts at $\chi_{r_0}=0$,
$\chi_{\theta_0}=0$, corresponds to the physical space orbit with
initial conditions $r_0=r_a=17.81477$ and
$\theta_0=\theta_{\text{min}}=1.220793$.  The dotted line orbit with
intitial conditions $\chi_{r_0}=0$ and $\chi_{\theta_0}=0.7854$ has
physical space intial conditions of $r_0=r_a=17.81477$ and
$\theta_0=1.39579$.  Notice that any two adjacent line segments
belonging to a single orbit are separated in $\chi_{r}$ by
$\frac{2\pi}{\num}$ and in $\chi_{\theta}$ by $\frac{2\pi}{\den}$ but
are not traced out sequentially for general $\frac{p}{z}$.

In the same way that the rational numbers have zero measure on the
line, the set of resonant tori has zero measure in the 4D phase space.
To date, most of the literature on the adiabatic EMRI problem has
ignored resonant geodesics precisely for this reason.  Nevertheless,
as we will see, the judicious exploitation of this measure zero set
leads to significant computational efficiencies in adiabatic EMRI
calculations.

\section{Averaging in the adiabatic approximation}
\label{sec:adiab-insp-thru-flux-balance}

Given the background on geodesic dynamics, we now turn to the
adiabatic approximation of EMRIs, an approximation that has seen
substantial debate in the literature.  As we elucidate below, most of
that debate conflates the question of what kind of averaging procedure
to use in the equations of motion (\ref{eq:ELzQ_adiab_ODEs}) with
other related but logically independent questions about the adiabatic
approximation. In this section, we clarify why phase space averaging
(as opposed to time averaging) is the correct averaging procedure.  We
also establish the results we will need in Section \ref{sec:definite}
to exploit resonant orbits for computational savings.

\subsection{The adiabatic equations of motion}
\label{subsec:adiabatic_eom}
Let $\vec{X}$ denote the Boyer-Lindquist coordinates of the
inspiraling object along with its canonical radial and polar momenta.
In the absence of radiation reaction, the equations of motion are
\begin{subequations}
\label{eq:unperturbed_eom}
\begin{alignat}{1}
  \D{t}{\vec{X}} &= \vec{G} (\vec{X}, \elqvec) \\
  \D{t}{\elqvec} &= 0
  \quad ,
\end{alignat}
\end{subequations}
where the RHSs $\vec{G}$ of the positional equations are some form of
the equations for geodesic motion, e.g.\ Hamilton's equations for
free-particle Kerr motion.  Radiation reaction adds to the RHSs new
functions
\begin{subequations}
\label{eq:total_self-force_eom}
\begin{alignat}{1}
  \D{t}{\vec{X}} &=
  \vec{G} (\vec{X}, \elqvec) + \vec{F} (\vec{X}, \elqvec)
  \label{eq:total_self-force_eom_positional}
  \\
  \D{t}{\elqvec} &=
  0 + \vec{f} (\vec{X}, \elqvec)
  \label{eq:total_self-force_eom_ELzQ}
\end{alignat}
\end{subequations}
that are determined by the full gravitational self-force on the
particle.  Those unknown functions can be expanded in a perturbation
series in powers of a natural small parameter: the system's mass ratio
$\varepsilon \equiv \mu/M \ll 1$.  Furthermore, at each order in
$\varepsilon$, the functions above decompose into a sum of dissipative
and conservative pieces:
\begin{subequations}
\label{eq:forcing_terms_to_2nd_order}
\begin{alignat}{1}
  \begin{split}
    \vec{F} &=
    \varepsilon
    \left[
      \vec{F}^{(1)}_{\text{diss}} + \vec{F}^{(1)}_{\text{cons}} 
      \right]
    \\
    &\mrph{=}
    {}+ \varepsilon^{2}
    \left[
      \vec{F}^{(2)}_{\text{diss}} + \vec{F}^{(2)}_{\text{cons}} 
      \right]
    + \bigO \left( \varepsilon^{3} \right)
  \end{split}
  \label{eq:position_forcing_terms_thru_2nd_order}
  \\
  \vec{f} &=
  \varepsilon
  \left[
    \vec{f}^{(1)}_{\text{diss}} + \vec{f}^{(1)}_{\text{cons}} 
    \right]
  + \varepsilon^{2}
  \left[
    \vec{f}^{(2)}_{\text{diss}} + \vec{f}^{(2)}_{\text{cons}} 
    \right]
  + \bigO \left( \varepsilon^{3} \right)
  \label{eq:ELzQ_forcing_terms_thru_2nd_order}
  \quad .
\end{alignat}  
\end{subequations}
See \cite{hinderer2008} and references therein for a fuller account.

We expect a natural separation of timescales in this system.  The
``fast'' positional variables $\vec{X}$ will change substantially on a
short timescale equal to an orbital period $T_{\text{orb}} \sim M$,
while the ``slow'' orbital parameters $\elqvec$ only change
substantially on the much longer timescale $T_{\text{rad}} \sim
M/\varepsilon$.  Due to the coupling of the $\vec{X}$ and $\elqvec$
equations, both $\vec{X}(t)$ and $\elqvec(t)$ should exhibit
oscillations around a secularly trending central value, but the
oscillations in $\elqvec$ should be $\bigO (\varepsilon)$.  In such a
system, a first-order averaging procedure seeks an approximate and
hopefully more tractable set of equations for the slow variables from
which the dependence on the fast variables (and thus the source of the
small oscillations) is removed \cite{arnold,kevorkian,verhulst}.

Averaging must therefore decouple the $\elqvec$ equations from the
$\vec{X}$ equations in (\ref{eq:total_self-force_eom}) in order to
isolate the secular trend in the former.\footnote{Note that the
  converse is not possible, since the fast variables are coupled to
  the slow ones at zeroth order, where the slow variables appear as
  constant parameters.} One can even adopt the point of view that the
desideratum of a preliminary averaging procedure is to decouple the
equations for the slow and the fast variables from each other as much
as possible.  We represent those averaged, decoupled equations for the
orbital parameters (equivalent to equation (\ref{eq:ELzQ_adiab_ODEs}))
as
\begin{alignat}{1}
\label{eq:secular_ODE}
  \begin{split}
    \D{t}{\elqvec_{\text{secular}}} &= 
    \avg{ \D{t}{\elqvec} }
    \\
    &=
    \avg{ \vec{f} \left (\vec{X}, \elqvec \right) }
    \\
    &= \vec{\mathcal{F}} \left ( \elqvec \right)
    \quad .
  \end{split}
\end{alignat}
Throughout this paper, we will represent averages of all sorts with
angle brackets $\avg{}$ and use subscripts on the brackets to denote
the type of average implied.  Note that in (\ref{eq:secular_ODE}), we
have used $\avg{\cdots}$ to denote an average without yet specifiying
which variables that average is to be taken over.

\subsection{Flux balance and its relationship to averaging}
\label{subsec:flux_balancing}

Although we now have the general form of the adiabatic equations, we
cannot write them explicitly because we still do not know how to
evaluate the self-force.  Mino showed \cite{mino2003,mino2005} that,
under the assumption of non-resonance, the infinite time-averaged
values\footnote{The quantities of physical interest are averages over
  coordinate time $t$.} of the functions $\mathcal{F}_{E}$ and
$\mathcal{F}_{L_{z}}$ equal the sum of the infinite time-averaged
fluxes of the corresponding orbital parameters at radial infinity and
the central black hole horizon in GWs emitted by the
system\footnote{Mino's proof suggests that this equivalence is only
  true for non-resonant geodesics and possibly a small subset of
  resonant geodesics.}.  While there is no conserved $Q$-current to
associate with a GW $Q$-flux, Mino likewise showed that there are
analogous infinite time-averaged quantities at infinity and the
horizon that sum to the infinite time-averaged value of
$\mathcal{F}_{Q}$.  Though not strictly physically accurate, we will
henceforth refer to those quantities as fluxes of $Q$ for ease.
Subsequent work \cite{drasco2005,sago2005,ganz2007} has led to
explicit formulae for these $Q$-fluxes.

Fortunately, we do know how to calculate the aforementioned
time-averaged fluxes at infinity and the horizon via the
computationally mature Teukolsky formalism.  Various Teukolsky-based
(TB) codes can compute the required fluxes from equatorial orbits
\cite{glampedakis2002}, spherical orbits\footnote{Many authors refer
  to orbits of constant radial coordinate $r$ as ``circular'' even
  when they are nonequatorial.  We prefer the term ``spherical'' for
  such orbits (as used in \cite{wilkins1972}) and reserve the term
  circular for constant $r$ equatorial orbits.} (constant $r$)
\cite{hughes2001}, and now generic orbits of arbitrary inclination and
eccentricity \cite{drasco2006,ganz2007,fujita2009}.

These developments have led to the following two-stage implementation
of the adiabatic approximation.  The first stage, usually called the
radiative approximation\footnote{There is some dispute about whether
  the neglected secular effects of the averaged conservative piece of
  the self-force manifest themselves at the same order in the small
  expansion parameter $\mu/M$ as the dissipative pieces
  \cite{pound2005,pound2007:2,pound2007:1,hinderer2008}.  That dispute
  does not concern us here.  Whatever its limitations, the radiative
  approximation, is here to stay for at least the foreseeable future,
  if for no other reason than that it is the only relativistically
  correct approximation to the inspiral motion accessible to numerical
  calculation in the status quo.}, keeps only the lowest-order
contributions from the dissipative self-force
(since the conservative contributions will average to zero).  The
second stage, called the flux-balance method, uses the time-averaged
{\it nonlocal} fluxes (computable) as proxies for the averaged {\it
  local} contributions of the dissipative self-force (not currently
computable).  The RHSs of equation (\ref{eq:secular_ODE}) end up with
nonlocal fluxes inside the brackets and an interpretation of those
brackets as infinite time-averages.

There is a problem, however, with this perscription, which intertwines
two logically distinct facets of the adiabatic approximation to EMRIs:
\begin{enumerate}
\item Is it mathematically appropriate to interpret the angle brackets
  in equation (\ref{eq:secular_ODE}) as a time-average, or is some
  other sort of average required?
\item Given the answer to 1, can we evaluate the RHS of
  (\ref{eq:secular_ODE}), either directly or by finding a numerically
  equivalent proxy?
\end{enumerate}
After all, the fact that we \emph{can} compute a time-averaged proxy
does not imply that we \emph{should} be time averaging in the first
place.

The form of equation (\ref{eq:secular_ODE}) suggests two ways to
average the RHS in order to remove the dependence on the positional
variables: for fixed $\elqvec$, we can either phase space average over
the torus, or we can evaluate the RHS along a specific orbit on that
torus and then average over time.  In Section
\ref{sec:four-analys-summary}, we offer a definitive argument in favor
of torus averaging instead of time averaging.

To arrive at that conclusion, we must first distinguish between torus
functions and time functions.  Torus functions assign a value to every
point on a phase space torus, while time functions assign a value to
points along an individual orbit that are labeled by the value of an
evolution parameter (i.e.\ a time variable).  Our conclusions about
adiabatic averaging will be based on differences in how Fourier
analysis is done on these two domains --- a 2-dimensional compact
position space for the torus-functions and a 1-dimensional noncompact
time axis for the time-functions.  Moreover, numerically accurate flux
calculations require frequency-domain TB codes that separately compute
fluxes from individual Fourier modes, and the aforementioned different
domains also impact the details of the modewise flux calculation.

Before delving into those details, we must mention an important point.
Average values, whether in the torus or time sense, are coordinate
dependent, and in certain applications it matters which coordinates
the average is taken over.  The angle brackets in equation
(\ref{eq:secular_ODE}), for instance, will turn out to denote a torus
average not over $\angvec$ but over a different set of torus
coordinates $\vec{\gamma} \equiv \left( \gamma_{r}, \gamma_{\theta}
\right)$ described in Appendix \ref{subsec:which_time}.  However,
torus averages with respect to $\angvec$ are much easier to compute
than those over $\vec{\gamma}$, in much the same way as Mino time
averages are easier to compute than are coordinate time averages.
Luckily, for every torus function $U(\vec{\gamma})$ and every time
function $u(t)$, we can always construct different functions $V
(\angvec)$ and $v(\lambda)$ such that
\begin{alignat}{1}
\label{eq:torus_and_time_fcn_avgs_for_different_coords}
  \avg{U}_{\vec{\gamma}} &= \avgtor{V} \\
  \avg{u}_{t} &= \avgti{v}
  \quad .
\end{alignat}
The relationship between $U$ and $V$ (or between $u$ and $v$) is
highlighted in Appendix
\ref{sec:mino-time-vs-coord-time-fourier-coeffs}.  Sections
\ref{sec:four-analys-torus-fcns} and \ref{subsec:time_avg_FA} present
the necessary Fourier analysis details.

We will always avail ourselves of the simplication in equation
(\ref{eq:torus_and_time_fcn_avgs_for_different_coords}).  Accordingly,
throughout the rest of the paper, we focus exclusively on torus
averages over $\angvec$ and time averages over Mino time $\lambda$
with the understanding that they may merely be computation-friendly
proxies for averages of different but related functions over different
torus or time coordinates.

\subsection{Torus averaging and Fourier analysis of torus-functions}
\label{sec:four-analys-torus-fcns}

We will call a \termdef{torus function} $f(\angvec; \elqvec) $ any
rule that assigns a complex number to every point on a phase space
torus.  Note that $\elqvec$ specifies both the torus function and the
phase-space torus that serves as its domain.  Usually, we will be
discussing properties of torus functions evaluated at some definite
value of $\elqvec$.  We thus omit the explicit dependence on the
orbital parameters $\elqvec$ for brevity, except where it might lead
to confusion.

We assume that every such torus-function is continuous and
differentiable in all its arguments (including $\elqvec$).  We also
require it to be single-valued on the torus, which implies $2\pi$
periodicity in each of the angle variables:
\begin{equation}
\label{eq:periodicity_of_torus_fcns}
  f(\angr, \angth) =
  f(\angr + 2\pi, \angth) =
  f(\angr, \angth + 2\pi)
  \quad .
\end{equation}

Like any function that is independently periodic in two independent
variables, a torus-function has a double Fourier series representation
\begin{alignat}{1}
\label{eq:spatial_Fourier_expnsn_of_torus_fcn}
  f (\angvec) &=
  \sum_{\n,\kr}A_{\n\kr}
  e^{-i \kr \angr} e^{-i \n \angth}
\end{alignat}
with the $A_{kn}$'s given as usual by
\begin{alignat}{1}
\label{eq:spatial_double_coeffs}
  A_{\n\kr} &=
  \frac{1}{\left(2\pi\right)^{2}}
  \int^{2\pi}_{0}d\angr\,
  \int^{2\pi}_{0}d\angth\,
  f (\angr,\angth)
  e^{+i \kr \angr} e^{+i \n \angth}
  \quad .
\end{alignat}
In order to distinguish them from another set of double-indexed
quantities we introduce later, we will refer to the $A_{\n\kr}$'s as
\termdef{spatial Fourier coefficients} or \termdef{torus Fourier
  coefficients}.

We now define, in the usual way, the following useful quantities.  The
\termdef{torus average of} $f$ is
\begin{alignat}{1}
\label{eq:tor_avg_func}
  \begin{split}
    \avgtor{f(\angvec)} &\equiv
    \frac{1}{\left(2\pi\right)^{2}}
    \int^{2\pi}_{0}d\angr\,
    \int^{2\pi}_{0}d\angth\,
    f (\angr,\angth)
    \\
    &=
    A_{00}
    \quad .
  \end{split}
\end{alignat}
The \termdef{torus averaged Fourier power of $f$} is
\begin{alignat}{1}
\label{eq:torus_avg_Fourier_power_spacedom}
  \pwrtor &\equiv
  \frac{1}{\left(2\pi\right)^{2}}
  \int^{2\pi}_{0}d\angr
  \int^{2\pi}_{0}d\angth
      {\absval{f (\angr, \angth)}}^{2}
      \quad .
\end{alignat}
By Parseval's theorem, the torus-averaged power must also equal
\begin{equation}
  \label{eq:torus_avg_Fourier_power_freqdom}
  \pwrtor = \sum_{\n,\kr} {\absval{A_{\n\kr}}}^{2}
  \quad .
\end{equation}
The \termdef{2D power spectrum of $f$} is the contribution to the
torus-averaged Fourier power from each pair of spatial frequencies or
wavenumbers $(\wvnr, \wvnth)$.  Note that because the period in each
of the $\angr$ and $\angth$ directions is $2\pi$, the corresponding
fundamental spatial frequencies are $\wvnr = \wvnth = 1$, so we see
power only at integer lattice points $(\n,\kr)$ in the 2D wavenumber
space.

All statements above are standard results from the Fourier analysis of
functions on a compact 2D spatial domain.  They apply equally well on
resonant and non-resonant tori.

\subsection{Time averaging and Fourier analysis of time-functions}
\label{subsec:time_avg_FA}

We can evaluate any torus function along a curve
(\ref{eq:chi_of_lambda}) on its associated torus that corresponds to
an orbit. Since each orbit is specified by its initial position
$\offvec$ on the torus, each torus function naturally induces a
2-parameter family of time functions, one for each $\offset$ pair.
Time functions, then, are grouped into 5-parameter families -- 3
parameters $\elqvec$ to specify a torus, and 2 parameters $\offvec$ to
specify an orbit on that torus.

We denote one memeber of such a family as $f \left( \angvec(\lambda);
\offvec; \elqvec \right) $.  We will sometimes omit the explicit
$\offvec$ and $\elqvec$-dependence of a time function and simply write
$f(\lambda)$, again except where clarity would suffer. Throughout this
paper, we adopt the notational convention that a time function and the
torus function from which it is derived are denoted by the same symbol
($f$, in the examples so far).

For time functions, non-resonant and resonant tori must be treated
separately.

\subsubsection{Non-reosnant tori}
\label{sec:four-analys-time-nonres}

When $\omth/\omr$ is irrational, every $\n,\kr$ pair leads to a
distinct frequency
\begin{alignat}{2}
\label{eq:defn_omega_kn}
  &\omkn \equiv \kr \omr +\n \omth  \,,
  \quad & \quad
  \n,\kr \in \mathbb{Z}
  \quad .
\end{alignat}
Such a biperiodic time-function is not periodic: it is bounded on
$(-\infty, \infty)$, but there is no finite time interval over which
the function exactly repeats itself.  Still, every biperiodic function
has a unique discrete Fourier representation \cite{samoilenko}
\begin{alignat}{1}
\label{eq:biperiodic_temp_Fourier_expnsn}
  f (\lambda; \offvec) &=
  \sum_{\n,\kr} A_{\n\kr; \lambda}
  e^{-i (\kr \omr + \n \omth) \lambda}
  \quad .
\end{alignat}
Note that the harmonics are \emph{not} equally spaced in frequency.
The temporal Fourier coefficients $A_{\kr\n; \lambda}$ (which we have
named suggestively) are given by the limit
\begin{alignat}{1}
\label{eq:temp_double_coeffs_APF_defn}
  A_{\n\kr; \lambda} &= \lim_{\Lambda\rightarrow\infty}
  \frac{1}{\Lambda}
  \int^{b + \Lambda/2}_{b - \Lambda/2} d\lambda\,
  f (\lambda; \offvec)
  e^{i ( \kr \omr + \n \omth ) \lambda}
  \quad ,
\end{alignat}
which exists and is independent of $b$ \cite{samoilenko} (henceforth,
we set $b = 0$ for convenience).  Equivalently, we could say that the
Fourier transform of $f(\lambda; \offvec)$ consists of a series
\begin{equation}
\label{eq:FT_of_nonres_time_fcn}
  \widetilde{f}(\om) =
  \sum_{\n,\kr} A_{\n\kr; \lambda} \,
  \delta \left( \om - (\kr \omr + \n \omth) \right)
\end{equation}
of delta-function impulses unequally spaced in frequency.

Paralleling the Fourier discussion of torus functions, we now define
the time-averaged value, time-averaged Fourier power, and the 1D power
spectrum of a time function associated with a non-resonant orbit.
Biperiodic functions offer no single period over which to time-average
in a natural way.  Given the existence\footnote{Technically, the
  logical presentation of Fourier coefficients and time-averages for
  biperiodic (or more general multiperiodic) functions goes in the
  reverse order.  First, the existence of the infinite time-average in
  (\ref{eq:temp_double_coeffs_APF_defn}) is established for a
  biperiodic function $f(\lambda)$.  The existence of the Fourier
  coefficients in (\ref{eq:temp_double_coeffs_APF_defn}) then follows
  from the existence of the average value and the fact that
  $f(\lambda) e^{i (\kr\omr + \n\omth) \lambda}$ is also biperiodic.
  We have chosen this order to parallel the presentations in Sections
  \ref{sec:four-analys-torus-fcns} and
  \ref{sec:four-analys-time-res}.}  of expressions like
(\ref{eq:temp_double_coeffs_APF_defn}), averaging over all time seems
like a sensible choice.  The theory of almost-periodic functions
states that such an infinite time-average indeed exists
\cite{corduneanu}, so we define
\begin{alignat}{1}
\label{eq:defn_nonres_time_avg}
  \avgti{f(\lambda; \offvec)} &=
  \lim_{\Lambda\rightarrow\infty}
  \frac{1}{\Lambda}\int^{\infty}_{-\infty} d\lambda\,
  f (\lambda; \offvec)
  \quad .
\end{alignat}
We will refer to $\avgti{f}$ simply as \emph{the} time-average of $f$
rather than as the infinite time-average value, as it is sometimes
called.  Comparing equations
(\ref{eq:temp_double_coeffs_APF_defn}) and
(\ref{eq:defn_nonres_time_avg}), the time-average equals
\begin{equation}
\label{eq:nonres_time_avg_equals_A00}
  \avgti{f(\lambda; \offvec)} \equiv A_{00; \lambda}
  \quad .
\end{equation}

We define the time-averaged Fourier power of $f(\lambda)$ as a special
case of (\ref{eq:defn_nonres_time_avg}) by
\begin{alignat}{1}
 \label{eq:nonres_time_avg_Fourier_power_defn}
    \pwrti &\equiv
    \lim_{\Lambda\rightarrow\infty}\frac{1}{\Lambda}
    \int^{\frac{\Lambda}{2}}_{-\frac{\Lambda}{2}}d\lambda {\absval{f
        (\lambda; \offvec)}}^{2} 
    \quad .
\end{alignat}
Parseval's theorem also applies to biperiodic time-functions
\cite{corduneanu}, so
\begin{equation}
\label{eq:nonres_time_avg_Fourier_power_Parseval}
  \mathcal{P}_{\ti} =
  \sum_{\n,\kr} {\absval{A_{\n\kr; \lambda}}}^{2}
  \quad .
\end{equation}
That fact allows us to define a 1D power spectrum for $f(\lambda)$ as
the contribution to the time-averaged power from each temporal
frequency $\om$.  The graph of ${\absval{A_{\n\kr; \lambda}}}^{2}$
over the 1D $\om$-space would show power only at the discrete and
unequally spaced set of frequencies (\ref{eq:defn_omega_kn}).

The question now is how to evaluate these time averages in practice.
Though equation (\ref{eq:temp_double_coeffs_APF_defn}) defines the
$A_{\n\kr; \lambda}$'s, such integrals over infinite intervals divided
by infinite quantities do not lend themselves to simple evaluation,
either analytically or numerically\footnote{To evaluate equation
  (\ref{eq:temp_double_coeffs_APF_defn}) numerically, larger and
  larger values of $\Lambda$ would be required before converging to
  some accuracy.  This is computationally impractical because such a
  process will in general converge extremely slowly.  Thus, as the
  size of the integration interval grows, so will the required number
  of evaluations of the integrand, a particularly problematic
  development if the integrand is expensive to calculate.  Moreover,
  the prefactor of $1/\Lambda$ can eventually become so small that
  there is loss of significance in the final answer, thus compromising
  accuracy.  }.

To compute the temporal Fourier coefficients of $f(\lambda; \offvec)$,
we must instead proceed circuitously.  Consider the spatial Fourier
representation (\ref{eq:spatial_Fourier_expnsn_of_torus_fcn}) of the
torus-function $f(\angvec)$ evaluated along the orbit
(\ref{eq:chi_of_lambda}), which yields
\begin{alignat}{1}
\label{eq:temporal_Fourier_expnsn_intermediate}
  f (\lambda; \offr,\offth) &=
  \sum_{\n,\kr} A_{\n\kr}
  e^{-i \kr \offr + \n \offth}
  e^{-i (\kr \omr + \n \omth) \lambda}
  \quad .
\end{alignat}
By uniqueness\footnote{The set of complex exponential functions $e^{-i
    \om \lambda}$ for all $\om$ are a basis for absolutely
  integrable functions on the space $\lambda \in (-\infty, \infty)$.
  $f (\lambda; \offvec)$ inherits absolute integrability from the
  associated torus function $f(\angvec)$, which has a spatial double
  Fourier series representation and thus is absolutely integrable by
  assumption.  Equation (\ref{eq:biperiodic_temp_Fourier_expnsn}) is
  therefore a projection onto the complex exponential basis, and
  projections onto basis sets are unique.} of the Fourier
representation of $f(\lambda; \offvec)$, and comparing equations
(\ref{eq:biperiodic_temp_Fourier_expnsn}) and
(\ref{eq:temporal_Fourier_expnsn_intermediate}), we conclude that the
temporal Fourier coefficients of $f (\lambda; \offvec)$ and the
spatial Fourier coefficients of $f (\angvec)$ are related\footnote{We
  denoted the $\n\kr$th temporal Fourier coefficient by $A_{\n\kr;
    \lambda}$ in anticipation of its close relationship to the
  $\n\kr$th spatial Fourier coefficient $A_{\n\kr}$ of the associated
  torus function and added the $\lambda$ subscript to remind us of
  when we are dealing with spatial vs.\ temporal Fourier
  coefficients.} by
\begin{alignat}{1}
\label{eq:temp_double_coeffs_are_spatials_times_phase}
  A_{\n\kr; \lambda} &\equiv
  A_{\n\kr}e^{-i \kr \offr + \n \offth}
  \quad .
\end{alignat}

We note that each temporal coefficient differs from the corresponding
spatial coefficient in (\ref{eq:spatial_double_coeffs}) only by a
complex phase determined by the initial conditions $\offvec$ of the
orbit.  Consequently, their magnitudes are identical, regardless of
the initial position of the orbit:
\begin{alignat}{1}
\label{eq:temp_double_coeffs_mods_are_spatial_mods}
  \absval{A_{\n\kr}; \lambda} &= \absval{A_{\n\kr}} \,,
  \quad
  \forall \offvec \in \mathbb{T}^{2}_{\elq}
  \quad .
\end{alignat}
This fact is consistent with the ergodic property of these orbits.
Every orbit eventually comes arbitrarily close to every point on the
torus, so shifting initial conditions leads to a new orbit that is
arbitrarily close to a time translation of the original orbit.  And,
of course, time translations only affect the complex phase of temporal
Fourier coefficients.

If we know the torus function $f(\angvec)$, its spatial Fourier
coefficients $A_{\n\kr}$ can be computed by any number of efficient
numerical routines, without any of the difficulties that beset
computation of the $A_{\n\kr; \lambda}$'s via direct evaluation of the
definition (\ref{eq:biperiodic_temp_Fourier_expnsn}).  This fact,
combined with equation
(\ref{eq:temp_double_coeffs_are_spatials_times_phase}), leads to the
only practical recipe for computing the $A_{\n\kr; \lambda}$'s of the
orbit with initial position $\offvec$, namely to compute instead the
$A_{\n\kr}$'s and then use equation
(\ref{eq:temp_double_coeffs_are_spatials_times_phase}).
Ref.\ \cite{drasco2004} introduced just such a technique in the
specific context of functions of Kerr geodesics.

All the other quantities mentioned in this section are likewise
determined from their torus function counterparts.  From equations
(\ref{eq:nonres_time_avg_equals_A00}) and
(\ref{eq:temp_double_coeffs_are_spatials_times_phase}), $A_{00;
  \lambda} = A_{00}$.  We thus conclude that on a non-resonant torus,
the time average of $f(\lambda)$ equals the torus average of its
associated torus-function $f(\angvec)$.  Moreover, since this is true
for every time function on that torus, the time average of such a
function is independent of the initial condition $\offvec$:
\begin{equation}
\label{eq:nonres_toravg_equals_timeavg}
  \avgti{f (\lambda; \offvec)} =
  \avgtor{f (\angvec)} \,,
  \quad
  \forall \offvec \in \mathbb{T}^{2}_{\elq}
  \quad .
\end{equation}
Likewise, equations
(\ref{eq:temp_double_coeffs_mods_are_spatial_mods}) and
(\ref{eq:torus_avg_Fourier_power_freqdom}) imply that, on a
non-resonant torus, the time-averaged Fourier power of $f(\lambda;
\offvec)$ equals the torus-averaged Fourier power of $f(\angvec)$ for
every $\offvec$:
\begin{equation}
\label{eq:nonres_timeavg_power_equals_toravg_power}
  \pwrti \equiv \pwrtor
  \quad
  \forall \offvec \in \mathbb{T}^{2}_{\elq}
  \quad .
\end{equation}
Equations (\ref{eq:torus_avg_Fourier_power_freqdom}),
(\ref{eq:temp_double_coeffs_mods_are_spatial_mods}) and
(\ref{eq:nonres_timeavg_power_equals_toravg_power}) together imply
equation
(\ref{eq:nonres_time_avg_Fourier_power_Parseval})\footnote{This proves
  Parseval's theorem for biperiodic functions.}.  By extension, the 1D
power spectrum of $f(\lambda)$ can be derived from the 2D power
spectrum of $f(\angvec)$ by mapping wavenumber pairs to frequencies
using eq. (\ref{eq:defn_omega_kn}).

All of the above relationships between torus-function quantities and
those of any biperiodic time function induced via
(\ref{eq:temporal_Fourier_expnsn_intermediate}) are well-established
and well-known in the literature on almost-periodic functions
\cite{samoilenko,corduneanu} and on integrable Hamiltonian systems
\cite{arnold}.  Many of these facts, however, are used but not so
clearly delineated in this way in the literature relating to EMRI
calculations.  We have gone through the trouble of including them here
not only for completeness and clarity but also to emphasize that we
can only execute the above recipes if we know the corresponding torus
function $f(\angvec)$.

This leads us to a crucial observation.  If all we know is
$f(\lambda)$, either as some closed-form expression in terms of
$\lambda$ or as a numerical time-series, there is no practical scheme
for computing its temporal Fourier coefficients $A_{\n\kr; \lambda}$,
even though those coefficients are perfectly well-defined.  In
addition to the initial conditions $\offvec$ associated with
$f(\lambda)$, we \emph{must} also know the torus-function $f(\angvec)$
(or at least its spatial Fourier coefficients $A_{\n\kr}$) in order to
compute the $A_{\n\kr; \lambda}$'s.  We summarize the implications of
this fact for flux balancing in section \ref{sec:four-analys-summary}.

All of the equivalences noted between torus-function quantities and
their time-function counterparts followed from the assumption of
non-resonance.  On resonant tori, all of these equivalences break
down, as we now show.

\subsubsection{Resonant tori}
\label{sec:four-analys-time-res}

Unlike time functions evaluated along non-resonant orbits, time
functions on resonant orbits are singly periodic, with (possibly very
long) period $\Tnot$ and corresponding fundamental frequency
$\omnot=2\pi/\Tnot$.  The single periodicity of time functions of
resonant orbits means that all frequency-domain quantities have
straightforward and familiar definitions.

Any time function evaluated on a resonant orbit has a Fourier series
representation
\begin{alignat}{1}
\label{eq:temporal_Fourier_expnsn_resonant}
  f \left( \lambda; \offvec  \right) &=
  \sum_{\jo} C_{\jo;\lambda}
  e^{-i \jo \omnot \lambda}
\end{alignat}
whose coefficients\footnote{Periodicity of $f(\lambda)$ implies that
  the integral in equation (\ref{eq:single_coeff_time_series}) has the
  same value taken over any interval of length $\Tnot$.  We choose the
  symmetric interval $[-\Tnot/2, \Tnot/2]$ solely for aesthetic
  reasons.} are single-index objects
\begin{alignat}{1}
\label{eq:single_coeff_time_series}
  C_{\jo;\lambda}  &=
  \frac{1}{\Tnot} \int_{-\Tnot / 2}^{\Tnot / 2} d\lambda \,
  f \left(\lambda; \offvec   \right)
  e^{+i \jo \omnot \lambda}
  \quad .
\end{alignat}
Like the $A_{\n\kr; \lambda}$'s, each $C_{\jo;\lambda}$ varies with
the initial condition $\offvec$.  Unlike the $A_{\n\kr; \lambda}$'s,
the Fourier transform of $f(\lambda; \offvec)$ is a sequence
\begin{equation}
\label{eq:FT_of_res_time_fcn}
  \widetilde{f}(\om) =
  \sum_{\jo} C_{\jo; \lambda} \,
  \delta \left( \om - \jo \omnot \right)
\end{equation}
of equally spaced delta-function impulses in frequency space.

In the resonant case, we define the time average of $f(\lambda)$ 
straightforwardly as
\begin{equation}
\label{eq:defn_time_avg_res}
  \avgti{f(\lambda; \offvec)} \equiv 
  \frac{1}{\Tnot}
  \int^{\frac{\Tnot}{2}}_{-\frac{\Tnot}{2}} d\lambda\,
  f (\lambda; \offvec)
= C_{0; \lambda}
  \quad .
\end{equation}
Likewise, we define the time-averaged power as
\begin{alignat}{1}
\label{eq:time_avg_Fourier_power_res}
  \pwrti (\offvec) &\equiv
  \frac{1}{\Tnot}
  \int^{\frac{\Tnot}{2}}_{-\frac{\Tnot}{2}} d\lambda\,
  \absval{ f(\lambda; \offvec) }^{2} 
  \quad .
\end{alignat}
By Parseval's theorem, the time-averaged power is also given by
\begin{equation}
\label{eq:parseval_for_Cjs_res}
  \pwrti (\offvec) =
  \sum_{\jo} \absval{ C_{\jo;\lambda} }^{2}
  \quad .
\end{equation}

To flush out how the $C_{\jo; \lambda}$'s relate to the spatial
$A_{\n\kr}$'s and to the initial condition $\offvec$, we begin, as in
the non-resonant case, by inducing a time function
(\ref{eq:temporal_Fourier_expnsn_intermediate}) from a torus function.
In the resonant case, the frequency ratio $\omth/\omr$ is a rational
number $p/z$, where $p$ and $z$ are relatively prime and $p > z$.  (In
terms of integers in the definition $q_{r\theta}=w +\frac{v}{z}$,
$p=(w+1)z+v$.)  The individual $r$ and $\theta$ frequencies and
periods are therefore related to the fundamental frequency and total
period of the periodic orbit by
\begin{subequations}
\begin{alignat}{1}
\label{eq:relate_omnot_to_omr_and_omth}
  \omr &= z \omnot \\
  \omth &= p \omnot
\end{alignat}
\end{subequations}
and
\begin{subequations}
\begin{alignat}{1}
\label{eq:relate_Tnot_to_Tr_and_Tth}
  \Tr &= \frac{\Tnot}{z} \\
  \Tth &= \frac{\Tnot}{p}
  \quad .
\end{alignat}
\end{subequations}
As a result, all $\n\kr$ combinations for which
\begin{equation}
\label{eq:kzplusnp_with_same_j}
  \kr \den + \n \num = \jo
\end{equation}
lead to identical frequencies
\begin{equation}
\label{eq:kn_freqs_equal_to_j_omnot}
  \kr \omr + \n \omth =
  \kr \den \omnot + \n \num \omnot =
  \jo \omnot
\end{equation}
in the arguments of the exponential functions on the RHS of equation
(\ref{eq:temporal_Fourier_expnsn_intermediate}).

The selection rule (\ref{eq:kzplusnp_with_same_j}) maps every $\n\kr$
pair to some $\jo$.  By the uniqueness of Fourier represenations, we
conclude from equations
(\ref{eq:temporal_Fourier_expnsn_intermediate}) and
(\ref{eq:temporal_Fourier_expnsn_resonant}) that
\begin{equation}
\label{eq:temp_single_coeff_is_sum_of_phased_spatials}
  C_{\jo; \lambda} (\offvec) =
  \sum_{\substack{\n,\kr:\\ \kr\den+\n\num=\jo}}
  A_{\n\kr} e^{-i (\kr \offr + \n \offth)}
  \quad .
\end{equation}
Note that equation
(\ref{eq:temp_single_coeff_is_sum_of_phased_spatials}) is really only
a summation over a single index since the value of $\n$ in any term is
determined by the value of $\kr$ and the (fixed) value of $\jo$.

It is tempting to rewrite each term on the RHS of equation
(\ref{eq:temp_single_coeff_is_sum_of_phased_spatials}) as $A_{\n\kr;
  \lambda}$, mimicking the notation for the non-resonant temporal
Fourier coefficients.  We refrain from doing so because we seek a
clear distinction between spatial and temporal Fourier coefficients,
and temporal double-index coefficients are not defined in the resonant
case \cite{schilder2006,corduneanu}.  Fourier representations are
unique, so the familiar single-index representation
(\ref{eq:temporal_Fourier_expnsn_resonant}) is the only such
projection of $f(\lambda)$ onto a set of mutually orthogonal basis
functions.  If we were to write an expression like
(\ref{eq:biperiodic_temp_Fourier_expnsn}) on a resonant orbit, the
different harmonics on the RHS would not all be orthogonal, and we
would not have a bona fide Fourier expansion in hand until we
collapsed all terms corresponding to the same frequency into a single
term.

\begin{figure}[t]
  \centering
  \includegraphics[width=90mm]{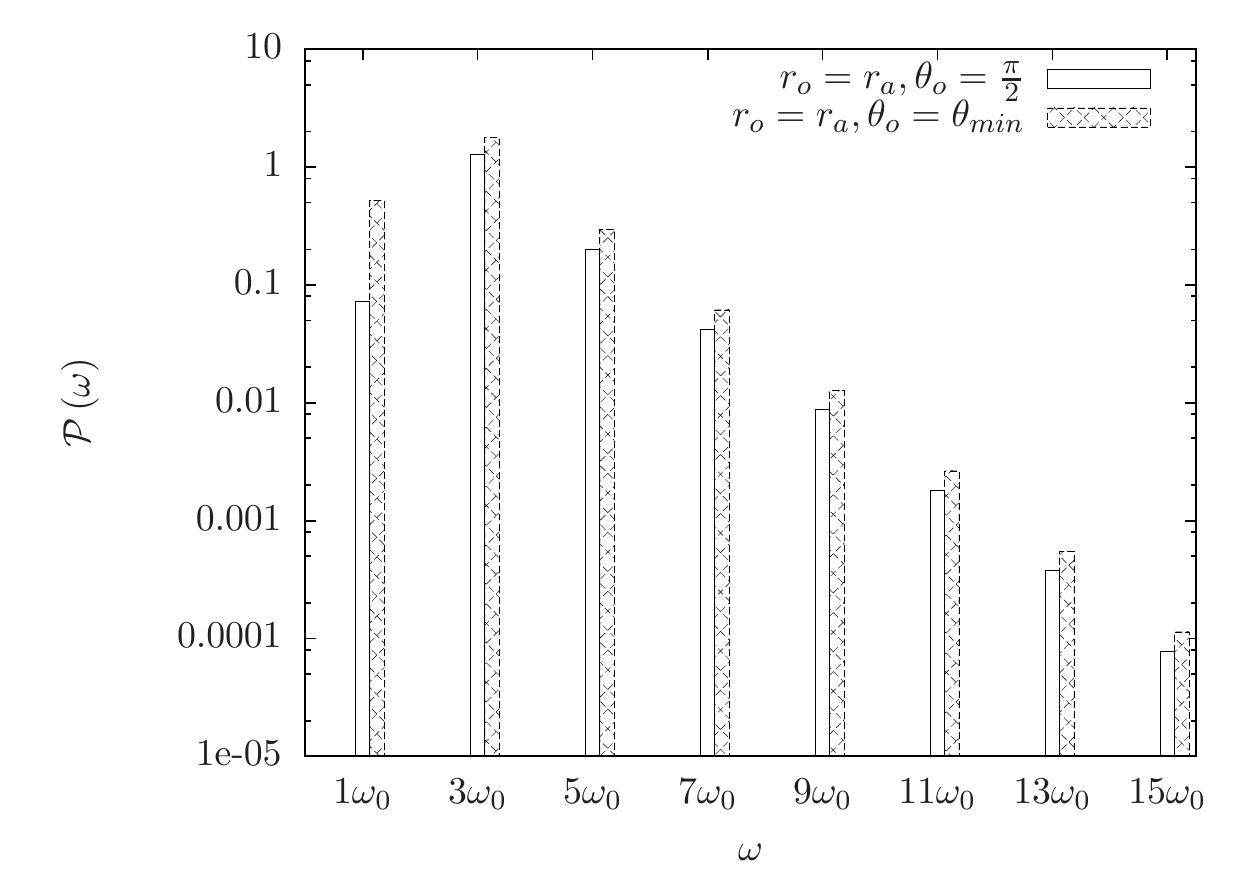}
  \includegraphics[width=90mm]{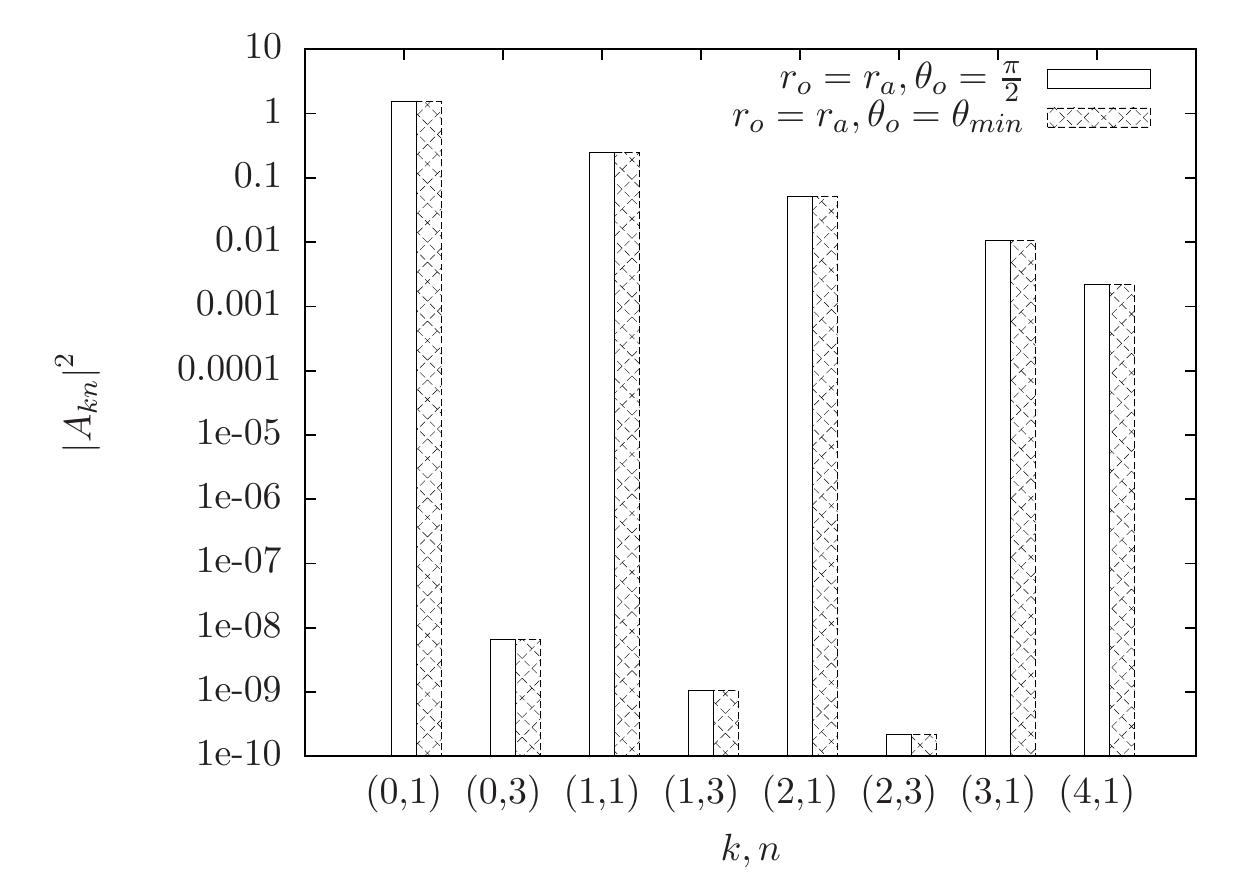}
  \hfill
  \caption{Top: The Fourier power spectrum of the function
    $r\cos{\theta}$ for a $q_{r\theta}=\frac{1}{2}$ periodic orbit for
    two different sets of initial coordinates, $\left(r_{0}=r_{a},
    \theta_{0}=\frac{\pi}{2}\right)$ and $\left(r_{0}=r_{a},
    \theta_{0}=\theta_{min}\right)$, but the same sets of orbital
    parameters, $a=0.9$, $E=0.954788$, $L_{z}=2.65115$ and
    $Q=0.944969$. Bottom: The magnitudes of some spatial Fourier
    coefficients for the same orbits in the top panel.}
  \label{fig:periodic}
\end{figure}

Equation (\ref{eq:temp_double_coeffs_are_spatials_times_phase})
implied that, on non-resonant orbits, several quantities one can
compute for a time-function $f(\lambda; \offvec)$ turn out to be
independent of $\offvec$: the magnitudes of its Fourier coefficients,
its time-averaged value, its time-averaged Fourier power, and its
power spectrum.  In contrast, equation
(\ref{eq:temp_single_coeff_is_sum_of_phased_spatials}) implies that,
on resonant orbits, each of those quantities \emph{does} depend on the
initial condition $\offvec$.  Each $C_{\jo; \lambda}$ is a sum of
spatial $A_{\n\kr}$'s with $\offvec$-dependent phases rather than just
one such term (cf.\ equation
(\ref{eq:temp_double_coeffs_are_spatials_times_phase})), so both the
magnitudes
\begin{equation}
\label{eq:temp_sing_coeff_modulus}
  \absval{C_{\jo; \lambda}} (\offvec) =
  \absval{
    \sum_{\substack{\n,\kr:\\ \kr\den+\n\num=\jo}}
    A_{\n\kr} e^{-i (\kr \offr + \n \offth)}
  }
\end{equation}
and time-averaged value
\begin{alignat}{1}
\label{eq:timeavg_of_f_depends_on_ICs}
  \begin{split}
    \avgti{f(\lambda; \offvec)} &=
    C_{0; \lambda} (\offvec)
    \\
    &=
    \sum_{\substack{\n,\kr:\\ \kr\den+\n\num=0}}
    A_{\n\kr} e^{-i (\kr \offr + \n \offth)}
    \\
    &=
    A_{00} +
    \sum_{\substack{\kr \neq 0,\n \neq 0:\\ \kr\den+\n\num=0}}
    A_{\n\kr} e^{-i (\kr \offr + \n \offth)}
  \end{split}
\end{alignat}
retain $\offvec$-dependence.  The squared magnitudes
\begin{widetext}
\begin{eqnarray}
\label{eq:c_j_mod}
\left|C_{j}\left(\chi_{r_o},\chi_{\theta_o} \right) \right|^{2} &=&
\sum_{\substack{kn:\\\jo=\kr\den+\n\num}}A_{kn}e^{-i\left(\kr\chi_{r_o}+\n\chi_{\theta_o}
  \right)}\sum_{\substack{k'n':\\\jo=\kr'\den+\n'\num}}A^{*}_{k'n'}e^{i\left(\kr'\chi_{r_o}+\n'\chi_{\theta_o}
  \right)}
\\
\nonumber
&=& \sum_{\substack{k=k',n=n':\\\jo=\kr\den+\n\num}}\left|A_{kn}
\right|^{2} + \sum_{\substack{k\neq k', n\neq n':
    \\ \jo=\kr\den+\n\num, \\ \jo=\kr'\den+\n'\num}}
A_{kn}A^{*}_{k'n'}e^{-i\left(\kr-\kr'\right)\chi_{r_{o}}}e^{-i\left(\n-\n' \right)\chi_{\theta_o}}
\quad .
\end{eqnarray}
\end{widetext}
also depend on $\offvec$ through cross terms, and the time-averaged
power and power spectra inherit this dependence via
(\ref{eq:parseval_for_Cjs_res}).  Figure \ref{fig:periodic}
illustrates this point for the test function $r\cos{\theta}$.

Via its $\offvec$-dependence, equation
(\ref{eq:timeavg_of_f_depends_on_ICs}) defines a torus function in the
variables $\offr, \offth$.  Complex exponentials have a zero average
value, so averaging that torus function over all $\offr, \offth$ kills
every term in the summation on the RHS of
(\ref{eq:timeavg_of_f_depends_on_ICs}), leaving only $A_{00}$.  But
$A_{00}$ is the torus averaged value of the associated torus function
$f (\angvec)$.  We conclude that the torus-average over all initial
conditions of the time average of a time function equals the
torus-average of the underlying torus-function.  An identical argument
applies if we torus average the squared-magnitudes (\ref{eq:c_j_mod})
of the coefficients over all $\offvec$ and, by extension, if we
likewise torus-average the time-averaged power
(\ref{eq:time_avg_Fourier_power_res}).

The upshot is that the parallels between torus functions and time
functions obtained in the non-resonant case break down in the resonant
case:
\begin{alignat}{1}
  \avgti{f (\lambda; \offvec)} &\neq
  \avgtor{f (\angr, \angth)} 
  \label{eq:res_toravg_notequal_timeavg}
  \\
  \pwrti (\offvec) &\neq \pwrtor
  \label{eq:pwr_toravg_notequal_pwr_timeavg}
  \quad .
\end{alignat}
However, torus averages over initial conditions and torus averages of
time averaged \emph{are} equal for both a function $f$ and its Fourier
power:
\begin{alignat}{1}
  \avgtor{\avgti{f (\lambda; \offvec)}} &=
  \avgtor{f (\angr, \angth)} 
  \label{eq:res_toravg_equal_toravg_of_timeavg}
  \\
  \avgtor{\pwrti (\offvec)} &= \pwrtor
  \label{eq:pwr_toravg_equal_toravg_of_pwr_timeavg}
  \quad .
\end{alignat}

To clarify, equations (\ref{eq:res_toravg_notequal_timeavg}) and
(\ref{eq:pwr_toravg_notequal_pwr_timeavg}) state that the time average
of a time function and the torus average of its associated
torus-function are not \emph{identically equal} as they are in the
non-resonant case.  That does not, of course, preclude the possibility
that the two could be \emph{circumstantially equal} for some
particular choice of initial condition $\offvec$.  In fact, for
real-valued functions $f$, the mean-value theorem guarantees that
$\avgti{f(\lambda; \offvec)} = \avgtor{f(\angvec)}$ for at least one
$\offvecmvt{} \in \torelq$.  In general, $f$ will be complex-valued,
and we have no such guarantee.  The time-averaged Fourier power,
however, is strictly real, so there is at least one $\offvecmvt{}$
such that $\pwrti(\offvecmvt{}) = \pwrtor$.  We explore some
implications of this fact for adiabatic EMRI calculations in Section
\ref{sec:spec}.

\subsection{And the winner is\ldots torus averaging}
\label{sec:four-analys-summary}

To summarize, time averaging is equivalent to torus averaging for
non-resonant orbits. Furthermore, torus averaging is the only
practical recipe for computing Fourier coefficients and so torus
averaging is the explicit computation
instituted in practice. 

However, time averaging is {\it inequivalent} to torus averaging for
resonant orbits. Thus torus averaging wins out for two reasons.
First, torus averaging along a resonant torus crucially washes away
any $\offvec$ positional dependence, while time averaging does not.
The $\offvec$-dependence violates the spirit of averaging, namely to
remove all dependence on the fast variables.  Second, and more
seriously, though the time-averaged equations are continuous in
$\offvec$ for fixed $\elqvec$, they are in general discontinuous in
$\elqvec$ for fixed $\offvec$.  The situation will resemble that of
Thomae's modified Dirichlet function
\begin{equation}
\label{eq:modified_Dirichlet_fcn}
  D_{M}(x) =
  \begin{cases}
    0 & \text{if $x$ is irrational} \\
    \frac{1}{z} & \text{if $x = p/z$, with $p$ and $z$ coprime} \\
    1 & \text{if $x = 0$}
  \end{cases}
  \quad ,
\end{equation}
which is continuous at the irrationals, discontinuous on the
rationals, and nowhere differentiable\footnote{Such a function
  certainly seems unphysical.  It violates the hypotheses of
  continuity and differentiability in all arguments required by the
  theorems bounding the error in a solution to time-averaged equations
  with almost-periodic dependence on time \cite{verhulst}.  More
  pathologically, it violates the hypotheses for the well-posedness of
  an initial value problem and for the existence of solutions to
  systems of ODEs \cite{Boyce}.  Such a function would not be Riemann
  integrable, and it is unclear whether a system of such functions
  would even be Lebesgue integrable.}  \cite{beanland}.  Pragmatically
speaking, even if a set of ODEs with such pathologically discontinuous
and non-differentiable equations had a solution, it is unclear how one
would numerically integrate them.  Furthermore, the continuity
furnished by torus-averaged fluxes is absolutely essential for the
proper construction of a grid through which adiabatic trajectories are
to be interpolated, as discussed in the introduction.

In short, torus-averaged equations are well-behaved, while time-averaged
equations lose the continuity and differentiability that guarantee the
resulting equations are well-posed and have unique solutions, the very
basis of every standard numerical integration scheme.  

The arguments made in favor of torus averaging apply to the radiative
approximation, based on an average of the dissipative piece of the
\emph{local} self-force on the inspiraling particle.  But when it
comes to flux-balance as a specific implementation of the radiative
approximation, this now leaves a logical gap.  As acknowledged in
\cite{mino2003,mino2005}, the flux-balance arguments that allow the
nonlocal fluxes of conserved quantities to be used as proxies for the
local dissipative self-force have been derived on a time-averaged
basis and under the assumption of non-resonance.  Since time and torus
averages agree for non-resonant orbits/tori, the time-averaged
nonlocal fluxes are still good proxies for the torus-averaged local
dissipative self-force in the non-resonant case.

What has never been made explicit is whether flux-balance is also
valid in the resonant case, on either a time averaged or
torus-averaged basis.  We resolve this issue now: time averaged
flux-balancing may not be true on resonant orbits in general, but that
will be irrelevant since it will be true on a torus-averaged basis.
Mino showed that, under the assumption of non-resonance, the
time-averaged fluxes of $\elqvec$ at infinity and the horizon furnish
proxies for the time averaged RHSs of equations
(\ref{eq:secular_ODE}).  But then, by the arguments we have heavily
exploited, the corresponding torus-averaged versions must also be
equal.  And since torus averages are insensitive to the resonance or
non-resonance of the underlying torus {\it and are continuous in
  $\elqvec$}, the flux-balance prescription is valid in a
torus-averaged sense for all orbits. If the torus-averaged fluxes were
not good proxies for the torus-averaged local equations only at
resonances, a discrete set of measure zero, then they could not be
continuous in $\elqvec$.  But we have shown torus averages are
everywhere continuous. {\it De facto}, then, Mino's argument
establishes the validity of torus-averaged flux-balancing generally.

It would thus seem that both flux-balancing as a general procedure and
its specific implementation in a frequency-domain application of the
Teukolsky formalism treat non-resonant and resonant tori equally, as
stated in \cite{hinderer2008}.  Flux-balancing is, in fact, thusly
impartial, but interestingly, the Teukolsky formalism is not.  As we
will show below, a TB torus-averaged flux calculation can achieve
computational savings of an order of magnitude or more on low-order
resonant tori that are simply not available on non-resonant tori.
These efficiencies follow from a simple observation about the Fourier
integrals a TB code must evaluate and are independent of the specific
implementation in code of the Teukolsky formalism.  Thus, rather than
disfavoring resonances, as has been commonly assumed, the Teukolsky
formalism actually shows favoritism for resonances, and properly
leveraged, that favoritism can substantially accelerate adiabatic
inspiral calculations.

\section{Computational savings along resonances}
\label{sec:definite}

We now explain computational efficiences that exploit resonant
tori. Although specific to the Teukolsky formalism, the computational
expedience can be understood without all details of that formalism.
We simply assert some features and formulae from a TB flux calculation
that we require to make our argument.  For reference, Appendix
\ref{sec:tuek_formalism} gives a somewhat more detailed overview of
the Teukolsky formalism and offers at least skeletal derviations of
the formulae listed below.  For a fuller treatment of the Teukolsky
formalism, which is beyond the scope of this work, we direct the
reader to the references listed in Appendix \ref{sec:tuek_formalism}.

As a reminder, our argument is specific to \emph{frequency-domain}
Teukolsky calculations and corresponding codes. In
the context of the EMRI problem, such codes compute a combined
multipole and Fourier decomposition of the metric perturbations at
infinity and the black hole horizon due to a geodesic source.

\subsection{The fluxes of $\elqall$}
\label{sec:fluxes-elqall}

The fluxes of the conserved quantities $\elqvec$ are usually reported
as quantities averaged over \emph{coordinate} time $t$ on non-resonant
orbits\footnote{See Appendix \ref{sec:using_TSN} for the time-averaged
  fluxes from resonant orbits.}.  By the arguments of Section
\ref{sec:four-analys-time-nonres} and Appendix
\ref{subsec:which_time}, this is equivalent to the torus-averages of
the fluxes on all tori over the torus coordinates $\vec{\gamma}$.  We
therefore report those same expressions here as torus-averaged fluxes.
For $E$ and $L_{z}$, those expressions are
\cite{drasco2006,drasco2005}
\begin{subequations}
\label{eq:torus_avgd_TB_fluxes}
\begin{alignat}{1}
  \left\langle{\D{t}{E}}\right\rangle^{H/\infty}_{\vec{\gamma}} &=
  \sum_{lmkn}
  \frac{\alpha_{lmkn}^{\infty/H}}{4 \pi \coorom_{mkn}^{2}}
  \absval{ Z^{\infty/H}_{lmkn} }^{2}
  \label{eq:torus_avgd_TB_E_flux}
  \\
  \left\langle{\D{t}{L_{z}}}\right\rangle^{H/\infty}_{\vec{\gamma}} &=
  \sum_{lmkn}
  \frac{\alpha_{lmkn}^{\infty/H} m}{4 \pi \coorom_{mkn}^{3}}
  \absval{ Z^{\infty/H}_{lmkn} }^{2}
  \label{eq:torus_avgd_TB_Lz_flux}
  \quad .
\end{alignat}
Based on Mino's argument in \cite{mino2003},
Refs.\ \cite{Sago:2005gd,sago2005} worked out the corresponding
expression for the time-averaged $Q$ flux for non-resonant orbits,
which we also report as the torus-averaged flux
\begin{alignat}{1}
\label{eq:torus_avgd_TB_Q_flux}
  \begin{split}
    \left\langle\frac{dQ}{dt}\right\rangle_{\vec{\gamma}}^{H/\infty} &=
    - 2\left\langle a^2 E\cos^{2}{\theta}\right\rangle_{\vec{\gamma}}
    \left\langle\frac{dE}{dt}\right\rangle_{\vec{\gamma}}^{H/\infty}
    \\
    &{}
    +2\left\langle\cot^2{\theta}\right\rangle_{\vec{\gamma}}
    \left\langle\frac{dL_z}{dt}\right\rangle_{\vec{\gamma}}^{H/\infty}
    \\
    &{}
    -\sum_{lmkn}
    \frac{\n\coorom_{\theta}}{2\pi\coorom_{mkn}^{3}}
    \alpha_{lmkn}^{\infty/H}
    \left|Z^{\infty/H}_{lmkn}\right|^{2}
    \quad .
  \end{split}
\end{alignat}
\end{subequations}
The prefactors in the first two lines of
(\ref{eq:torus_avgd_TB_Q_flux}) are computed only once for the entire
torus.  Thus, substituting equations (\ref{eq:torus_avgd_TB_E_flux})
and (\ref{eq:torus_avgd_TB_Lz_flux}) into the RHS of
(\ref{eq:torus_avgd_TB_Q_flux}) and combining like terms with those in
the summation of the last line, the flux for $Q$ has the same general
form as the fluxes of $E$ and $L_{z}$ have.  Our savings arguments
will be based on that form, so although we will speak about $E$ and
$L_{z}$ for concreteness, those arguments will apply to $Q$ as well.
Appendices \ref{sec:tuek_formalism} and \ref{sec:using_TSN} summarize
the derivations of these expressions.

Before proceeding with those arguments, we clarify the notation in
equations (\ref{eq:torus_avgd_TB_fluxes}).  First, the apparent
discrepancy between the ordering of the $H/\infty$ superscripts on the
left- and righthand sides of the equations is not a typographical
error.  On the LHS, the superscript denotes fluxes at the black hole
horizon and radial infinity, respectively.  The somewhat backward
notational choice to have the fluxes at infinity depend on a quantity
labeled $Z^{H}_{lmkn}$ and the horizon fluxes on $Z^{\infty}_{lmkn}$
is, at this point, ingrained in the literature.  To maintain a modicum
of notational uniformity, we have labeled the weighting factors
$\alpha^{H/\infty}$ with the same backward superscript convention.
The exact form of those weighting factors will not concern us.  What
matters for our purposes is that every factor $\alpha_{lmkn}^{H}$ for
the fluxes at infinity is equal to $1$ and that every factor
$\alpha_{lmkn}^{\infty}$ for the fluxes at the horizon is real and
depends on $\n,\kr$ only through $\coorom_{\n\kr}$.  All the arguments
to follow apply equally to fluxes at infinity and at the horizon.  We
borrow the notation $\Hinf$ from Ref.\ \cite{drasco2006} to denote
either of $H/\infty$.

Continuing, the indices $l,m$ are standard multipole
indices\footnote{The values $l=0,1$ are not relevant in GW
  calculations, for which the lowest non-vanishing moment is the $l=2$
  quadrupole.}, with $l \geq 2, -l \leq m \leq l$.  Our argument will
focus on the Fourier analysis of each $l,m$ term individually, so
that, unless explicitly stated otherwise, $l,m$ are taken to be fixed
everywhere in this section, while $\n,\kr$ each run from $-\infty$ to
$\infty$.  The frequencies
\begin{equation}
\label{eq:omega_mkn}
  \coorom_{mkn} \equiv m\coorom_{\varphi} + \coorom_{\n\kr} = m\coorom_{\varphi} +
  \kr\coorom_r + \n\coorom_{\theta}
\end{equation}
are the combined harmonics $\coorom_{\n\kr}$ of the $r$ and $\theta$
fundamental frequencies (the coordinate time version of equation
(\ref{eq:defn_omega_kn})) and the fundamental azimuthal frequency
$\coorom_{\varphi}$.  Note that the integer $m$ is both a multipole
index and the relative contribution of $\coorom_{\varphi}$ to each
frequency $\coorom_{m\n\kr}$.  Other than attaching itself as a label
to frequencies in this way, however, $m$ will not appear as a Fourier
index in any sense below.

Finally, Appendix \ref{sec:using_TSN} explains why we have written the
fluxes as average values over the $\vec{\gamma}$ torus coordinates
mentioned in Section \ref{subsec:flux_balancing} and in Appendix
\ref{subsec:which_time}.  We note here simply that if we seek
adiabatic solutions in the form $\elqvec (t)$ (as opposed to $\elqvec
(\lambda)$), then the angle brackets in (\ref{eq:secular_ODE}) should
also be averages over $\vec{\gamma}$, so that
(\ref{eq:torus_avgd_TB_fluxes}) have the correct form to be proxies
for $\vec{\mathcal{F}} (\elqvec)$.  The representations of the LHSs of
the flux equations as averages over $\vec{\gamma}$ is otherwise
irrelevant, since in light of
(\ref{eq:torus_and_time_fcn_avgs_for_different_coords}), we will
always seek equivalent and easier to compute $\angvec$-averaged
quantities.

\subsection{$Z^{\Hinf}_{lm\n\kr}$ as Fourier coefficients
  of a torus function}
\label{sec:zhinf_lmnkr-as-fourier-torus-coeffs}

With these preliminaries out of the way, we are ready to list the
features of the RHSs of (\ref{eq:torus_avgd_TB_fluxes}) that we will
need for our savings arguments both in this section and in Section
\ref{sec:spec}.  For our principal argument, what matters is that for
fixed $l,m$ values, each $Z^{\Hinf}_{lm\n\kr}$ takes the form of a
Fourier coefficient of some torus function,
\begin{multline}
\label{eq:Zkn_as_torus_fcn_Fourier_coeff}
  Z_{lm\n\kr}^{\Hinf} =
  \\
  \int^{2\pi}_{0} d\angr\,
  \int^{2\pi}_{0} d\angth\,
  e^{i\left(\kr\angr +\n\angth \right)}
  f^{\Hinf}_{lm; \coorom = \coorom_{mkn}} \angset
  \quad .
\end{multline}
This form of the $Z^{\Hinf}_{lm\n\kr}$'s associated with a geodesic
source of arbitrary eccentricity and inclination is detailed in
several references (see, for instance,
\cite{drasco2005,drasco2006,drasco2004, Sago:2005gd, sago2005,
  ganz2007}) and summarized in Appendix \ref{sec:tuek_formalism}.

Equation (\ref{eq:Zkn_as_torus_fcn_Fourier_coeff}) parallels the form
of equation (\ref{eq:spatial_double_coeffs}) from Section
\ref{sec:four-analys-torus-fcns}, but there is one critical
difference.  For fixed $l,m$, the function $f^{\Hinf}_{lm; \coorom}
\angset$ further depends on a continuous parameter $\coorom$ that must
be set to $\coorommkn$ when evaluating $Z^{\Hinf}_{lm\n\kr}$ for a
given multipole mode.  Postponing for the moment any details of the
function $f^{\Hinf}_{lm; \coorom} \angset$ or its derivation, we
remark that this dependence on the coordinate time harmonic
frequencies of the source as an external parameter persists despite
the fact that equation (\ref{eq:Zkn_as_torus_fcn_Fourier_coeff}) is a
spatial Fourier integral.

Thus, for fixed $l,m$, the $Z^{\Hinf}_{lm\n\kr}$ are not the Fourier
coefficients of a single function but rather isolated Fourier
coefficients of several different functions\footnote{This is part of
  the reason why we cannot compute all the $Z^{\Hinf}_{lm\n\kr}$
  coefficients for a given $l,m$ at once with, for instance, a
  2-dimensional Fast Fourier Transform (FFT).}.  On a non-resonant
torus, every $\n,\kr$ pair leads to a different value of
$\coorom_{mkn}$, and every coefficient computed has a distinct
function $f^{\Hinf}_{lm; \coorom} \angset$ in the integrand.  On a
resonant torus with associated frequency ratio $\coorom_{\theta} /
\coorom_r =\omth/\omr = \num/\den$, all $\n,\kr$ pairs that satisfy
the selection rule (\ref{eq:kzplusnp_with_same_j}) for the same $\jo$
lead to identical values of $\ommkn$, and some coefficients with
different values of $\n,\kr$ will share the same integrand function
$f^{\Hinf}_{lm; \coorom} \angset$.  The practical implications of this
asymmetry for a TB flux calculation constitute the basis of our
savings argument.

In anticipation of later arguments, we also note that for each fixed
value of the pair $l,m$, the resulting doubly infinite sum over
$\n,\kr$ in (\ref{eq:torus_avgd_TB_fluxes}) has the appearance of a
torus-averaged Fourier power in the sense of Section
\ref{sec:four-analys-torus-fcns} with the identification
\begin{alignat}{1}
\label{eq:fluxes_as_toravgd_Fourier_powers}
  \begin{split}
    \absval{A_{\n\kr}}^{2} &=
    \text{prefactor } \times 
    \absval{Z^{\Hinf}_{lmkn}}^{2}
    \\
    A_{\n\kr} &=
    \sqrt{\text{prefactor}} \times 
    Z^{\Hinf}_{lmkn}
    \quad .
  \end{split}
\end{alignat} 
The prefactors in front of $\absval{Z^{\Hinf}_{lmkn}}^{2}$ turn out to
be real-valued and non-negative for all values of the indices, so it
is valid to subsume them into some new coefficients $A_{kn}$.

\subsection{Recycling computations between Fourier modes}
\label{sec:recycl-comp-between-modes}

The complex-valued quantities $Z^{\Hinf}_{lm\n\kr}$ are the backbone
of a frequency-domain Teukolsky calculation, and a code that
implements such a calculation spends by far the lion's share of its
CPU budget on computing them.  To explain how resonances can be
leveraged to optimize that budget, we must look a bit more closely at
the integrand functions $f^{\Hinf}_{lm; \coorom} \angset$.

The main ingredients in $f^{\Hinf}_{lm; \coorom} \angset$ are two
separate functions that have the same sort of $\coorom$ dependence
described above: a radial Teukolsky function $R^{\Hinf}_{lm;\coorom}
(r)$ and a spin-weighted spheroidal harmonic
${}_{-2}S^{a\coorom}_{lm}\left(\theta\right)$.  We imagine
re-expressing the former as a torus function of $\angr$ alone and the
latter as a torus function of $\angth$ alone but will continue to
write them as functions of $r$ and $\theta$, as they are in the rest
of the literature.  $f^{\Hinf}_{lm; \coorom} \angset$ consists of a
somewhat messy assortment of terms and factors involving these two
functions, several of their derivatives, the coordinates and
velocities of the particle (both of these are absorbed into the torus
coordinates $\angr,\angth$), and other elementary functions.

Each of $R^{\Hinf}_{lm;\coorom}$ and
${}_{-2}S^{a\coorom}_{lm}\left(\theta\right)$ satisfies an ODE that
depends on $l,m$ and $\coorom$ in a nontrivial and partly implicit way
(see Appendix \ref{sec:tuek_formalism}).  No simple closed-form
solutions to these equations exist that make the functional dependence
of the solutions on those parameters explicit.  As a result, for every
distinct set of values $(l,m,\coorom)$, those ODEs must be solved from
scratch to obtain the numerical representations of
$R^{\Hinf}_{lm;\coorom}$ and ${}_{-2}S^{a\coorom}_{lm}$ needed to
evaluate the integrand.  Particularly in the case of
$R^{\Hinf}_{lm;\coorom}$, this operation is computationally costly.

Schematically, then, one calculates each $Z^{\Hinf}_{lm\n\kr}$ for
fixed $l,m$ via the following steps:
\begin{enumerate}
\item Determine the frequency $\coorom = \coorom_{mkn}$
\item Obtain a representation of ${}_{-2}S^{a\coorom}_{lm}$
\item Obtain a representation of $R^{\Hinf}_{lm;\coorom}$ for $\Hinf =
  H,\infty$ (this step requires first determining an eigenvalue of the
  $\SHspin$ ODE)
\item Evaluate $f^{\Hinf}_{lm; \coorom} (\angvec_{i})$ at whatever
  abscissae $\angvec_{i}$ are required by the specific numerical
  integration algorithm chosen
\item Compute whatever weights $w_{i}$ the integration algorithm may
  require for those function values and tabulate the integral
  (\ref{eq:Zkn_as_torus_fcn_Fourier_coeff}) as $\sum_{i} w_{i}
  f^{\Hinf}_{lm; \coorom} (\angvec_{i})$ .
\end{enumerate}

On a non-resonant torus, each $\n,\kr$ pair produces a different
answer to step 1 and requires the execution from scratch of all the
remaining steps as well.  On a resonant torus, in contrast, the
$\n,\kr$ pairs can be grouped by a common value of $\jo$ in the
selection rule (\ref{eq:kzplusnp_with_same_j}).  Steps 1--3 need only
be performed once for an entire $\jo$-group.  Depending on the
integration algorithm selected, steps 4 and 5 may also only need to be
performed once or a small number of times per $\jo$-group, with a
total number of reusable function evaluations set by the
$Z^{\Hinf}_{lm\n\kr}$ in the group requiring the greatest number of
sample points to attain some target accuracy.  We will make the
reasonable assumption that steps 1 and 5, even if done several times
per $\jo$-group, are a small fraction of the total cost of evaluating
all the coefficients in that group, and we will take the cost of steps
2-4 as an estimate of the total cost of computing any single
coefficient.

Consider now evaluating all the $Z^{\Hinf}_{lm\n\kr}$ on a low-order
resonant torus with $\coorom_{\theta}/\coorom_{r}=\omth/\omr = \num/\den = 1 + q_{r\theta}$ and on
a neighboring non-resonant torus with nearly identical orbital
parameters.  By the continuity of the $Z^{\Hinf}_{lm\n\kr}$ with
respect to $\elqvec$, coefficient values will be nearly identical on
those two tori.  The integer values of $\kr_{\text{max}},
\n_{\text{max}}$ determined should also be identical or nearly identical on the two tori
(we assume for simplicity that they are identical).  Let $\numdoub$
and $\numsing$ denote, respectively, the number of separate times
steps 2--4 above must be executed on the non-resonant torus and
resonant torus.  To make a more apples to apples comparison, one can
instead let $\numdoub$ represent the total number of distinct
executions of steps 2--4 on the resonant torus if the resonance of
that torus is not acknowledged from the outset.  Roughly speaking,
generating all the $Z^{\Hinf}_{lm\n\kr}$ with $\absval{\kr} \leq
\kr_{\text{max}}, \absval{\n} \leq \n_{\text{max}}$ on the
non-resonant torus will require $\numdoub/\numsing$ times more
computation than it will on the neighboring resonant torus.
Symmetries in the underlying equations imply that the value of
$Z^{\Hinf}_{l(-m)(-\n)(-\kr)}$ is uniquely determined by the value of
$Z^{\Hinf}_{lm\kr\n}$.  Thus, in practice, one of the indices $\kr$
and $\n$ can be restricted to run over only nonnegative values, and
the value of $\numdoub/\numsing$ should take that fact into account.

\begin{figure}
  \centering
  \includegraphics[width=90mm]{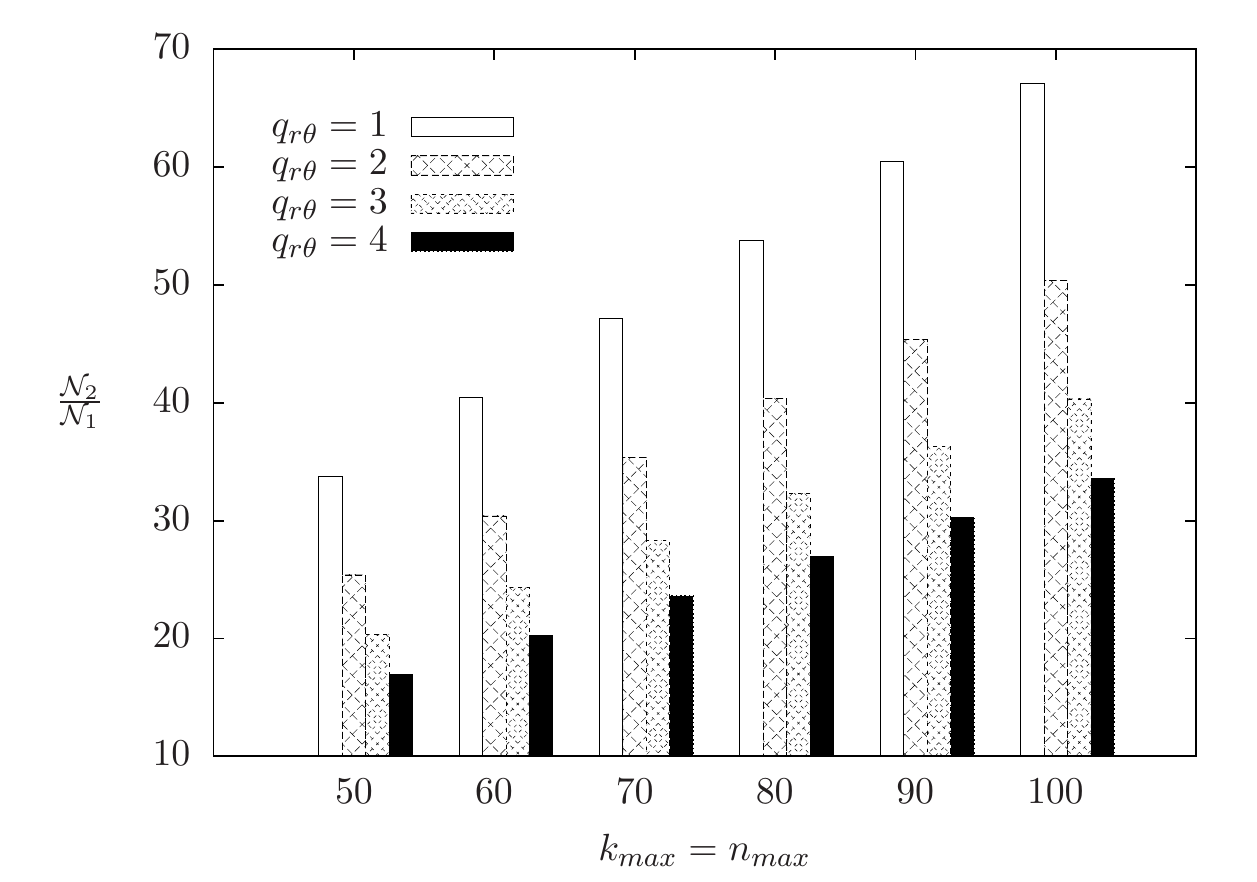}
  \includegraphics[width=90mm]{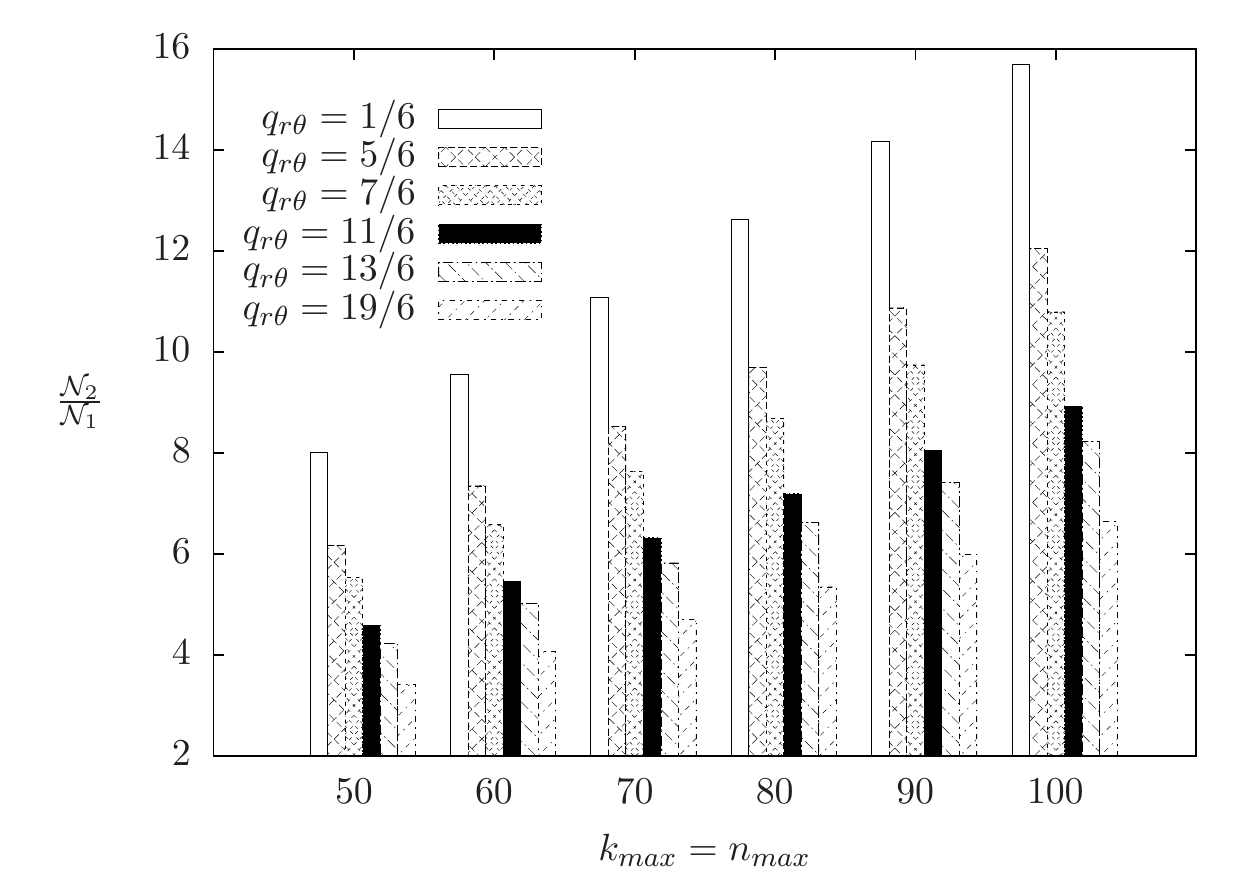}
  \includegraphics[width=90mm]{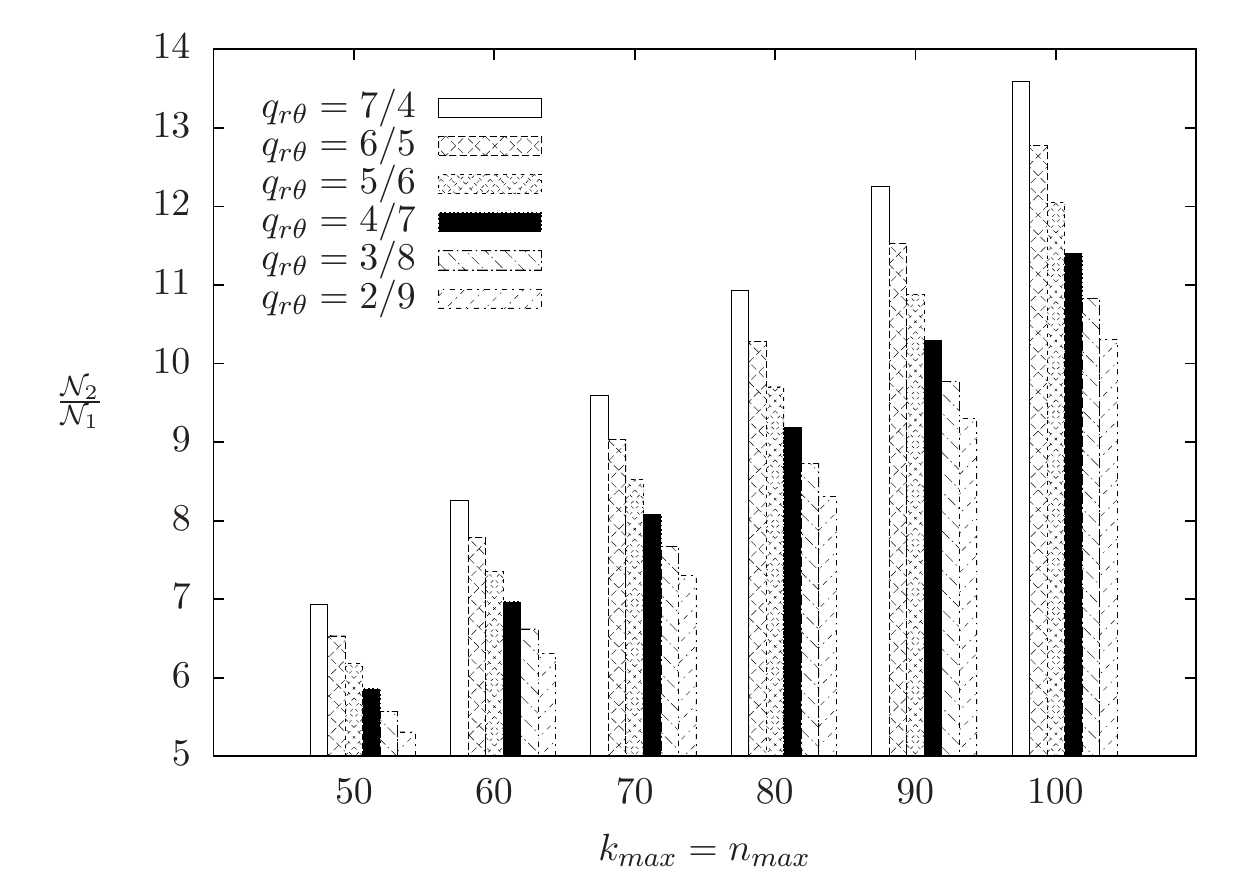}
  \hfill

  \caption{The three histograms show the average number of
    $Z^{\Hinf}_{lm\kr\n}$ coefficients that pertain to a single
    frequency on a resonant torus, a number that corresponds to the
    savings factor $\mathcal{N}_{2} / \mathcal{N}_{1}$.  We show the
    savings factor for a variety of $q_{r\theta}$ geodesics and a
    variety of ${\kr}_{\text{max}}$ and ${\n}_{\text{max}}$.  Top:
    $q_{r\theta}=\text{Integers}$.  Middle: $q_{r\theta}$ is a variety
    of values all with the same denominator, $z=6$.  Bottom:
    $q_{r\theta}$ is a variety of non-integer values all with $\num
    =11$ but different $\den$.}

  \label{fig:histogram_torus}
\end{figure}

Figure \ref{fig:histogram_torus} estimates the savings factor
$\numdoub/\numsing$ for resonant tori with various values of
$q_{r\theta}$ and for several representative hypothetical values of
$\kr_{\text{max}}$ and $\n_{\text{max}}$ consistent with the reported
performance of the TB code for arbitrary eccentricities and
inclinations described in Ref.\ \cite{drasco2006}\footnote{We estimate
  $\kr_{\text{max}}$ and $\n_{\text{max}}$ based on the code in
  \cite{drasco2006} rather than the similar code in \cite{fujita2009}
  only because the truncation rules used in \cite{drasco2006} are more
  amenable to direct cost comparison with our proposal.  Both codes
  seem to need to compute a total number of modes of similar order of
  magnitude to achieve high flux accuracy, and both are apt to profit
  from our proposal.}.  For simplicity, we have taken $\kr_{\text{max}}
= \n_{\text{max}}$.

We can see the following trends in the histograms.  First, for a given
$\n_{\text{max}}, \kr_{\text{max}}$, if we fix the value of $\num$ and
increase $\den$ or vice versa, the savings factor drops.  Thus, the
savings factor is largest when both $p$ and $z$ are as low as
possible.  The greatest savings (over an order of magnitude) accrue
when $z = 1$.  Second, the larger the values of $\n_{\text{max}},
\kr_{\text{max}}$, i.e.\ the more slowly converging the expressions
for the fluxes, the greater the savings factor for a given
$\num/\den$.  Generally speaking, the most slowly converging fluxes
are for orbits with moderate to high eccentrities \cite{drasco2006,
  ganz2007}, which typically have higher associated values of
$q_{r\theta}$ since they are closer to the separatrix between plunging
and non-plunging motion \cite{grossman2011}.  Thus, for instance, a
rough approximation of the true savings factor in the top two panels
of Figure \ref{fig:histogram_torus} would be given by a roughly
horizontal or slightly downward sloping line connecting the histogram
bar with the lowest $q_{r\theta}$ at the lowest $\kr_{\text{max}}$
with the highest $q_{r\theta}$ at the highest $\kr_{\text{max}}$.  A
good rough predictor for the expected savings would thus be the $z$
value of a torus, yielding a savings factor of $\sim 30$ for $z = 1$
and $\sim 7$ for $z = 6$.  The lowest savings factor on that graph of
$\sim 3$ (corresponding to the not-so-low-order resonance with
$\num/\den = 25/6$) is nothing to sneeze at, and more typically the
savings factor from acknowledging resonance would appear to be around
an order of magnitude on average.

While a detailed audit of comparative cost would have to be done on a
code-specific basis, the potential payoff of these observations makes
a case for testing our proposal in existing codes.

\subsection{Numerical EMRI grids}
\label{sec:numerical-emri-grids}

Even if we stipulate that fluxes can be computed more efficiently on
low-order resonant tori than on non-resonant tori, is this fact
necessarily useful?  After all, points in $\elqvec$-space
corresponding to resonant tori, let alone low-order ones, are already
a measure zero set, so to construct the inspiral curve $\elqvec(t)$,
wouldn't the RHSs of the adiabatic ODEs have to be evaluated in
general (and, formally, infinitely more often) on non-resonant tori
than on resonant ones?  Interestingly, while the answer to that
question is ``yes'', the actual calculation of TB fluxes itself need
only ever be done on resonant tori, and at least predominantly (and
possibly exclusively) on low-order ones, at least for the foreseeable
future.

The reason has to do with the \emph{absolute} computational cost of
those fluxes, even on resonant tori.  Simply inserting a TB
frequency-domain flux routine into the RHS of, say, a standard
Runge-Kutta ODE solver to generate inspiral curves in real-time is
untenable, even with a large number of processors at one's disposal to
parallelize the TB calculation.  Instead, solution of the adiabatic
ODEs will proceed as follows.  For each value of the black hole
parameters, one would build a numerical grid of flux values on some
dense mesh of points in $\elqvec$-space and then interpolate off of
that grid to obtain the fluxes for arbitrary values of $\elqvec$.
Once handed such a grid, those interpolated flux values would go into
a standard ODE solver which could presumably generate inspiral curves
very efficiently.  The main expense to consider, then, is the
construction of the grid.

Our proposal is that such a grid should be built using exclusively
resonant grid points.  More specifically, we propose a hierarchical
population of such a grid, beginning with the low-order resonant
points and then increasing the order of the resonance (or just
increasing $z$, if our loose conjecture about horizontal lines in the
top panels of Figure \ref{fig:histogram_torus} proves to be correct)
until some requisite grid density is obtained to minimize
interpolation error.  Those grid density requirements may force the
evaluation of fluxes on some higher-order resonances, but no resonant
grid point (whether low- or high-order) will ever be more expensive to
populate with Fourier flux data than a nearby non-resonant grid point
will be.  The worst-case scenario near certain locations in the space
would be to break even by using a resonant versus a non-resonant grid
point.  Our hierarchical approach would seem to at least lower if not
minimize the total computational cost of such a grid.

We remark on two features of our proposal.  First, the savings
factor discussed depends only on the decision to use resonant tori for
TB calculations and no other implementation-specific features of that
calculation.  Thus, those savings will multiply any additional savings
that may stem from other algorithmic improvements in any such
implementation or from the availability of more or faster processors
to perform the TB calculations.

Second, its efficacy has nothing to do with interesting physical
effects that may occur in the neighborhood of resonant orbital
parameter values during a real inspiral \cite{flanagan2010}.  The
flux-balance method, and in fact the adiabatic approximation in
general, may fail to capture these effects.  Any such failure is
immaterial to our argument, which rests not on \emph{physical}
properties of resonant tori but rather \emph{mathematical} ones they
have in specific relation to frequency-domain TB calculations.  In
other words, despite the fact that the adiabatic approximation might
be least faithful to reality in and around resonances, leveraging
resonances is nonetheless the most efficient means of attaining an
adiabatic approximation for those regimes where it is likely to be
faithful.

\subsection{Gravitational waveform snapshots}
\label{sec:grav-wavef-snapshots}

As already argued, we are free to interpret the coefficients
$Z^{\Hinf}_{lm\n\kr}$ either as spatial Fourier coefficients of a
torus function of $\angvec$ or of a different torus function of
$\vec{\gamma}$.  As shown in Appendix \ref{sec:tuek_formalism}, the
$t$-function versions of the $Z^{\Hinf}_{lm\n\kr}$ coefficients are
used to build the Weyl scalar $\psi_{4}$ at radial infinity, from
which the two polarizations of the waveform $h$ are constructed.
These waveform ``snapshots'' from geodesic sources \cite{drasco2006}
are useful for exploring how a known orbital motion impacts GW
signals, and though they will quickly go out of phase with a true
inspiral signal, they are still likely to play a pivotal role in
hierarchical searches for GWs from EMRIs.

More specifically, with the $\vec{\gamma}$-coefficients
$Z^{\Hinf}_{lm\n\kr}$ in hand, $h$ can be reconstructed from the
associated $t$-coefficients.  By analogy to equation
(\ref{eq:temp_double_coeffs_are_spatials_times_phase}), we get
\begin{alignat}{1}
\label{eq:Z_t_coeffs_nonres}
  Z^{\Hinf}_{lm\n\kr; t} &\equiv
  Z^{\Hinf}_{lm\n\kr}
  e^{-i (\kr \gamma_{r_{0}} + \n \gamma_{\theta_{0}})}
  \quad ,
\end{alignat}
in the nonresonant case, and by analogy to equation
(\ref{eq:temp_single_coeff_is_sum_of_phased_spatials}) get
\begin{alignat}{1}
\label{eq:Z_t_coeffs_res}
  Z^{\Hinf}_{lm\jo; t} (\vec{\gamma}_{0}) =
  \sum_{\substack{\n,\kr:\\ \kr\den+\n\num=\jo}}
  Z^{\Hinf}_{lm\n\kr}
  e^{-i (\kr \gamma_{r_{0}} + \n \gamma_{\theta_{0}})}
\end{alignat}
in the resonant case if we know the initial conditions.  Since the
waveforms (or, rather, their Fourier representations) depend on the
$Z^{\Hinf}_{lm\n\kr}$ coefficients, then like the fluxes, they will
also probably need to be interpolated from a grid that stores the
$Z^{\Hinf}_{lm\n\kr}$ values themselves instead of or in addition to
the fluxes.  The same arguments made above for the fluxes thus
cross-apply to waveform snapshots.

\section{Speculations on further savings}
\label{sec:spec}

In this section, we sketch a speculative but tantalizing possibility
for further efficiencies in adiabatic EMRI grid construction beyond
those discussed in Section \ref{sec:definite}.  The idea centers
around calculating time-averaged rather than torus-averaged fluxes on
resonant tori.  At first glance, that suggestion seems to fly in the
face of earlier arguments that the RHSs of the adiabatic equations
should be torus-averaged fluxes and that torus averages and
time averages are not identical on resonant tori.  The apparent
incongruity disappears, however, in light of two facts:
\begin{enumerate}
\item On any resonant torus, the mean value theorem guarantees that
  torus-averaged fluxes equal time-averaged fluxes on certain special
  orbits.
\item For low-order resonances, those time-averaged fluxes are more
  accurate and cheaper to compute.
\end{enumerate}
The additional savings are beyond the cost benefit of incorporating
the proposal of Section \ref{sec:definite}.

We substantiate these claims below in turn.  We caution the reader
that, in contrast to the savings of Section \ref{sec:definite}, those
discussed in this section may prove more elusive in practice because
determining the special orbits mentioned in step 1 above could prove
so difficult as not to be net-beneficial.  We discuss such limitations
and suggest fruitful avenues of numerical investigation to help
further reduce the cost of generating adiabatic inspirals.

\subsection{Using time-averages to compute torus-averages}
\label{sec:using-time-for-torus}

The time-averaged fluxes from a single resonant orbit do not appear
elsewhere in the literature.  As we explain in Appendix
\ref{sec:using_TSN}, the arguments of Section
\ref{sec:four-analys-time-res} imply that those fluxes are (note,
these are single-index objects in $j$)
\begin{alignat}{1}
  \left\langle{\D{t}{E}}\right\rangle_{t}^{\Hinf} &= \sum_{lm\jo}
  \frac{\alpha_{lm\jo}^{\Hinf}}{4 \pi \coorom_{m\jo}^{2}} \absval{
    Z^{\Hinf}_{lm\jo;\lambda} }^{2}
  \label{eq:res_time_avgd_TB_E_flux}
  \\
  \left\langle{\D{t}{L_{z}}}\right\rangle_{t}^{\Hinf} &=
  \sum_{lm\jo}
  \frac{\alpha_{lm\jo}^{\Hinf} m}{4 \pi \coorom_{m\jo}^{3}}
  \absval{ Z^{\Hinf}_{lm\jo;\lambda} }^{2}
  \label{eq:res_time_avgd_TB_Lz_flux}
  \quad .
\end{alignat}
It remains to be shown whether the following would translate to
$\left\langle\frac{dQ}{dt}\right\rangle_{t}$.  We restrict attention
in this section to $E$ and $L_{z}$ fluxes for the sake of exposition.

As before, we assume fixed $l,m$ in everything below.  In the fluxes,
the frequencies
\begin{equation}
\label{eq:omega_mj}
  \coorom_{m\jo} \equiv m\coorom_{\varphi} + \jo\coorom_{P}
  \quad ,
\end{equation}
the real-valued weight factors $\alpha^{\Hinf}_{lm\jo}$, and the
temporal Fourier coefficients
\begin{alignat}{1}
\label{eq:res_Zj_defn}
  Z_{lm\jo;\lambda}^{\Hinf} &=
  \frac{1}{\Tnot}
  \int^{\Tnot}_{0} d\lambda\,
  e^{i\jo\omnot\lambda}
  f_{lm\jo;\lambda}^{\Hinf}
  \left(
  \angvec (\lambda; \offvec)
  \right)
\end{alignat}
all become single-index quantities by the arguments of Section
\ref{sec:four-analys-time-res}.  We recall from equation
(\ref{eq:fluxes_as_toravgd_Fourier_powers}) that the torus-averaged
fluxes have the form of a torus-averaged power of some unspecified
torus-function.  Likewise, for fixed values of $l,m$, the
time-averaged fluxes (\ref{eq:res_time_avgd_TB_E_flux}) and
(\ref{eq:res_time_avgd_TB_Lz_flux}) have the appearance of a
time-averaged Fourier power in the sense of Section
\ref{sec:four-analys-time-res} with the identification
\begin{alignat}{1}
\label{eq:fluxes_as_timeavgd_Fourier_powers}
  \begin{split}
    \absval{C_{\jo; \lambda}}^{2} &=
    \text{prefactor } \times 
    \absval{Z^{\Hinf}_{lm\jo; \lambda}}^{2}
    \\
    C_{\jo; \lambda} &=
    \sqrt{\text{prefactor}} \times 
    Z^{\Hinf}_{lm\jo; \lambda}
    \quad .
  \end{split}
\end{alignat}

Each time-averaged flux, like any time-averaged Fourier power, is
real-valued.  Therefore, by equations
(\ref{eq:fluxes_as_timeavgd_Fourier_powers}) and
(\ref{eq:pwr_toravg_equal_toravg_of_pwr_timeavg}) and the mean-value
argument made at the end of Section \ref{sec:four-analys-time-res},
there exist initial positions $\offvecmvtE, \offvecmvtLz$ on the torus
such that
\begin{alignat}{1}
  \left\langle{\D{t}{E}}\right\rangle^{\Hinf}_{t}
  \left( \offvecmvtE \right) &=
  \left\langle{\D{t}{E}}\right\rangle^{\Hinf}_{\vec{\gamma}}
  \label{eq:mvt_for_E_flux}
  \\
  \left\langle{\D{t}{L_{z}}}\right\rangle^{\Hinf}_{t}
  \left( \offvecmvtLz \right) &=
  \left\langle{\D{t}{L_{z}}}\right\rangle^{\Hinf}_{\vec{\gamma}}
  \label{eq:mvt_for_Lz_flux}
  \quad .
\end{alignat}
Actually, there must be at least two continuous $1$-parameter families
of special initial values $\offvecmvtE$ (one for each of $\Hinf =
H,\infty$) and two such families for $\offvecmvtLz$: any two initial
conditions that lie on the same orbit simply time-translate that
orbit, and time-translation does not change time-averaged function
values or time-averaged powers.

None of the values $\offvecmvtE$ and $\offvecmvtLz$ need agree.  Thus,
if we sought to determine the torus-averaged fluxes indirectly by
instead evaluating time-averaged fluxes, we might need to evaluate
each coefficient $Z^{\Hinf}_{lm\jo; \lambda}$ as many as
four\footnote{Or, rather, six times, since each $\offvecmvtQ$ would
  likely also be different from $\offvecmvtE$ and $\offvecmvtLz$.}
times, once for each of the initial conditions $\offvecmvtE$ and
$\offvecmvtLz$.

We can, however, also apply the mean-value argument individually to
each real-valued $\absval{Z^{\Hinf}_{lm\jo; \lambda}}^{2}$.  In this
case, we would obtain a sequence of special initial conditions
$\offvecmvtj$ that cause each $\absval{Z^{\Hinf}_{lm\jo;
    \lambda}}^{2}$ to attain its torus-averaged value over all
possible initial conditions.  The different $\offvecmvtj$ would not
necessarily agree for different values of $\jo$.  Since the prefactors
in (\ref{eq:fluxes_as_timeavgd_Fourier_powers}) are independent of
initial position, each $\offvecmvtj$ would simultaneously set the
$\jo$th term in the power spectrum of every flux to its torus-averaged
value.  Evaluating the time-averaged fluxes for any of the individual
initial conditions $\offvecmvtj$ would not produce a torus-averaged
flux.  However, since the average of a sum of terms must equal the sum
of the individual averages of those terms, the sum of the resulting
$\absval{C_{\jo;\lambda}}^{2}$ (each evaluated at a possibly different
$\offvecmvtj$) would yield the torus-averaged value of all fluxes
simultaneously.  Recalling that each integrand in
(\ref{eq:res_Zj_defn}) is different anyway, there is no further waste
in evaluating each one using a different initial condition
$\offvecmvtj$.

It is important to note that we have simply made an existence argument
for $\offvecmvtE, \offvecmvtLz$ and every $\offvecmvtj$.  What those
values actually are would vary from problem to problem, and finding
them for the Teukolsky problem may not be practical.  The integrands
in (\ref{eq:res_Zj_defn}) are not especially analytically transparent,
so it may be that they can only be determined by evaluating those
integrands for several initial conditions $\offvec$, which would
defeat the purpose of invoking the mean-value theorem in the first
place.  Still, we believe the potential added savings from knowing the
$\offvecmvtj$ merits exploring whether the Teukolsky calculation
harbors some structure or symmetries that would allow those initial
conditions to be determined with little or no added expense.  We turn
to those additional potential savings now.

\subsection{Relative cost of time-averaged vs.\ torus-averaged
  functions on low-order resonant tori}
\label{sec:relative-cost-time-vs-torus-on-res-tori}

Assume that we have in hand the $\offvecmvtj$ for each $\jo$ and agree
to evaluate the coefficients $Z^{\Hinf}_{lm\jo; \lambda}$ using those
special initial conditions. The added efficiency is twofold: each
$C_{\jo;\lambda}$ should potentially be less expensive to compute than
any given $A_{\n\kr}$ (by reducing a double integral to a single
integral), and fewer such $C_{\jo;\lambda}$'s than $A_{\n\kr}$'s will
have to be computed in order to achieve a given target accuracy in the
torus-averaged fluxes (by reducing a double sum to a single sum).  In
fact, the more efficient calculation might even increase the resulting
flux accuracy.  We justify those claims in turn below.

\subsubsection{Cost of a coefficient}
\label{subsec:cost}

For ease of illustration, we will estimate the relative computational
costs of a single $C_{\jo;\lambda}$ and of any single $A_{\n\kr}$ for
which $\n,\kr$ satisfy the selection rule
(\ref{eq:kzplusnp_with_same_j}).  To make the comparison more stark,
we remap the integral (\ref{eq:res_Zj_defn}) to the interval
$[0,2\pi]$ via a linear change of variable
\begin{equation}
\label{eq:defn_periodic_angle_var}
  \angnot \equiv \omnot \lambda
\end{equation}
to obtain
\begin{alignat}{1}
\label{eq:res_Zj_with_chi_not}
  Z_{lm\jo;\lambda}^{\Hinf} &=
  \frac{1}{2\pi}
  \int^{2\pi}_{0} d\angnot\,
  e^{i \jo \angnot}
  f_{lm\jo;\lambda}^{\Hinf}
  \left(
  \angvec (\angnot; \offvec)
  \right)
  \quad .
\end{alignat}

The relative cost of the single integral
(\ref{eq:res_Zj_with_chi_not}) and its double-index counterpart
(\ref{eq:Zkn_as_torus_fcn_Fourier_coeff}) will depend on the specific
numerical integration algorithms used to evaluate them and are
difficult to estimate. However, we can sketch a crude argument that
the single integral should be more cost efficient by considering the
Fast Fourier transform (FFT) as the algorithm.

Consider first the $1D$ integral (\ref{eq:res_Zj_with_chi_not}), which
we can interpret as the $\jo$th Fourier coefficient of a periodic
function on $[0,2\pi]$.  For a periodic function, an FFT will return
all the Fourier coefficients from $C_{-N_{1}}$ through $C_{N_{1}}$ by
sampling the integrand at $2 N_{1} + 1$ equally spaced
abscissae\footnote{To make the formulae that follow more intelligible,
  we are separately counting the value at $2\pi$, even though it is
  the same as the value at $0$}.  So to capture $C_{\jo}$, we would
need $2 \absval{\jo} + 1$ evaluations of the integrand.  However, the
highest index coefficients computed via an FFT are heavily afflicted
by aliasing error, while the lowest index coefficients computed are
relatively free of such error.  To minimize aliasing effects, we
imagine increasing the number of sample points (and thus of
coefficients computed) by some integer safety
factor\footnote{Ref.\ \cite{press} recommends a factor of at least $4$
  for most applications.}  $\mathcal{S}$ so that $C_{\jo}$ will be one
of lowest index coefficients returned by the FFT and thus fairly free
of aliasing error.  The total number of integrand evaluations under
this scheme for computing $C_{\jo}$ would thus be $\mathcal{S} ( 2
\absval{\jo} + 1 )$.

Now imagine evaluating the double integral
(\ref{eq:Zkn_as_torus_fcn_Fourier_coeff}) using a $2D$ FFT, which we
(even more crudely) envision simply as nested $1D$ FFTs.  Assuming the
same safety factor $\mathcal{S}$ throughout, we would need
$\mathcal{S} \left( 2 \absval{\kr} + 1 \right) \mathcal{S} \left(2
\absval{\n} + 1 \right) = \mathcal{S}^{2} (2 \absval{\kr} + 1) (2
\absval{\n} + 1)$ function evaluations.  Re-expressing $\jo$ in terms
of $\kr$ and $\n$ via the selection rule and using the number of
integrand evaluations as a metric of numerical expense, the ratio of
the cost of $A_{\n\kr}$ to the cost of $C_{\jo}$ would be
\begin{equation}
\label{eq:relative_FFT_cost_of_Cj_and_Akn}
  \frac{\text{cost of }A_{\n\kr}}{\text{cost of }C_{\jo}} =
  \mathcal{S}
  \frac{(2 \absval{\kr} + 1)(2 \absval{\n} + 1)}
       {2 \absval{\kr\den + \n\num} + 1}
  \quad .
\end{equation}
Generally speaking, for small values of both $\num$ and $\den$, the
denominator in the cost ratio is smaller than the numerator since
$\kr$ and $\n$ more often than not have opposite signs for a given
$\jo$.  It is conceivable that a single $A_{\n\kr}$ could turn out
less costly to evaluate than $C_{\jo}$, but the likelihood of that
would become higher as both $\num$ and $\den$ became large, for which
case a resonant torus would be barely distinguishable from a
non-resonant torus in terms of all the aspects discussed in this
paper.

The argument above artificially increases the true cost of evaluating
both integrals and is not intended even to be fully convincing, let
alone a proof.  Rather, it is a heuristic illustration of a rule of
thumb in numerical integration that, with similarly behaved
integrands, $1D$ integrals are less costly to compute than $2D$
integrals.

\subsubsection{Number of coefficients}
\label{subsec:num_coeffs}

In contrast to the relative cost of computing a coefficient, we can
say more definitively that the total number of single-index
coefficients needed to achieve some specified accuracy in the
torus-averaged fluxes will be less than the number of double-index
coefficients needed to obtain the same accuracy.

Suppose achieving a certain flux accuracy for a given $l,m$ pair
requires computing all $A_{\n\kr}$ with indices up to
${\kr}_{\text{max}}$ and ${\n}_{\text{max}}$.  Denote the total number
of torus coefficients computed by $\mathcal{N}_{\tor}$.  The
$Z^{\Hinf}$ coefficients satisfy
$\left|Z^{\Hinf}_{lm\omega}\right|^{2} =
\left|Z^{\Hinf}_{l(-m)(-\omega)}\right|^{2}$ \cite{chrzanowski1976},
so one of $\kr$ and $\n$ need only run over non-negative values to
obtain all the coefficients with $\absval{\kr} \leq
{\kr}_{\text{max}}, \absval{\n} \leq {\n}_{\text{max}}$.  The total
number of coefficients actually computed is therefore (having $\n$ run
only non-negative)
\begin{eqnarray}
\label{eq:N_2}
\mathcal{N}_{\tor} &=&
\left(2{\kr}_{\text{max}}+1\right)
\left({\n}_{\text{max}}+1\right)-{\kr}_{\text{max}}
\quad .
\end{eqnarray}

For comparison, we determine the number $\mathcal{N}_{\ti}$ of
$C_{\jo}$ coefficients (evaluated at $\offvecmvtj$) that we would have
to calculate so that, in light of the arguments of subsection
\ref{sec:using-time-for-torus}, every $\absval{A_{\n\kr}}^{2}$ above
would automatically be included in the sum of all
$\absval{C_{\jo}}^{2}$.  As we showed in subsection
\ref{sec:using-time-for-torus}, the maximum $\jo$ index that needs to
be included in the single-index series that will thusly catch every
$\n,\kr$ pair is
\begin{eqnarray}
\label{eq:j_max}
{\jo}_{\text{max}} &=& \den {\kr}_{\text{max}}+ \num {\n}_{\text{max}}
\quad .
\end{eqnarray}
The symmetry of $Z^{\Hinf}$ implies that $\jo$ need not run both
positive and negative, and the number of $C_{\jo}$'s we would need to
calculate to ensure at least the same level of flux convergence as
that attained with the $A_{\n\kr}$ coefficients is
\begin{eqnarray}
\label{eq:up_bound_N_1}
\mathcal{N}_{\ti} &=& \den {\kr}_{\text{max}}+ \num
{\n}_{\text{max}} + 1
\quad .
\end{eqnarray}
Comparing equations (\ref{eq:N_2}) and (\ref{eq:up_bound_N_1}), we see
that we need a factor of
\begin{eqnarray}
\label{eq:coeff_reduc}
\mathcal{N}_{\text{savings}} &=&
\frac{\mathcal{N}_{\tor}}{\mathcal{N}_{\ti}} 
\\ 
\nonumber 
&=& \frac{\left(2k_{\text{max}}+1\right)\left(n_{\text{max}}+1\right)-k_{\text{max}}}{\den\kr_{\text{max}}+\num\n_{\text{max}}+1}
\end{eqnarray}
fewer coefficients.  The reduction in the number of coefficients
therefore depends on the order of the periodic orbit as well as on
$n_{\text{max}}$ and $k_{\text{max}}$.   The lower the values of $\num$
and $\den$, the greater the reduction factor.

Figure \ref{fig:histogram_torus} showed the average number of $kn$
modes on a resonant torus per distinct frequency.  It also gives a
general sense of how $\mathcal{N}_{\text{savings}}$ varies with
$k_{\text{max}}$ and $n_{\text{max}}$.  The agreement between the two
is not exact because, when computing all $\jo$ coefficients up to the
maximum $\jo_{\text{max}}$, some additional frequencies will be
included that do not correspond to any of the included $kn$
frequencies with $\absval{k} \leq k_{\text{max}}, \absval{n} \leq
n_{\text{max}}$.  Therefore, Figure \ref{fig:histogram_torus}
overestimates $\mathcal{N}_{\text{savings}}$ but only slightly and
gives a better estimate for the larger values of $k_{\text{max}},
n_{\text{max}}$.  For example, for $q_{r\theta}=\frac{1}{6}$, equation
(\ref{eq:coeff_reduc}) gives $\mathcal{N}_{\text{savings}} \approx
7.84$ for $n_{\text{max}}=k_{\text{max}}=50$ and
$\mathcal{N}_{\text{savings}} \approx 12.45$ for
$n_{\text{max}}=k_{\text{max}}=80$, both of which agree with the
values in the histogram of Figure \ref{fig:histogram_torus} within a
few percent.  On the basis of Figure \ref{fig:histogram_torus}, we can
therefore conclude that focusing on temporal rather than spatial
Fourier coefficients and invoking the above mean-value arguments could
reduce by a factor of an order of magnitude or so the total number of
$Z^{\Hinf}$ coefficients required to obtain accurate torus-averaged
fluxes.

\section{Conclusion}
\label{sec:con}

Computation of adiabatic inspirals with a grid of resonant orbits
could be an order of magnitude more efficient than the same
computation with a non-resonant grid.  If our speculations are
verified and double sums can be collapsed to single sums (and double
integrals to single integrals), there may be substantial additional
savings since fewer and simpler coefficients will be required.  To
date, no accurate adiabatic EMRIs have been computed.  Such a dramatic
boost in speed would bring EMRIs more within computational reach.

\acknowledgments

We gratefully acknowledge valuable discussions with Scott Hughes.
This work was supported by an NSF grant AST-0908365.  JL gratefully
acknowledges support of a KITP Scholarship, under Grant no. NSF
PHY05-51164.

\vfill\eject

\bigskip 
\bigskip 
\bigskip

\appendix
\section{Torus coordinates and time coordinates}
\label{subsec:which_time}

\subsection{Mino time vs.\ coordinate time Fourier coefficients}
\label{sec:mino-time-vs-coord-time-fourier-coeffs}

Suppose $A_{kn; t}$ is a Fourier coefficient of some biperiodic
coordinate time function $f(t)$,
\begin{equation}
\label{eq:Aknt_defn}
  A_{kn; t} =
  \lim_{T \to \infty} \frac{1}{T}
  \int_{-\frac{T}{2}}^{\frac{T}{2}} dt\,
  e^{i (\n\cooromth + \kr\cooromr) t}
  f(t)
  \quad .
\end{equation}
Then $A_{kn; t}$ is also the $kn$th Fourier coefficient of a different
biperiodic Mino time function $g(\lambda)$,
\begin{alignat}{1}
\label{eq:Aknt_equals_Aknlambda}
  \begin{split}
    A_{kn; t} &=
    \lim_{\Lambda \to \infty} \frac{1}{\Lambda}
    \int_{-\frac{\Lambda}{2}}^{\frac{\Lambda}{2}} d\lambda\,
    e^{i (\n\omth + \kr\omr) \lambda}
    g(\lambda)
    \\
    &\equiv A_{kn; \lambda}
    \quad .
  \end{split}
\end{alignat}

We prove equation (\ref{eq:Aknt_equals_Aknlambda}) by constructing the
function $g$ from $f$ explicitly.  We will need the fact (see
Ref.\ \cite{drasco2004} for details) that $\txtD{\lambda}{t}$ depends
on $r$ and $\theta$ and is thus biperiodic when evaluated on a
trajectory $r(\lambda), \theta(\lambda)$.  $\txtD{\lambda}{t}$ also
has a nonzero average value $\Gamma$ defined by
\begin{equation}
\label{eq:defn_t_to_lambda_Gamma_factor}
  \Gamma \equiv 
  \lim_{\Lambda \to \infty}
  \frac{1}{\Lambda}
  \int_{-\Lambda/2}^{\Lambda/2}
  d\lambda\, \D{\lambda}{t}
  \quad .
\end{equation}
Consequently, the function $t(\lambda)$ takes the form
\begin{equation}
\label{eq:t_of_lambda}
  t(\lambda) = \Gamma\lambda + \Delta t(\lambda)
\end{equation}
where $\Delta t(\lambda)$ is biperiodic in $\lambda$ and has zero
average value.

We can now construct $g(\lambda)$.  Since $\Delta t(\lambda)$ is
biperiodic, it is also bounded.  Thus, if we define $T \equiv
t(\Lambda)$, then in the limit $T \to \infty$, we get $T \to \Gamma
\Lambda$.  We can now change variables in the integral in
(\ref{eq:Aknt_defn}):
\begin{widetext}
\begin{alignat}{1}
\label{eq:inf_t_integral_as_inf_lambda_integral}
  \begin{split}
    \lim_{T \to \infty} \frac{1}{T}
    \int_{-\frac{T}{2}}^{\frac{T}{2}} dt\,
    e^{i (\n\cooromth + \kr\cooromr) t}
    f(t)
    &=
    \lim_{\Gamma\Lambda \to \infty} \frac{1}{\Gamma\Lambda}
    \int_{-\frac{\Gamma\Lambda}{2}}^{\frac{\Gamma\Lambda}{2}}
    d\lambda\,
    e^{i (\n\cooromth + \kr\cooromr)
      \left( \Gamma\lambda + \Delta t(\lambda) \right)}
    \D{\lambda}{t}(\lambda)
    f \left( t(\lambda) \right)
    \\
    &=
    \lim_{\Lambda \to \infty} \frac{1}{\Lambda}
    \int_{-\frac{\Lambda}{2}}^{\frac{\Lambda}{2}}
    d\lambda\,
    e^{i (\n\omth + \kr\omr) \lambda}
    e^{i (\n\cooromth + \kr\cooromr) \Delta t(\lambda)}
    \D{\lambda}{t}(\lambda)
    f \left( t(\lambda) \right)
    \quad .
  \end{split}
\end{alignat}
\end{widetext}
In the second line above, we have absorbed\footnote{All we need is for
  the denominator of the prefactor and the size of the integration
  interval to agree.  Since there is no preferred size for that
  interval (the function $f$ is not periodic) and we are taking the
  infinite limit, we are free to call the size of that interval
  $\Gamma\Lambda$ or $\Lambda$.} the $\Gamma$ into the $\Lambda$ and
used the fact that the coordinate-time and Mino-time frequencies will
be related \cite{drasco2004} by
\begin{subequations}
\label{eq:rel_between_Minotime_and_coordtime_freqs}
\begin{alignat}{1}
  \coorom_{r} &= \frac{1}{\Gamma} \omr
  \label{eq:rel_between_Minotime_and_coordtime_r_freqs}
  \\
  \coorom_{\theta} &= \frac{1}{\Gamma} \omth
  \label{eq:rel_between_Minotime_and_coordtime_th_freqs}
  \quad ,
\end{alignat}
\end{subequations}
to convert to Mino frequencies in the argument of the exponential.

Comparing (\ref{eq:Aknt_equals_Aknlambda}) to
(\ref{eq:inf_t_integral_as_inf_lambda_integral}), we see that due to
the dependence on $k$ and $n$ in the argument of the exponential (the
coordinate time frequencies $\cooromth$ and $\cooromr$ appear here
simply as parameters), there is actually a different function
\begin{equation}
\label{eq:gkn_functions_defined}
  g_{kn}(\lambda) =
  e^{i (\n\cooromth + \kr\cooromr) \Delta t(\lambda)}
  \D{\lambda}{t}(\lambda)
  f \left( t(\lambda) \right)
\end{equation}
for each $A_{kn; \lambda}$.  In other words, the $A_{kn;t}$ are
Fourier coefficients of a single function $f$ while each $A_{kn;
  \lambda}$ is the $kn$th Fourier coefficient of a different function
$g_{kn}(\lambda)$.  But that poses no problem --- we only sought to
show that every $t$-Fourier coefficient is also the $\lambda$-Fourier
coefficient of \emph{some} function of $\lambda$.  The pragmatic
importance of this fact has to do with the evaluation of coefficients
of torus functions, which, though stated in slightly different
language, is the crux of the original argument in
Ref.\ \cite{drasco2004} and which we discuss in Section
\ref{sec:lambda-based-vs-t-based-torus-coords}.

Analogous reasoning to the above in the resonant case with
\begin{equation}
\label{eq:defn_resonant_freq_ratio}
  \frac{\cooromth}{\cooromr} =
  \frac{\omth}{\omr} =
  \frac{p}{z}
\end{equation}
and $p,z$ relatively prime shows that each coefficient $C_{\jo; t}$ of
$f(t)$ can likewise be considered a Mino time coefficient $C_{\jo;
  \lambda}$ of some different function
\begin{equation}
\label{eq:gj_functions_defined}
  g_{\jo}(\lambda) =
  e^{i \jo \cooromnot \Delta t(\lambda)}
  \D{\lambda}{t}(\lambda)
  f \left( t(\lambda) \right)
  \quad .
\end{equation}
The difference is that now each of $\Delta t(\lambda),
\txtD{\lambda}{t} (\lambda)$ and $f \left( t(\lambda) \right)$ is a
singly periodic function of $\lambda$ with period
\begin{equation}
\label{eq:Lambda_naught_defn}
  \Tnot = z\Tr = p\Tth = \frac{2\pi}{\cooromnot}
  \quad .
\end{equation}
Temporal Fourier coefficients can be calculated in the resonant case
without first having to convert to Mino time.  We will nonetheless
express and evaluate coefficients $C_{\jo; t}$ as coefficients
$C_{\jo; \lambda}$ in order to parallel the non-resonant case.

\subsection{$\lambda$-based vs.\ $t$-based torus coordinates}
\label{sec:lambda-based-vs-t-based-torus-coords}

Figure \ref{fig:res_tor} represents $\torelq$ as a compact
$2\pi$-by-$2\pi$ square in the $\angr,\angth$ angle coordinates
defined in Section \ref{sec:phase-space-tori}.  Kerr geodesics trace
out lines on this torus-square at constant velocity.  With respect to
any other time parameter, geodesic curves continue to be lines on the
torus-square, but their parametric representations are not in general
linear in time, nor are their velocities on the torus constant.  

For any choice of time parameter, however, there is always some set of
coordinates on the torus such that geodesic motion on that torus is
linear in that time parameter and has constant
velocity\footnote{Darboux's theorem gaurantees that there is a way to
  write Hamilton's equations with respect to any evolution parameter.
  If the system is integrable, there will then exist a transformation
  to angle variables on the torus that increase linearly with respect
  to that evolution parameter \cite{arnold}.}.  For instance, with
respect to coordinate time $t$, there will be coordinates
$\vec{\gamma} \equiv \left( \gamma_{r}, \gamma_{\theta} \right)$ such
that
\begin{subequations}
\label{eq:gamma_of_t}
\begin{alignat}{1}
  \gamma_{r} (t) &= \coorom_{r} t + \gamma_{r_{0}}
  \label{eq:gamma_r_of_t}
  \\
  \gamma_{\theta} (t) &= \coorom_{\theta} t + \gamma_{\theta_{0}}
  \label{eq:gamma_theta_of_t}
  \quad .
\end{alignat}
\end{subequations}

Note that in any set of angle coordinates in which the trajectory
velocities are constant with respect to some time parameter, orbit
trajectories will all be lines with the same slope $1 + q_{r\theta}$.
Such coordinate systems are nevertheless distinct: identical ordered
pairs in two such coordinate systems will not, in general, correspond
to the same point on the torus.

Though not unique, the $\angr$-$\angth$ coordinate system on $\torelq$
is nevertheless uniquely \emph{useful}.  Since each of $\gamma_{r},
\gamma_{\theta}$ would be a combination of $\angr$ and $\angth$, each
point on the projected $r$-$p_{r}$ curve of an orbit would be labelled
by a pair of values $(\gamma_{r}, \gamma_{\theta})$ rather than by a
single value $\angr$, and likewise for the projected
$\theta$-$p_{\theta}$ curve.  This mixing of radial and polar motions
in each torus coordinate makes most calculations harder than they need
to be, and the impetus behind $\angr$-$\angth$ coordinates is
precisely the convenience that flows from torus coordinates that
separately shadow radial and polar motion.

Still, we sometimes are interested in values of quantities averaged
over the $\vec{\gamma}$ coordinates.  Luckily, by the correspondence
between temporal Fourier coefficients of biperiodic functions and
spatial Fourier coefficients of torus functions, equations
(\ref{eq:Aknt_equals_Aknlambda}) and (\ref{eq:gkn_functions_defined})
further establish that if $A_{kn; \vec{\gamma}}$ is the $kn$th
coefficient of the torus function $f(\vec{\gamma})$ associated with
$f(t)$, then it is also the $kn$th coefficient $A_{kn; \angvec}$ of
the torus function $g_{kn}(\angvec)$ corresponding to
$g_{kn}(\lambda)$.  This is important for evaluating torus
coefficients in practice: it is usually very difficult to go from a
function $f \left( r(t), p_{r} (t), \theta(t), p_{\theta} (t) \right)$
to a form $f(\vec{\gamma})$ explicitly while it is straightforward to
go from $g \left( r(\lambda), p_{r} (\lambda), \theta(\lambda),
p_{\theta} (\lambda) \right)$ to $g(\angvec)$.

\section{A synopsis of the Teukolsky formalism}
\label{sec:tuek_formalism}

Here we summarize some relevant aspects of the Teukolsky formalism as
applied to the EMRI problem.  More details of this application can be
found in numerous references, including \cite{glampedakis2002,
  hughes2001, hughes2005, drasco2006, ganz2007, sago2005,
  Sago:2005gd}.  Our goal in this aapendix is to justify the
expressions for the coefficients $Z^{H/\infty}_{lmkn}$ and
$Z^{H/\infty}_{lm\jo; \lambda}$ in equations
(\ref{eq:Zkn_as_torus_fcn_Fourier_coeff}) and (\ref{eq:res_Zj_defn}),
respectively.

\subsection{The Weyl scalar, $\psi_{4}$}
\label{sec:weyl-scalar-psi_4}

In 1972 Teukolsky derived the master equation
\cite{teukolsky1972,teukolsky1973}, a separable partial differential
equation (PDE) whose solution describes the propagation in the Kerr
spacetime of small perturbations to fields of different spin-weights
$s$: scalar, electromagnetic and gravitational.  Each solution to the
master equation is a separable function which can be written as a
multipole expansion.  There are two computational approaches to
solving the master equation for each multipole mode: the time-domain
approach, which solves the resulting PDE directly, and the
frequency-domain approach, which further Fourier expands the
solutions.  For the purposes of extracting flux information from
gravitational perturbations, frequency-domain codes are the accuracy
standard and the ones to which our savings proposal applies.  We thus
restrict our attention to the frequency-domain approaches to solving
the master equation.

The combined multipole-Fourier expanded perturbations take the form
\begin{multline}
\label{eq:psi_four}
{}_{s}\psi\left(t,r,\theta,\varphi \right) = \\
\sum_{lm}\int^{\infty}_{-\infty} d\coorom R_{lm\coorom}\left(r\right){}_{s}S^{a\coorom}_{lm}\left(\theta\right)e^{-i\coorom
  t+im\varphi}
\quad ,
\end{multline}
where $\coorom$ denotes the coordinate-time frequency of the
perturbations at the field point due to the source.  Each ${}_{s}\psi$
is a function of the field point ($t$, $r$, $\theta$, $\varphi$) at which
we wish to evaluate the perturbation.  The $s$ marker in equation
(\ref{eq:psi_four}) is a ``spin-weight parameter''
\cite{teukolsky1972} which denotes the perturbation type.  For
gravitational radiation, $s=-2$, and ${}_{-2}\psi=\psi_{4}\rho^{-4}$
where
\begin{eqnarray}
\label{eq:teuk_rho}
\rho &=& -\left( r-ia\cos{\theta} \right)^{-1}
\quad .
\end{eqnarray}

The functions $R_{lm\coorom}(r)$ and $\SHspin (\theta)$ (described in
the next subsections) each depend on the parameter $\coorom$ as a
consequence of the separation of variables procedure.  When the source
is a geodesic, $\coorom$ turns out to be a discrete variable composed
of harmonics of the radial, polar and azimuthal frequencies of that
geodesic.  That discrete dependence can be expressed differently for
non-resonant orbits,
\begin{subequations}
\label{eq:omega_mkn}
\begin{alignat}{1}
\label{eq:omega_non_res}
\coorom &= \coorom_{mkn} = m\coorom_{\varphi}+\n\coorom_{r} +\kr\coorom_{\theta}
\intertext{and for resonant orbits,}
\label{eq:omega_res}
\coorom &= \coorom_{mj} = m\coorom_{\varphi}+\jo\cooromnot
\quad .
\end{alignat}
\end{subequations}
Once $\coorom$ becomes a discrete variable, we can replace the
integral over all possible $\coorom$ in equation (\ref{eq:psi_four})
with a sum over either $m,k,n$ or $m,j$ for non-resonant and resonant
sources, respectively.

Because everything in this paper deals with gravitational
spin-weighting, we henceforth omit all the $-2$ subscripts.  The net
result is that equation (\ref{eq:psi_four}) becomes
\begin{subequations}
\label{eq:psi_4_quad}
\begin{alignat}{1}
\label{eq:psi_4_quad_non_res}
\psi_{4}\left(t,r,\theta,\varphi\right) &=  {\rho}^{4}\sum_{lmnk}
R_{lm\coorom_{mkn}}\left(r\right)S^{a\coorom_{mkn}}_{lm}\left(\theta\right) e^{-i\coorom_{mkn} t+im\varphi}
\quad ,
\intertext{for a non-resonant source and}
\label{eq:psi_4_quad_res}
\psi_{4}\left(t,r,\theta,\varphi\right) &=  {\rho}^{4}\sum_{lmj}
R_{lm\coorom_{mj}}\left(r\right)S^{a\coorom_{mj}}_{lm}\left(\theta\right) e^{-i\coorom_{mj}t+im\varphi}
\end{alignat} 
\end{subequations}
for a resonant source.

\subsection{The Spheroidal Harmonics}
\label{subsubsec:spheroidal}

The functions $S^{a\coorom}_{lm}\left(\theta \right)$ with a
spin-weight of $s=-2$ are the gravitational (tensor) spheriodal
harmonics, a generalization of the likewise spin-weighted spherical
harmonics.  These functions satisfy \cite{teukolsky1973}
\begin{alignat}{1}
\label{eq:spheriodal}
& \left[\left(a\coorom \right)^{2}\cos^{2}{\theta}
  +4a\coorom\cos{\theta} - \left(\frac{m^{2}-4m\cos{\theta}+4}{\sin^{2}{\theta}}\right) +
  \mathcal{C}_{lm}\right]S^{a\coorom}_{lm}\left(\theta\right)
\nonumber
\\
&+\frac{1}{\sin{\theta}}\frac{d}{d\theta}\left(\sin{\theta}
\frac{dS^{a\coorom}_{lm}\left(\theta\right)}{d\theta}\right) =0
\quad .
\end{alignat}

$\mathcal{C}_{lm}$ are the eigenvalues for which equation
(\ref{eq:spheriodal}) has solutions.  Solving for $\SH (\theta)$ for
given $l,m,\coorom$ requires simultaneously determining an eigenvalue
$\mathcal{C}_{lm}$ and the associated spheroidal harmonic.  These
eigenvalue-eigenfunction pairs can be computed in several different
ways (see Refs.\ \cite{hughes2001,ganz2007,fujita2009}).

The spheroidal harmonics satisfy several orthogonality relations.  The
one we will need in this paper is that, for fixed $m$ and $\omega$,
\begin{alignat}{1}
\label{eq:spheroidal_delta}
\int^{\pi}_{0}S^{a\coorom}_{lm}\left(\theta\right)
    \bar{S}^{a\coorom}_{l'm}\left(\theta\right)
    \sin{\theta}d\theta &=\frac{1}{2\pi}
\delta_{ll'}
\quad ,
\end{alignat}
where the overbar denotes complex conjugation.  We have chosen a
normalization of $\frac{1}{2\pi}$, as in Ref.\ \cite{drasco2006}.

\subsection{The radial Teukolsky functions}
\label{subsubsec:rad_funcs}

Solving for the radial functions $R_{lm\coorom}\left(r\right)$ is more
difficult.  $R_{lm\coorom}\left( r\right)$ satisfy the inhomogeneous
radial Teukolsky equation \cite{{teukolsky1973}}
\begin{alignat}{1}
\label{eq:teuk_eq}
\mathcal{T}_{lm\coorom}\left(r\right) &= \Delta^{2}\frac{d}{dr}\left(\frac{1}{\Delta}\frac{dR_{lm\coorom}\left(r\right)}{dr}
\right) -V_{lm\coorom}\left(r\right)R_{lm\coorom}\left(r\right) 
\quad  .
\end{alignat}
The potential $V_{lm\coorom}\left(r\right)$ depends in part on the
eigenvalue $\mathcal{C}_{lm}$ of $\SH (\theta)$, so equation
(\ref{eq:spheriodal}) must be solved before the homogeneous or
inhomogeneous version of equation (\ref{eq:teuk_eq}) can be.

The source term $\mathcal{T}_{lm\coorom}$ is built by, among other
things, evaluating $\SH (\theta)$ and two homogeneous
solutions\footnote{Other basis solutions to the homogeneous equation
  exist, e.g.\ the out/down basis
  $R^{\text{out/down}}_{lm\coorom}\left(r\right)$.  For a summary, see
  \cite{drasco2005} and references therein.}
$R^{\text{in/up}}_{lm\coorom}\left(r\right)$ to (\ref{eq:teuk_eq})
along the geodesic source.  Two general methods are described for
constructing $R^{\text{in/up}}_{lm\coorom}\left(r\right)$.  One
approach integrates the homogeneous Teukolsky equation (or,
equivalently, the better numerically behaved Sasaki-Nakamura equation
\cite{hughes2000}) outward from the horizon.  The other expands
$R^{\text{in/up}}_{lm\coorom}\left(r\right)$ in terms of
hypergeometric functions (see \cite{fujita2009} and references
therein) and evaluates $R^{\text{in/up}}_{lm\coorom}\left(r\right)$
directly at certain points, possibly extrapolating its values to
nearby points with series expansions.  Both approaches are fairly
computationally costly.  For detailed explanations on these different
approaches and how various numerical problems are circumvented, see
\cite{press1973, teukolsky1974, chandrasekhar1975, leaver1985,
  hughes2000, fujita2009}.  We elaborate a bit more on the structure
of the source term below.

\subsection{The quantities  $Z^{H/\infty}_{lm\coorom}$}
\label{subsec:Z_coefs}

With the homogeneous radial solutions in hand, the inhomogeneous
Teukolsky equations can be solved using the method of variation of
parameters\footnote{Most references use the method of Green functions,
  but variation of parameters works as well.} \cite{haberman}.  The
radial functions $R_{lm\coorom}\left(r\right)$ can be written
as \text{\cite{{drasco2004},{drasco2006},{Glampedakis:2005hs},{teukolsky1972},{teukolsky1973},{sasaki1982}}}
\begin{alignat}{1}
\label{eq:gen_sol}
R_{lm\coorom} (r) &=
Z^{H}_{lm\coorom} (r) R^{\text{up}}_{lm\coorom}(r) +
Z^{\infty}_{lm\coorom}(r) R^{\text{in}}_{lm\coorom}(r)
\quad ,
\end{alignat}
where $Z^{H}_{lm\coorom}\left(r\right)$ and
$Z^{\infty}_{lm\coorom}\left(r\right)$ are defined as
\begin{alignat}{1}
\label{eq:Z_lmomega}
  \begin{split}
    Z^{H}_{lm\coorom}(r) &= \int^{r}_{r_{+}} dr'\,
    \frac{R^{\text{in}}_{lm\coorom}(r')
      \mathcal{T}_{lm\coorom}(r')}{c}
    \\
    Z^{\infty}_{lm\coorom} (r) &= \int^{\infty}_{r} dr'\,
    \frac{R^{\text{up}}_{lm\coorom}(r')
      \mathcal{T}_{lm\coorom}(r')}{c}
    \quad .
  \end{split}
\end{alignat}
The constant $c$ is related to the Wronskian of
$R^{\text{in}}_{lm\coorom}\left(r\right)$ and
$R^{\text{up}}_{lm\coorom}\left(r\right)$. $r_{+}$ is the larger root
of $\Delta$ and is the radial coordinate of the black hole horizon
(the smaller root is denoted $r_{-}$).

Because we are only interested in the radiation going into the black
hole and being carried away to infinity, we are only concerned about
the asymptotic behavior of $R_{lm\coorom}\left(r\right)$ as $r\to
r_{+}$ and $r\to\infty$.  In fact, the homogeneous basis solutions
have been chosen to have the simplifying feature that
\label{eq:Z_asy_behave} 
\begin{alignat}{1}
  \begin{split}
    Z^{\infty}_{lm\coorom} (r\to r_{+}) &= Z^{\infty}_{lm\coorom}
    \\
    Z^{H}_{lm\coorom}(r\to r_{+}) &= 0
    \\
    Z^{\infty}_{lm\coorom}(r\to\infty) &= 0
    \\
    Z^{H}_{lm\coorom}(r\to\infty) &= Z^{H}_{lm\coorom}
    \quad .    
  \end{split}
\end{alignat}
Note that we have used the same notation for the \emph{functions}
$Z^{H/\infty}_{lm\coorom}(r)$ and for the \emph{constants}
$Z^{H/\infty}_{lm\coorom}$ representing their asymptotic values at
$\infty$ and $r_{+}$, respectively.  As mentioned in Section
\ref{sec:fluxes-elqall}, the literature seems stuck with the rather
backward notational convention that $Z^{H}_{lm\coorom}$ is
nonvanishing at $\infty$ while $Z^{\infty}_{lm\coorom}$ is
nonvanishing at $r_{+}$.

The radial functions as $r\to\infty$ and $r\to r_{+}$ thus become
\begin{subequations}
\label{eq:radial_funcs}
\begin{alignat}{1}
  \begin{split}
    R_{lm\coorom}^{\infty} &= R_{lm\coorom} (r\rightarrow\infty)
    \\
    &= Z_{lm\coorom}^{H}R^{\text{up}}_{lm\coorom}(r\rightarrow\infty)
    \\
    &= Z_{lm\coorom}^{H}r^{3}e^{i\coorom r^{*}}
  \end{split}
  \\
  \begin{split}
    R_{lm\coorom}^{H} &= R_{lm\coorom}(r\rightarrow r_{+}) 
    \\
    &= Z_{lm\coorom}^{\infty}R^{\text{in}}_{lm\coorom}
    (r\rightarrow r_{+})
    \\ 
    &= Z_{lm\coorom}^{\infty} \Delta^{2} e^{-ikr^{*}}
    \quad ,
  \end{split}
\end{alignat}
\end{subequations}
where and $k \equiv \coorom - ma/(2 r_{+})$ and $r^{*}$ is the Kerr
tortoise coordinate defined by
\begin{alignat}{1}
\label{eq:tortoise_coord_defn}
  \begin{split}
    r^{*}(r) &= r +
    \frac{2r_{+}}{r_{+} - r_{-}} \ln \frac{r - r_{+}}{2} -
    \frac{2r_{-}}{r_{+} - r_{-}} \ln \frac{r - r_{-}}{2}
    \\
    \D{r}{r^{*}} &= \frac{r^{2} + a^{2}}{\Delta}
    \quad .
  \end{split}
\end{alignat}

The coefficients $Z^{H/\infty}_{lm\coorom}$ are found from
\begin{alignat}{1}
\label{eq:Z_asym}
Z^{H/\infty}_{lm\coorom} &=
\int^{\infty}_{r_{+}} dr'\,
\frac{R^{\text{in/up}}_{lm\coorom}(r')
\mathcal{T}_{lm\coorom}(r')}{c}
\quad .
\end{alignat}
The source function $\mathcal{T}_{lm\coorom}\left(r\right)$ in
(\ref{eq:Z_asym}) is an integral of the form
\begin{multline}
\label{eq:source_fn}
\mathcal{T}_{lm\coorom}(r) = 
\\
\int^{\infty}_{-\infty} dt\,
\int d\Omega\, B_{m\coorom} (t,r,\theta,\varphi)
\SH e^{i \coorom t} e^{-i m \varphi}
\quad .
\end{multline}
The source term is derived in \cite{breuer}, but we have written it
borrowing the notation $B_{m\coorom}$ from \cite{drasco2006}.  All
that matters for our purposes is that $B_{m\coorom}$ is built out of a
series of operations on the null tetrad components of the
energy-momentum tensor of the orbiting particle and thus contains
delta functions and derivatives of delta functions centered on the
source geodesic.  Thus, the $d\Omega$ integral can be evaluated,
resulting in every $\theta$ and $\varphi$ in (\ref{eq:source_fn})
being replaced with the source trajectories $\theta_{s}(t),
\varphi_{s}(t)$ (the subscript $s$ here denotes ``source'' as opposed
to a spin-weight as earlier in this Appendix).

Delta functions $\delta \left( r - r_{s}(t) \right)$ and derivatives
thereof still remain in (\ref{eq:source_fn}), along with the
integration over $t$.  When we plug (\ref{eq:source_fn}) into equation
(\ref{eq:Z_asym}), we can switch the order of integration for $r'$ and
$t$ and use those remaining delta functions in $r$ (we rename $r'$ to
$r$ now, for simplicity) to replace every $r$ with $r_{s}(t)$.  The
net result is that $Z^{H/\infty}_{lm\coorom}$ takes the form
\begin{alignat}{1}
\label{eq:source_fourier_integral}
Z^{H/\infty}_{lm\coorom} &=
\int^{\infty}_{-\infty}dt\, 
e^{i\coorom t}e^{-im\varphi_{s}\left(t\right)}
\mathcal{I}_{lm\coorom}^{H/\infty}
\left(r_{s}\left(t\right),\theta_{s}\left(t\right)\right)
\quad .
\end{alignat}
The functions $\mathcal{I}_{lm\coorom}^{H/\infty}$ depend on $r$ and
$\theta$ both directly and via a combination of elementary functions,
the spheroidal harmonics, the homogeneous radial Teukolsky functions,
and various derivatives thereof.  Explicit expressions can be found in
several sources (see, for instance, \cite{drasco2006,fujita2009}).

We will now use this form for $Z^{H/\infty}_{lm\coorom}$ to define the
coefficients $Z^{H/\infty}_{lmkn}$ and $Z^{H/\infty}_{lmj;\lambda}$ to
which the efficiency arguments of Sections \ref{sec:definite} and
\ref{sec:spec} apply, respectively.

\subsection{The quantities $Z^{H/\infty}_{lmkn}$}
\label{sec:quant-zhinfty_lmkn}

We now show that, when the source is a non-resonant orbit, the
$\coorom$-dependence of $Z^{H/\infty}_{lm\coorom}$ takes the form
\begin{equation}
\label{eq:Zlmkn_as_weights_of_delta_fcn_comb}
  Z^{H/\infty}_{lm\coorom} =
  \sum_{kn} Z^{H/\infty}_{lmkn; \lambda}
  \delta \left( \coorom - \coorommkn \right)
  \quad ,
\end{equation}
where $\coorommkn$ are the coordinate-time harmonic frequencies
defined in equation (\ref{eq:omega_non_res}).  Note that even though
equation (\ref{eq:Zlmkn_as_weights_of_delta_fcn_comb}) has the form of
the Fourier transform of an almost-periodic function with respect to
coordinate time $t$, we are free to interpret the coefficients of the
delta functions either as Fourier coefficients of a coordinate time
function or as Fourier coefficients of a (different) Mino time
function.

For the reasons stated in Appendix \ref{subsec:which_time}, we opt for
the latter and begin by rewriting (\ref{eq:source_fourier_integral})
as an integral over Mino time.  Treating all source coordinates as
functions of $\lambda$ and using equation (\ref{eq:t_of_lambda}), we
get
\begin{multline}
\label{eq:source_fourier_integral_over_lambda}
Z^{H/\infty}_{lm\coorom} =
\\
\int^{\infty}_{-\infty} d\lambda\, 
e^{i\coorom \left( \Gamma\lambda + \Delta t(\lambda) \right)}
e^{-im\varphi_{s} (\lambda)}
\mathcal{I}_{lm\coorom}^{H/\infty}
\left( r_{s}(\lambda),\theta_{s}(\lambda) \right)
\,.
\end{multline}

We now use the fact (see Ref.\ \cite{drasco2004}) that, like
$\txtD{\lambda}{t}$, $\txtD{\lambda}{\varphi}$ depends on $r$ and
$\theta$ and is thus biperiodic when evaluated on a trajectory
$r(\lambda), \theta(\lambda)$.  $\txtD{\lambda}{\varphi}$ has a
nonzero average value $\omphi$ defined by
\begin{equation}
\label{eq:defn_Omega_phi}
  \omphi \equiv 
  \lim_{\Lambda \to \infty}
  \frac{1}{\Lambda}
  \int_{-\Lambda/2}^{\Lambda/2}
  d\lambda\, \D{\lambda}{\varphi}
  \quad .
\end{equation}
Consequently, the function $\varphi_{s}(\lambda)$ takes the form
\begin{equation}
\label{eq:phi_of_lambda}
  \varphi_{s}(\lambda) = \omphi\lambda +
  \Delta \varphi_{s}(\lambda)
\end{equation}
where $\Delta \varphi_{s}(\lambda)$ is biperiodic in $\lambda$ and has
zero average value.  Like its radial and polar counterparts, $\omphi$
is related to the coordinate-time frequency $\cooromphi$ via
\begin{equation}
\label{eq:rel_between_Minotime_and_coordtime_phi_freqs}
  \cooromphi = \frac{1}{\Gamma} \omphi
  \quad .
\end{equation}
More generally, coordinate-time and Mino-time frequencies are related
by
\begin{equation}
\label{eq:rel_between_Minotime_and_coordtime_gen_freqs}
  \coorom = \frac{1}{\Gamma} \om
  \quad .
\end{equation}
In light of equations
(\ref{eq:phi_of_lambda})--(\ref{eq:rel_between_Minotime_and_coordtime_gen_freqs}),
equation (\ref{eq:source_fourier_integral_over_lambda}) becomes
\begin{multline}
\label{eq:source_fourier_integral_over_lambda_with_omphi}
Z^{H/\infty}_{lm\coorom} =
\\
\int^{\infty}_{-\infty} d\lambda\, 
e^{i\om \lambda} e^{-i m \omphi \lambda}
e^{i \coorom \Delta t(\lambda)}
e^{-i m \Delta\varphi_{s} (\lambda)}
\mathcal{I}_{lm\coorom}^{H/\infty}
\left( r_{s}(\lambda),\theta_{s}(\lambda) \right)
\,.
\end{multline}
Note that the coordinate-time frequency $\coorom$ still appears as a
parameter in both $\mathcal{I}_{lm\coorom}^{H/\infty}$ and in the
argument of $e^{i \coorom \Delta t(\lambda)}$.

Like $\Delta t(\lambda)$ and $\Delta\varphi(\lambda)$, the functions
$\mathcal{I}_{lm\coorom}^{H/\infty}$ are biperiodic in $\lambda$.  We
define the biperiodic function
\begin{equation}
\label{eq:f_lmomega_defn}
  f^{H/\infty}_{lm\coorom} (\lambda) \equiv
  e^{i \coorom \Delta t(\lambda)}
  e^{-i m \Delta\varphi_{s} (\lambda)}
  \mathcal{I}_{lm\coorom}^{H/\infty}
  \left( r_{s}(\lambda),\theta_{s}(\lambda) \right)
\end{equation}
and Fourier expand it as
\begin{equation}
\label{eq:f_lmomega_time_Fourier_expnsn}
  f^{H/\infty}_{lm\coorom} (\lambda) \equiv
  \sum_{k,n}
  Z^{H/\infty}_{lm\coorom;kn;\lambda}
  e^{-i (\n\omth + \kr\omr) \lambda}
  \quad .
\end{equation}

Inserting (\ref{eq:f_lmomega_time_Fourier_expnsn}) into equation
(\ref{eq:source_fourier_integral_over_lambda_with_omphi}) yields
\begin{alignat}{1}
  \label{eq:Zlmomega_as_delta_fcn_comb}
  \begin{split}
    Z^{H/\infty}_{lm\coorom} &=
    \sum_{k,n}
    Z^{H/\infty}_{lm\coorom;kn;\lambda}
    \int^{\infty}_{-\infty} d\lambda\,
    e^{i\om \lambda}
    e^{-i (m \omphi + \n\omth + \kr\omr) \lambda}
    \\
    &=
    \sum_{k,n}
    Z^{H/\infty}_{lm\coorom;kn;\lambda}
    2 \pi
    \delta (\coorom - \coorommkn)
    \\
    &=
    \sum_{k,n}
    Z^{H/\infty}_{lm\coorom=\coorommkn;kn;\lambda}
    2 \pi
    \delta (\coorom - \coorommkn)
    \quad .
  \end{split}
\end{alignat}
The multipole index $m$ and the Fourier indices $k,n$ all do double
duty by helping to specify the value of the parameter $\coorom$ in
$Z^{H/\infty}_{lm\coorom=\coorommkn;kn;\lambda}$.  Consequently, each
$Z^{H/\infty}_{lm\coorom=\coorommkn;kn;\lambda}$ is fully specified by
the four integers $l,m,k,n$ and we can define
\begin{equation}
\label{eq:Z_lmkn_lambda_defn}
  Z^{H/\infty}_{lmkn;\lambda} \equiv
  2\pi
  Z^{H/\infty}_{lm\coorom=\coorommkn;kn;\lambda}
  \quad .
\end{equation}
We will absorb the factor of $2 \pi$ into the function
$f^{H/\infty}_{lm\coorom}$ in the integral defining
$Z^{H/\infty}_{lmkn;\lambda}$.

Since each $Z^{H/\infty}_{lmkn;\lambda}$ is a temporal Fourier
coefficient, then by the arguments of Section
\ref{sec:four-analys-time-nonres}, we determine it by instead
computing the corresponding spatial Fourier coefficient
\begin{multline}
\label{eq:Z_lmkn_as_torus_coeff}
  Z^{H/\infty}_{lmkn} \equiv
  \\
  \frac{1}{(2\pi)^{2}}
  \int_{0}^{2\pi} d\angr\,
  \int_{0}^{2\pi} d\angth\,
  e^{i \n\angth} e^{i \kr\angr}
  f^{H/\infty}_{lm\coorom=\coorommkn} \angset
\end{multline}
of the associated torus function in the $\angvec$ torus coordinates.
Despite our earlier notation, the function
$\mathcal{I}_{lm\coorom}^{H/\infty}$ actually depends not only on $r$
and $\theta$ but also on their conjugate momenta (since it depends on
the $r$ and $\theta$ velocities via the energy-momentum tensor of the
particle).  $ f^{H/\infty}_{lm\coorom}$ inherits this dependence, and
it is thus appropriate to write it as a function of the torus
coordinates $\angvec$ that need not have any special symmetries on the
torus.

As we show in Appendix \ref{sec:using_TSN}, the averaged fluxes
required to evaluate the RHS of the adiabatic equations for $\elqvec$
depend on the $Z^{H/\infty}_{lmkn}$ defined in
(\ref{eq:Z_lmkn_as_torus_coeff}).  Thus, equation
(\ref{eq:Z_lmkn_as_torus_coeff}) verifies equation
(\ref{eq:Zkn_as_torus_fcn_Fourier_coeff}), on which the savings
arguments of Section \ref{sec:zhinf_lmnkr-as-fourier-torus-coeffs} are
based.

\subsection{The quantities $Z^{H/\infty}_{lm\jo; \lambda}$}
\label{sec:quant-zhinfty_lmjo-lambda}

We now show that when the source is a resonant orbit, the
$\coorom$-dependence of $Z^{H/\infty}_{lm\coorom}$ instead takes the
form
\begin{equation}
\label{eq:Zlmj_as_weights_of_delta_fcn_comb}
  Z^{H/\infty}_{lm\coorom} =
  \sum_{\jo} Z^{H/\infty}_{lm\jo; \lambda}
  \delta \left( \coorom - \coorommj \right)
  \quad ,
\end{equation}
where $\coorommj$ are now the coordinate-time harmonic frequencies
defined in equation (\ref{eq:omega_res}).  As before, we are free to
interpret the coefficients of the delta functions either as Fourier
coefficients of a coordinate time function or as Fourier coefficients
of a (different) Mino time function and take the computationally
tractable latter option.

Equations
(\ref{eq:source_fourier_integral_over_lambda})--(\ref{eq:f_lmomega_defn})
carry over to the resonant case with the difference that each of
$\Delta t(\lambda), \Delta \varphi (\lambda), r(\lambda), \theta
(\lambda)$ and thus $\mathcal{I}^{H/\infty}_{lm\coorom}$ is now singly
periodic with period $\Tnot$.  Thus $f^{H/\infty}_{lm\coorom}$ can be
Fourier expanded in harmonics of a single fundamental frequency
$\omnot$,
\begin{equation}
\label{eq:f_lmomega_time_res_Fourier_expnsn}
  f^{H/\infty}_{lm\coorom} (\lambda; \offvec) \equiv
  \sum_{\jo}
  Z^{H/\infty}_{lm\coorom;\jo;\lambda} (\offvec)
  e^{-i \jo\omnot \lambda}
\end{equation}
with
\begin{equation}
\label{eq:Zlmomegaj_defn}
  Z^{H/\infty}_{lm\coorom;\jo;\lambda} (\offvec) =
  \frac{1}{\Tnot}
  \int_{0}^{\Tnot} d\lambda\,
  e^{i \jo\omnot \lambda}
  f^{H/\infty}_{lm\coorom} (\lambda; \offvec)
  \quad .
\end{equation}
The functions $f^{H/\infty}_{lm\coorom} (\lambda; \offvec)$ are
induced from some torus function, so by the arguments of Section
\ref{sec:four-analys-time-res}, both they and the coefficients
$Z^{H/\infty}_{lm\coorom;\jo;\lambda} (\offvec)$ depend on initial
positions, which we can represent compactly as a dependence on initial
position $\offvec$ on the phase space torus.

Inserting (\ref{eq:f_lmomega_time_res_Fourier_expnsn}) into the
resonant version of equation
(\ref{eq:source_fourier_integral_over_lambda_with_omphi}) yields
\begin{alignat}{1}
  \label{eq:Zlmomega_as_res_delta_fcn_comb}
  \begin{split}
    Z^{H/\infty}_{lm\coorom} &=
    \sum_{\jo}
    Z^{H/\infty}_{lm\coorom;\jo;\lambda}
    \int^{\infty}_{-\infty} d\lambda\,
    e^{i\om \lambda}
    e^{-i (m \omphi + \jo\omnot) \lambda}
    \\
    &=
    \sum_{\jo}
    Z^{H/\infty}_{lm\coorom;\jo;\lambda}
    2 \pi
    \delta (\coorom - \coorommj)
    \\
    &=
    \sum_{\jo}
    Z^{H/\infty}_{lm\coorom=\coorommj;\jo;\lambda}
    2 \pi
    \delta (\coorom - \coorommj)
    \quad .
  \end{split}
\end{alignat}
The relevant quantities, then, are those in equation
(\ref{eq:Zlmomegaj_defn}) with the parameter $\coorom$ in
$f^{H/\infty}_{lm\coorom}$ set to $\coorommj$.

Paralleling the non-resonant case, each
$Z^{H/\infty}_{lm\coorom=\coorommj;\jo;\lambda}$ is fully specified by
the three integers $l,m,\jo$.  By absorbing the factor of $2 \pi$ into
the functions $f^{H/\infty}_{lm\coorom}$ in the integrand of
(\ref{eq:Zlmomegaj_defn}), we can define the notationally more compact
coefficients
\begin{equation}
\label{eq:Z_lmj_lambda_defn}
  Z^{H/\infty}_{lm\jo;\lambda} \equiv
  2\pi
  Z^{H/\infty}_{lm\coorom=\coorommj;\jo;\lambda}
  \quad .
\end{equation}
By the construction above, each such $Z^{H/\infty}_{lm\jo;\lambda}$ is
given by
\begin{equation}
\label{eq:Zlmj_defn}
  Z^{H/\infty}_{lm\jo;\lambda} =
  \frac{1}{\Tnot}
  \int_{0}^{\Tnot} d\lambda\,
  e^{i \jo\omnot \lambda}
  f^{H/\infty}_{lm\coorom=\coorommj} (\lambda; \offvec)
  \quad .
\end{equation}

As we show in Appendix \ref{sec:using_TSN}, the time-averaged fluxes
for $\elqvec$ from resonant orbits depend on the $Z^{H/\infty}_{lm\jo;
  \lambda}$ defined in (\ref{eq:Zlmj_defn}).  Thus, equation
(\ref{eq:Zlmj_defn}) verifies equation (\ref{eq:res_Zj_defn}), on
which the more speculative savings arguments of Section \ref{sec:spec}
are based.

\section{Fluxes from the Teukolsky formalism}
\label{sec:using_TSN}

In this appendix, we review how the apparatus of Appendix
\ref{sec:tuek_formalism} yields fluxes of conserved quantities.
Several authors
\cite{{hughes2001},{drasco2006},{glampedakis2002},{drasco2005},ganz2007,fujita2009}
implement this Fourier-domain formalism in TB codes to calculate the
radiative $\elqvec$ fluxes at radial infinity and the horizon to
determine how the inspiral evolves.  We show how expressions for
time-averaged (as opposed to torus-averaged) fluxes differ between
non-resonant and resonant orbits.

\subsection{Overview of flux calculation}
\label{subsec:gen_flux_calc}

We will not refer to the $Q$ flux, but restrict our discussion to $E$
and $L_{z}$.  To determine the evolution of an inspiral in orbital
parameter space, we use the the $E$ and $L_{z}$ fluxes at radial
infinity and the horizon as proxies for the local self-force.  This
subsection gives an overview of the calculation for finding these
fluxes.

From the Weyl scalar $\psi_{4}$, the gravitational waveform and the
$E$, $L_{z}$ and $Q$ radiation fluxes can be calculated.
Specifically, we can calculate the polarizations $h_{+}$ and
$h_{\times}$ of the metric perturbations at infinity (i.e.\ the GWs)
via
\begin{eqnarray}
\label{eq:psi_4_grav_wave}
\psi_{4} &=& \frac{1}{2}\frac{\partial^{2}}{\partial t^{2}}\left(h_{+}+ih_{\times}\right)
\quad .
\end{eqnarray}
After integrating equation (\ref{eq:psi_4_grav_wave}) twice to get
$h_{+}+ih_{\times}$ (the integration constants are set to zero), we
can calculate an effective GW stress-energy tensor at infinity
\cite{isaacson1968} as
\begin{eqnarray}
\label{eq:stress_eng_tensor}
T^{\infty}_{\alpha\beta} &=& \frac{1}{16\pi}\left\langle \frac{\partial
  h_{+}}{\partial x^{\alpha}}\frac{\partial{h_{+}}}{\partial
  x^{\beta}}+\frac{\partial h_{\times}}{\partial
  x^{\alpha}}\frac{\partial{h_{\times}}}{\partial x^{\beta}}
\right\rangle_{\substack{\text{several}\\ \text{wavelengths}}}
\!\!\!\!\!\!.
\end{eqnarray}  
The average over several wavelengths signifies the following
\cite{maggiore}.  The stress-energy tensor of GWs contributes to
curvature in the analogus way that the stress-energy tensor of matter
does.  However, since GWs are fluctuations in the metric itself, we
cannot define their stress-energy tensor at a field point because,
with only information at one point, we cannot distinguish between the
curvature of the background spacetime and the contributions to that
curvature from fluctuations on the spacetime.  To distinguish between
the background and the effect of the fluctuations, there needs to be
either a length or frequency scale separation between the two.  In our
case, as $r\rightarrow\infty$, the background curvature scale is much
greater than the wavelength of the fluctuations.  Therefore, we
average over several gravitational wavelengths in order to smooth out
the fluctuations and determine only the secular contribution of such
GWs to the curvature.

While the background curvature scale is much greater than the GW
wavelength, GW detectors look not for spatial fluctuations to the
metric but rather temporal ones.  Therefore, in the spirit of
applicability to actual experiments, it is more useful to distingiush
between a background frequency (i.e.\ a reciprocal of a curvature
scale in the timelike direction) and the much larger frequency of the
fluctuations.  Therefore, rather then average over several
wavelengths, we average over several periods to isolate the net effect
of the GWs \cite{maggiore}.

The energy and angular momentum fluxes carried to radial infinity by
GWs are related to the components of $T^{\infty}_{\alpha\beta}$ by,
\begin{eqnarray}
\label{eq:fluxes_from_T}
T^{\infty}_{tt} &=& \frac{dE}{dtdA}
\\
\nonumber
T^{\infty}_{t\varphi} &=& \frac{dL_{z}}{dtdA}
\quad .
\end{eqnarray}
The $E$ and $L_{z}$ fluxes are calculated by integrating equation
(\ref{eq:fluxes_from_T}) over a $2$-sphere of radius $r$ on a constant
$t$ spacelike hypersurface
\begin{eqnarray}
\label{eq:E_L_fluxes_from_T}
\left(\frac{dE}{dt}\right)^{\infty} &=& \int T^{\infty}_{tt}r^{2} d\Omega
\\
\nonumber
\left(\frac{dL_{z}}{dt}\right)^{\infty} &=& \int T^{\infty}_{t\varphi} r^{2}d\Omega
\quad .
\end{eqnarray}
The time-averaged fluxes from a given geodesic can then be calculated
by
\begin{eqnarray}
\label{eq:time_avg_flux}
\left\langle \frac{dE}{dt}\right\rangle_{t}^{\infty} &=&
\lim_{T\rightarrow\infty}\frac{1}{T}\int^{\frac{T}{2}}_{-\frac{T}{2}}\left(\frac{dE}{dt}\right)^{\infty}dt
\\ 
\nonumber 
\left\langle\frac{dL_{z}}{dt}\right\rangle_{t}^{\infty} &=&
\lim_{T\rightarrow\infty}\frac{1}{T}\int^{\frac{T}{2}}_{-\frac{T}{2}}\left(\frac{dL_{z}}{dt}\right)^{\infty}dt
\quad .
\end{eqnarray} 

Alternatively, we can calculate the torus-averaged fluxes by taking
the average of the time-averages for all geodesics with a given set of
orbital parameters over all possible initial conditions,
\begin{eqnarray}
\label{eq:time_avg_flux}
\left\langle \frac{dE}{dt}\right\rangle_{\vec{\gamma}}^{\infty} &=&
\frac{1}{\left(2\pi\right)^{2}}\int^{2\pi}_{0}d\gamma_{r_o}\int^{2\pi}_{0}d\gamma_{\theta_o}\left\langle\frac{dE}{dt}\right\rangle^{\infty}_{t}
\\ 
\nonumber 
\left\langle\frac{L_{z}}{dt}\right\rangle_{\vec{\gamma}}^{\infty}
&=&
\frac{1}{\left(2\pi\right)^{2}}\int^{2\pi}_{0}d\gamma_{r_o}\int^{2\pi}_{0}d\gamma_{\theta_o}\left\langle\frac{dL_{z}}{dt}\right\rangle^{\infty}_{t}
\quad .
\end{eqnarray} 
An analougus procedure can be performed to calculate the fluxes at the
horizon \cite{hawking1972,teukolsky1974}.

\subsection{Fluxes from $\psi_{4}$}
\label{sec:old-appx-c}

Combining equations (\ref{eq:psi_4_quad}) with (\ref{eq:radial_funcs})
we find that 
\begin{subequations}
\label{eq:psi_4_asymps}
\begin{alignat}{1}
\psi_{4}^{\infty} &= \rho^{-4}\sum_{lmkn}Z_{lmkn;\lambda}^{H}r^{3}e^{i\coorom_{mkn}r^*}S^{a\coorom_{mkn}}_{lm}\left(\theta\right)e^{-i\coorom_{mkn}t+im\varphi}
\\
\intertext{for non-resonant orbits, and}
\psi_{4}^{\infty} &= \rho^{-4}\sum_{lmj}Z_{lmj;\lambda}^{H}r^{3}e^{i\coorom_{mj}r^*}S^{a\coorom_{mj}}_{lm}\left(\theta\right)e^{-i\coorom_{mj}t+im\varphi}
\quad .
\end{alignat}
\end{subequations}
Similar expressions can be found for $\psi_{4}^{H}$, but for brevity,
we will only proceed with the detailed computations for
$r\rightarrow\infty$.  Also, we only work out the details for the $E$
flux, but the $L_{z}$ flux follows exactly the same perscription.

Following the prescription laid out in section
\ref{subsec:gen_flux_calc} we find that 
\begin{widetext}
\label{eq:e_l_dot_inst}
\begin{alignat}{1}
\label{eq:e_l_dot_inst_time_coefs}
\left(\frac{dE}{dt}\right)^{\infty}
&=\frac{1}{4\pi}\left\langle\sum_{lm\coorom}\sum_{l'm'\coorom'}\frac{1}{\coorom\coorom'}Z^{H}_{lm\coorom;\lambda}\bar{Z}^{H}_{l'm'\coorom';\lambda}e^{i\left(\coorom-\coorom'\right)r^{*}}
e^{-i\left(\coorom-\coorom'\right)t}\int^{\pi}_{0}\sin{\theta}S^{a\coorom}_{lm}{\left(\theta\right)}\bar{S}^{a\coorom'}_{l'm'}\left(\theta\right)d\theta\int^{2\pi}_{0}
d\varphi
e^{i\left(m-m'\right)\varphi}\right\rangle_{\substack{\text{several}\\\text{periods}}}
\quad , 
\end{alignat}
\end{widetext}
where $\coorom, \coorom'$ denote the discrete variables that are
two-indexed for resonant orbits and three-indexed for non-resonant
orbits.  Performing the $\varphi$ integration in equation
(\ref{eq:e_l_dot_inst_time_coefs}) yields $2\pi \delta_{mm'}$. We thus
set $m=m'$ everywhere.

For non-resonant orbits, equation (\ref{eq:e_l_dot_inst_time_coefs}) becomes
\begin{widetext}
\begin{alignat}{1}
\label{eq:e_l_dot_inst_time_coefs_non_res}
\left(\frac{dE}{dt}\right)^{\infty} &=\frac{1}{2}\left\langle\sum_{lmkn}\sum_{l'k'n'}\frac{1}{\coorom_{mkn}\coorom_{mk'n'}}Z^{H}_{lmkn;\lambda}\bar{Z}^{H}_{l'mk'n';\lambda}
e^{-i\left\{\left(\kr-\kr'\right)\coorom_{r}+\left(\n-\n'\right)\coorom_{\theta}\right\}\left(t-r^{*}\right)}\int_{0}^{\pi}d\theta\sin\theta
S^{a\coorom_{mkn}}_{lm}\bar{S}^{a\coorom_{mk'n'}}_{l'm}\right\rangle_{\substack{\text{several}\\\text{periods}}}
\quad .
\end{alignat}
\end{widetext}
We are interested in an infinite time-average of equation
(\ref{eq:e_l_dot_inst_time_coefs_non_res}).  Therefore, we can drop
the average over several periods because the time-averaging process
will smooth out the fluxes so that the period averaging will have no
further effect once we have time-averaged.  The time-average of a
function corresponds to the constant term in a Fourier expansion, so
the argument of the exponential in $t$ will need to be zero.
Therefore, performing the inifinite time-average yields the added
conditions $k=k'$ and $n=n'$.  This equates the frequencies
everywhere, including in the spheroidal harmonics.  We can therefore
now perform the $\theta$ integration using (\ref{eq:spheroidal_delta})
and get $l=l'$.  The result is that the infinite time average in the
non-resonant case is
\begin{alignat}{1}
\label{eq:e_l_dot_infinite_time_avg_non_res}
\left\langle\frac{dE}{dt}\right\rangle^{\infty}_{t} &=\sum_{lmkn}\frac{1}{4\pi\coorom_{mkn}^{2}}\left|Z^{H}_{lmkn;\lambda}\right|^{2}
\quad .
\end{alignat}

In Section \ref{sec:four-analys-time-nonres}, we saw that infinite
time averages over non-resonant orbits are the same as torus averages
over non-resonant tori with respect to the corresponding torus
coordinates.  The time average over $t$ on the LHS of
(\ref{eq:e_l_dot_infinite_time_avg_non_res}) thus corresponds to a
torus average over the $\vec{\gamma}$ torus coordinates.  We also saw
that
$\left|Z^{H}_{lmkn;\lambda}\right|^{2}=\left|Z^{H}_{lmkn}\right|^{2}$,
where the $Z^{H}_{lmkn}$ are spatial Fourier coefficients (for fixed
$l,m$) with respect to the $\angvec$ torus coordinates.  Therefore,
the torus averaged energy flux is
\begin{alignat}{1}
\label{eq:e_l_dot_tor_avg_all}
\left\langle\frac{dE}{dt}\right\rangle^{\infty}_{\vec{\gamma}}
&=
\sum_{lmkn}\frac{1}{4\pi\coorom_{mkn}^{2}}
\left|Z^{H}_{lmkn}\right|^{2}
\quad .
\end{alignat}
This expression is true for all tori, as torus averages are
insensitive to whether the orbits on that torus are resonant or
non-resonant.

Analogous arguments lead to the angular momentum flux at infinity.
Similar arguments to those above then lead to the corresponding fluxes
at the horizon.  The upshot is that all the torus-averaged fluxes are
given by
\begin{eqnarray}
\label{eq:tor_av_ti_av_non_res}
\left\langle\frac{dE}{dt}\right\rangle^{H/\infty}_{\vec\gamma}
&=& \sum_{lmkn}\frac{\alpha_{lmkn}^{H/\infty}}{4\pi\coorom^{2}_{mkn}}\left|Z^{H/\infty}_{lmkn}
\right|^{2} 
\\
\nonumber
\left\langle\frac{dL_z}{dt}\right\rangle^{H/\infty}_{\vec\gamma} &=& \sum_{lmkn}\frac{\alpha_{lmkn}^{H/\infty}m}{4\pi\coorom^{3}_{mkn}}\left|Z^{H/\infty}_{lmkn}
\right|^{2} \quad,
\end{eqnarray}
where $\alpha_{lmkn}^{\infty}\equiv 1$ and the details of
$\alpha_{lmkn}^{H}$ can be found in reference \cite{teukolsky1974}.
We note that there is no residual dependence on the initial conditions
$\offvec$.

We return to equation (\ref{eq:e_l_dot_inst_time_coefs_non_res}) and
evaluate it for a resonant orbit,
\begin{widetext}
\begin{alignat}{1}
\label{eq:e_l_dot_inst_time_coefs_res}
\left(\frac{dE}{dt}\right)^{\infty} &=\frac{1}{2}\left\langle
\sum_{lmj}\sum_{l'j'}\frac{1}{\coorom_{mj}\coorom_{mj'}}Z^{H}_{lmj;\lambda}Z^{*H}_{lmj';\lambda}
e^{-i\left(\jo-\jo'\right)\cooromnot\left(t-r^{*}\right)}\int_{0}^{\pi}d\theta\sin\theta
S^{a\coorom_{mj}}_{lm}\bar{S}^{a\coorom_{mj'}}_{l'm}\right\rangle_{\substack{\text{several}\\\text{periods}}}
\quad .
\end{alignat}
\end{widetext}
As was the case with the non-resonant infinite time average, we can
drop the averaging over several periods.  Additionally, the resonant
time-average picks out the constant term in the Fourier expansion,
which results when $j=j'$.  Therefore,
\begin{eqnarray}
\label{eq:e_flux_ti_res}
\left\langle\frac{dE}{dt}\right\rangle ^{\infty}_{t} &=&
\sum_{lmj}\frac{1}{4\pi\coorom_{mj}^{2}}\left|Z^{H}_{lmj;\lambda}\right|^{2}
\quad .
\end{eqnarray}

The rest of the $E$ and $L_{z}$ time-averaged fluxes can be found
similarly and are,
\begin{eqnarray}
\label{eq:el_flux_ti_res}
\left\langle\frac{dE}{dt}\right\rangle ^{H/\infty}_{t} &=&
\sum_{lmj}\frac{\alpha_{lmj}^{H/\infty}}{4\pi\coorom^{2}_{mj}}\left|Z^{H/\infty}_{lmj;\lambda}\right|^{2}
\\
\nonumber
\left\langle\frac{dL_{z}}{dt}\right\rangle ^{H/\infty}_{t} &=&
\sum_{lmj}\frac{\alpha_{lmj}^{H/\infty}m}{4\pi\coorom^{3}_{mj}}\left|Z^{H/\infty}_{lmj;\lambda}\right|^{2}
\quad .
\end{eqnarray}
We remark that unlike the torus-averaged fluxes, the time-averaged
fluxes of resonant orbits clearly depend on the initial conditions of
the orbit, since as we saw in Section
\ref{sec:quant-zhinfty_lmjo-lambda},
$\left|Z^{H/\infty}_{lmj;\lambda}\right|^{2}$ is not the same for all
initial conditions.

Alternatively, we can write the time-averaged fluxes of equation
(\ref{eq:el_flux_ti_res}) explicitly in terms of the torus
coefficients $Z_{lmkn}^{H/\infty}$.  From section
\ref{sec:four-analys-time-res} we know that
\begin{alignat}{1}
\label{eq:torus_Z_coeffs}
Z_{lmj;\lambda}^{H/\infty} &=
\sum_{\substack{kn:\\\num\kr+\den\n=j}}Z_{lmkn}^{H/\infty}e^{-i\kr\chi_{r_0}}e^{-i\n\chi_{\theta_0}}
\quad .
\end{alignat}
Therefore, we can rewrite equation (\ref{eq:el_flux_ti_res}) as
\begin{widetext}
\begin{eqnarray}
\label{eq:el_flux_ti_res_tor}
\left\langle\frac{dE}{dt}\right\rangle ^{H/\infty}_{t} &=&
\sum_{lmkn}\sum_{\substack{k'n':\\\den\kr_+\num\n=\\\den\kr'+\num\n'}}\frac{\alpha_{lmkn}^{H/\infty}}{4\pi\omega^{2}_{mkn}}Z^{H/\infty}_{lmkn}Z^{*H/\infty}_{lmk'n'}e^{-i\left\{\left(\kr-\kr'\right)\chi_{r_o}+\left(\n-\n'\right)\chi_{\theta_o}\right\}}
\\
\nonumber
\left\langle\frac{dL_{z}}{dt}\right\rangle ^{H/\infty}_{t} &=&
\sum_{lmkn}\sum_{\substack{k'n':\\\den\kr_+\num\n=\\\den\kr'+\num\n'}}\frac{\alpha_{lmkn}^{H/\infty}m}{4\pi\omega^{3}_{mkn}}Z^{H/\infty}_{lmkn}Z^{*H/\infty}_{lmk'n'}e^{-i\left\{\left(\kr-\kr'\right)\chi_{r_o}+\left(\n-\n'\right)\chi_{\theta_o}\right\}}
\quad .
\end{eqnarray}
\end{widetext}

The explicit initial condition dependence is now evident.  Notice that
if we average the time-averaged flux expressions of equation
(\ref{eq:el_flux_ti_res_tor}) over all possible initial conditions, we
reproduce the torus-averaged fluxes of equation
(\ref{eq:e_l_dot_tor_avg_all})
\begin{alignat}{1}
\label{eq:res_tor_from_time}
\left\langle\frac{dE}{dt}\right\rangle^{H/\infty}_{\gamma} &=
\frac{1}{\left(2\pi\right)^{2}}\int^{2\pi}_{0}d\chi_{r_o}\int^{2\pi}_{0}d\chi_{\theta_o}\left\langle\frac{dE}{dt}\right\rangle^{H/\infty}_{t}
\\
\nonumber
&=
\sum_{lmkn}\frac{\alpha^{H/\infty}_{lmkn}}{4\pi\omega_{mkn}^{2}}\left|Z_{lmkn}^{H/\infty}\right|^{2}
\quad .
\end{alignat}

\bibliography{gr}

\begin{thebibliography}{47}
\expandafter\ifx\csname natexlab\endcsname\relax\def\natexlab#1{#1}\fi
\expandafter\ifx\csname bibnamefont\endcsname\relax
  \def\bibnamefont#1{#1}\fi
\expandafter\ifx\csname bibfnamefont\endcsname\relax
  \def\bibfnamefont#1{#1}\fi
\expandafter\ifx\csname citenamefont\endcsname\relax
  \def\citenamefont#1{#1}\fi
\expandafter\ifx\csname url\endcsname\relax
  \def\url#1{\texttt{#1}}\fi
\expandafter\ifx\csname urlprefix\endcsname\relax\def\urlprefix{URL }\fi
\providecommand{\bibinfo}[2]{#2}
\providecommand{\eprint}[2][]{\url{#2}}

\bibitem[{\citenamefont{Mino}(2005)}]{mino2005}
\bibinfo{author}{\bibfnamefont{Y.}~\bibnamefont{Mino}}, \bibinfo{journal}{Prog.
  Theor. Phys.} \textbf{\bibinfo{volume}{113}}, \bibinfo{pages}{733}
  (\bibinfo{year}{2005}), \eprint{gr-qc/0506003}.

\bibitem[{\citenamefont{Flanagan and Hinderer}(2010)}]{flanagan2010}
\bibinfo{author}{\bibfnamefont{E.~E.} \bibnamefont{Flanagan}} \bibnamefont{and}
  \bibinfo{author}{\bibfnamefont{T.}~\bibnamefont{Hinderer}}
  (\bibinfo{year}{2010}), \eprint{1009.4923}.

\bibitem[{\citenamefont{S~Drasco and Hughes}(2005)}]{drasco2005}
\bibinfo{author}{\bibfnamefont{E.~F.} \bibnamefont{S~Drasco}} \bibnamefont{and}
  \bibinfo{author}{\bibfnamefont{S.~A.} \bibnamefont{Hughes}},
  \bibinfo{journal}{Class.\ Quant.\ Grav.} \textbf{\bibinfo{volume}{22}},
  \bibinfo{pages}{801} (\bibinfo{year}{2005}), \eprint{arXiv:gr-qc/0505075}.

\bibitem[{\citenamefont{Pound et~al.}(2005)\citenamefont{Pound, Poisson, and
  Nickel}}]{pound2005}
\bibinfo{author}{\bibfnamefont{A.}~\bibnamefont{Pound}},
  \bibinfo{author}{\bibfnamefont{E.}~\bibnamefont{Poisson}}, \bibnamefont{and}
  \bibinfo{author}{\bibfnamefont{B.~G.} \bibnamefont{Nickel}},
  \bibinfo{journal}{Phys. Rev.} \textbf{\bibinfo{volume}{D72}},
  \bibinfo{pages}{124001} (\bibinfo{year}{2005}), \eprint{gr-qc/0509122}.

\bibitem[{\citenamefont{Pound and Poisson}(2008{\natexlab{a}})}]{pound2007:2}
\bibinfo{author}{\bibfnamefont{A.}~\bibnamefont{Pound}} \bibnamefont{and}
  \bibinfo{author}{\bibfnamefont{E.}~\bibnamefont{Poisson}},
  \bibinfo{journal}{Phys. Rev.} \textbf{\bibinfo{volume}{D77}},
  \bibinfo{pages}{044012} (\bibinfo{year}{2008}{\natexlab{a}}),
  \eprint{0708.3037}.

\bibitem[{\citenamefont{Pound and Poisson}(2008{\natexlab{b}})}]{pound2007:1}
\bibinfo{author}{\bibfnamefont{A.}~\bibnamefont{Pound}} \bibnamefont{and}
  \bibinfo{author}{\bibfnamefont{E.}~\bibnamefont{Poisson}},
  \bibinfo{journal}{Phys. Rev.} \textbf{\bibinfo{volume}{D77}},
  \bibinfo{pages}{044013} (\bibinfo{year}{2008}{\natexlab{b}}),
  \eprint{0708.3033}.

\bibitem[{\citenamefont{Hinderer and Flanagan}(2008)}]{hinderer2008}
\bibinfo{author}{\bibfnamefont{T.}~\bibnamefont{Hinderer}} \bibnamefont{and}
  \bibinfo{author}{\bibfnamefont{E.~E.} \bibnamefont{Flanagan}},
  \bibinfo{journal}{Phys. Rev.} \textbf{\bibinfo{volume}{D78}},
  \bibinfo{pages}{064028} (\bibinfo{year}{2008}), \eprint{0805.3337}.

\bibitem[{\citenamefont{Levin and Perez-Giz}(2008)}]{levin2008}
\bibinfo{author}{\bibfnamefont{J.}~\bibnamefont{Levin}} \bibnamefont{and}
  \bibinfo{author}{\bibfnamefont{G.}~\bibnamefont{Perez-Giz}},
  \bibinfo{journal}{Phys. Rev. D} \textbf{\bibinfo{volume}{77}},
  \bibinfo{pages}{103005} (\bibinfo{year}{2008}), \eprint{0802.0459}.

\bibitem[{\citenamefont{Grossman et~al.}(2011)\citenamefont{Grossman, Levin,
  and Perez-Giz}}]{grossman2011}
\bibinfo{author}{\bibfnamefont{R.}~\bibnamefont{Grossman}},
  \bibinfo{author}{\bibfnamefont{J.}~\bibnamefont{Levin}}, \bibnamefont{and}
  \bibinfo{author}{\bibfnamefont{G.}~\bibnamefont{Perez-Giz}}
  (\bibinfo{year}{2011}), \eprint{1105.5811}.

\bibitem[{\citenamefont{Levin and Grossman}(2008)}]{levin2008:2}
\bibinfo{author}{\bibfnamefont{J.}~\bibnamefont{Levin}} \bibnamefont{and}
  \bibinfo{author}{\bibfnamefont{R.}~\bibnamefont{Grossman}},
  \bibinfo{journal}{gr-qc/08093838}  (\bibinfo{year}{2008}).

\bibitem[{\citenamefont{{Mino}}(2003)}]{mino2003}
\bibinfo{author}{\bibfnamefont{Y.}~\bibnamefont{{Mino}}},
  \bibinfo{journal}{\prd} \textbf{\bibinfo{volume}{67}},
  \bibinfo{pages}{084027} (\bibinfo{year}{2003}), \eprint{arXiv:gr-qc/0302075}.

\bibitem[{\citenamefont{Levin}(2009)}]{levin2009}
\bibinfo{author}{\bibfnamefont{J.}~\bibnamefont{Levin}},
  \bibinfo{journal}{Class. Quant. Grav.} \textbf{\bibinfo{volume}{26}},
  \bibinfo{pages}{235010} (\bibinfo{year}{2009}), \eprint{0907.5195}.

\bibitem[{\citenamefont{{Levin} et~al.}(2010)\citenamefont{{Levin},
  {McWilliams}, and {Contreras}}}]{2010arXiv1009.2533L}
\bibinfo{author}{\bibfnamefont{J.}~\bibnamefont{{Levin}}},
  \bibinfo{author}{\bibfnamefont{S.~T.} \bibnamefont{{McWilliams}}},
  \bibnamefont{and}
  \bibinfo{author}{\bibfnamefont{H.}~\bibnamefont{{Contreras}}},
  \bibinfo{journal}{ArXiv e-prints}  (\bibinfo{year}{2010}),
  \eprint{1009.2533}.

\bibitem[{\citenamefont{Drasco and Hughes}(2006)}]{drasco2006}
\bibinfo{author}{\bibfnamefont{S.}~\bibnamefont{Drasco}} \bibnamefont{and}
  \bibinfo{author}{\bibfnamefont{S.}~\bibnamefont{Hughes}},
  \bibinfo{journal}{Phys. Rev. D} \textbf{\bibinfo{volume}{73}},
  \bibinfo{pages}{024027} (\bibinfo{year}{2006}), \eprint{gr-qc/0509101}.

\bibitem[{\citenamefont{Arnold et~al.}(2002)\citenamefont{Arnold, Kozlov, and
  Neishtadt}}]{arnold}
\bibinfo{author}{\bibfnamefont{V.~I.} \bibnamefont{Arnold}},
  \bibinfo{author}{\bibfnamefont{V.~V.} \bibnamefont{Kozlov}},
  \bibnamefont{and} \bibinfo{author}{\bibfnamefont{A.~I.}
  \bibnamefont{Neishtadt}}, \emph{\bibinfo{title}{Mathematical Aspects of
  Classical and Celestial Mechanics}} (\bibinfo{publisher}{Springer},
  \bibinfo{year}{2002}), \bibinfo{edition}{3rd} ed.

\bibitem[{\citenamefont{Kevorkian and Cole}(1996)}]{kevorkian}
\bibinfo{author}{\bibfnamefont{J.}~\bibnamefont{Kevorkian}} \bibnamefont{and}
  \bibinfo{author}{\bibfnamefont{J.~D.} \bibnamefont{Cole}},
  \emph{\bibinfo{title}{Multiple Scale and Singular Perturbation Methods}}
  (\bibinfo{publisher}{Springer}, \bibinfo{year}{1996}).

\bibitem[{\citenamefont{Verhulst}(2005)}]{verhulst}
\bibinfo{author}{\bibfnamefont{F.}~\bibnamefont{Verhulst}},
  \emph{\bibinfo{title}{Methods and applications of singular perturbations}},
  vol.~\bibinfo{volume}{50} of \emph{\bibinfo{series}{Texts in Applied
  Mathematics}} (\bibinfo{publisher}{Springer}, \bibinfo{address}{New York},
  \bibinfo{year}{2005}), ISBN \bibinfo{isbn}{978-0387-22966-9; 0-387-22966-3},
  \bibinfo{note}{boundary layers and multiple timescale dynamics},
  \urlprefix\url{http://dx.doi.org/10.1007/0-387-28313-7}.

\bibitem[{\citenamefont{Sago et~al.}(2006)\citenamefont{Sago, Tanaka, Hikida,
  Ganz, and Nakano}}]{sago2005}
\bibinfo{author}{\bibfnamefont{N.}~\bibnamefont{Sago}},
  \bibinfo{author}{\bibfnamefont{T.}~\bibnamefont{Tanaka}},
  \bibinfo{author}{\bibfnamefont{W.}~\bibnamefont{Hikida}},
  \bibinfo{author}{\bibfnamefont{K.}~\bibnamefont{Ganz}}, \bibnamefont{and}
  \bibinfo{author}{\bibfnamefont{H.}~\bibnamefont{Nakano}},
  \bibinfo{journal}{Prog. Theor. Phys.} \textbf{\bibinfo{volume}{115}},
  \bibinfo{pages}{873} (\bibinfo{year}{2006}), \eprint{gr-qc/0511151}.

\bibitem[{\citenamefont{Ganz et~al.}(2007)\citenamefont{Ganz, Hikida, Nakano,
  Sago, and Tanaka}}]{ganz2007}
\bibinfo{author}{\bibfnamefont{K.}~\bibnamefont{Ganz}},
  \bibinfo{author}{\bibfnamefont{W.}~\bibnamefont{Hikida}},
  \bibinfo{author}{\bibfnamefont{H.}~\bibnamefont{Nakano}},
  \bibinfo{author}{\bibfnamefont{N.}~\bibnamefont{Sago}}, \bibnamefont{and}
  \bibinfo{author}{\bibfnamefont{T.}~\bibnamefont{Tanaka}},
  \bibinfo{journal}{Prog. Theor. Phys.} \textbf{\bibinfo{volume}{117}},
  \bibinfo{pages}{1041} (\bibinfo{year}{2007}), \eprint{gr-qc/0702054}.

\bibitem[{\citenamefont{Glampedakis and Kennefick}(2002)}]{glampedakis2002}
\bibinfo{author}{\bibfnamefont{K.}~\bibnamefont{Glampedakis}} \bibnamefont{and}
  \bibinfo{author}{\bibfnamefont{D.}~\bibnamefont{Kennefick}},
  \bibinfo{journal}{Phys. Rev. D} \textbf{\bibinfo{volume}{66}},
  \bibinfo{pages}{044002} (\bibinfo{year}{2002}), \eprint{gr-qc/0203086}.

\bibitem[{\citenamefont{Wilkins}(1972)}]{wilkins1972}
\bibinfo{author}{\bibfnamefont{D.~C.} \bibnamefont{Wilkins}},
  \bibinfo{journal}{Phys. Rev. D} \textbf{\bibinfo{volume}{5}},
  \bibinfo{pages}{814} (\bibinfo{year}{1972}).

\bibitem[{\citenamefont{Hughes}(2001)}]{hughes2001}
\bibinfo{author}{\bibfnamefont{S.~A.} \bibnamefont{Hughes}},
  \bibinfo{journal}{erratum-ibbid.d} \textbf{\bibinfo{volume}{63}},
  \bibinfo{pages}{049902} (\bibinfo{year}{2001}),
  \urlprefix\url{http://www.citebase.org/abstract?id=oai:arXiv.org:gr-qc/99100%
91}.

\bibitem[{\citenamefont{Fujita et~al.}(2009)\citenamefont{Fujita, Hikida, and
  Tagoshi}}]{fujita2009}
\bibinfo{author}{\bibfnamefont{R.}~\bibnamefont{Fujita}},
  \bibinfo{author}{\bibfnamefont{W.}~\bibnamefont{Hikida}}, \bibnamefont{and}
  \bibinfo{author}{\bibfnamefont{H.}~\bibnamefont{Tagoshi}},
  \bibinfo{journal}{Prog. Theor. Phys.} \textbf{\bibinfo{volume}{121}},
  \bibinfo{pages}{843} (\bibinfo{year}{2009}), \eprint{0904.3810}.

\bibitem[{\citenamefont{Samoilenko}(1991)}]{samoilenko}
\bibinfo{author}{\bibfnamefont{A.~M.} \bibnamefont{Samoilenko}},
  \emph{\bibinfo{title}{Elements of the Mathematical Theory of Multi-Frequency
  Oscillations}} (\bibinfo{publisher}{Springer-Verlag}, \bibinfo{year}{1991}).

\bibitem[{\citenamefont{Corduneanu}(1968)}]{corduneanu}
\bibinfo{author}{\bibfnamefont{C.}~\bibnamefont{Corduneanu}},
  \emph{\bibinfo{title}{Almost Periodic Functions}}
  (\bibinfo{publisher}{Interscience Publishers}, \bibinfo{year}{1968}).

\bibitem[{\citenamefont{{Drasco} and {Hughes}}(2004)}]{drasco2004}
\bibinfo{author}{\bibfnamefont{S.}~\bibnamefont{{Drasco}}} \bibnamefont{and}
  \bibinfo{author}{\bibfnamefont{S.~A.} \bibnamefont{{Hughes}}},
  \bibinfo{journal}{Phys.\ Rev.\ D} \textbf{\bibinfo{volume}{69}},
  \bibinfo{pages}{044015} (\bibinfo{year}{2004}),
  \eprint{arXiv:astro-ph/0308479}.

\bibitem[{\citenamefont{F~Schilder and Osinga}(2006)}]{schilder2006}
\bibinfo{author}{\bibfnamefont{S.~S.} \bibnamefont{F~Schilder},
  \bibfnamefont{W~Vogt}} \bibnamefont{and}
  \bibinfo{author}{\bibfnamefont{H.~M.} \bibnamefont{Osinga}},
  \bibinfo{journal}{International Journal for Numerical Methods in Engineering}
  \textbf{\bibinfo{volume}{67}}, \bibinfo{pages}{629} (\bibinfo{year}{2006}).

\bibitem[{\citenamefont{Boyce and DiPrima}(2005)}]{Boyce}
\bibinfo{author}{\bibnamefont{Boyce}} \bibnamefont{and}
  \bibinfo{author}{\bibnamefont{DiPrima}}, \emph{\bibinfo{title}{Elementary
  Differential Equations and Boundary Value Problems}}
  (\bibinfo{publisher}{Wiley}, \bibinfo{year}{2005}).

\bibitem[{\citenamefont{Kevin~Beanland and Stevenson}(2009)}]{beanland}
\bibinfo{author}{\bibfnamefont{J.~W.~R.} \bibnamefont{Kevin~Beanland}}
  \bibnamefont{and}
  \bibinfo{author}{\bibfnamefont{C.}~\bibnamefont{Stevenson}},
  \bibinfo{journal}{The American Mathematical Monthly}
  \textbf{\bibinfo{volume}{116}}, \bibinfo{pages}{531} (\bibinfo{year}{2009}),
  \urlprefix\url{http://www.jstor.org/stable/40391145}.

\bibitem[{\citenamefont{Sago et~al.}(2005)\citenamefont{Sago, Tanaka, Hikida,
  and Nakano}}]{Sago:2005gd}
\bibinfo{author}{\bibfnamefont{N.}~\bibnamefont{Sago}},
  \bibinfo{author}{\bibfnamefont{T.}~\bibnamefont{Tanaka}},
  \bibinfo{author}{\bibfnamefont{W.}~\bibnamefont{Hikida}}, \bibnamefont{and}
  \bibinfo{author}{\bibfnamefont{H.}~\bibnamefont{Nakano}},
  \bibinfo{journal}{Prog. Theor. Phys.} \textbf{\bibinfo{volume}{114}},
  \bibinfo{pages}{509} (\bibinfo{year}{2005}), \eprint{gr-qc/0506092}.

\bibitem[{\citenamefont{William H.~Press and Flannery}(2007)}]{press}
\bibinfo{author}{\bibfnamefont{W.~T.~V.} \bibnamefont{William H.~Press},
  \bibfnamefont{Saul A.~Teukolsky}} \bibnamefont{and}
  \bibinfo{author}{\bibfnamefont{B.~P.} \bibnamefont{Flannery}},
  \emph{\bibinfo{title}{Numerical Recipes}} (\bibinfo{publisher}{Cambridge
  University Press}, \bibinfo{year}{2007}), \bibinfo{edition}{3rd} ed.

\bibitem[{\citenamefont{Chrzanowski}(1976)}]{chrzanowski1976}
\bibinfo{author}{\bibfnamefont{P.~L.} \bibnamefont{Chrzanowski}},
  \bibinfo{journal}{Phys. Rev.} \textbf{\bibinfo{volume}{D13}},
  \bibinfo{pages}{806} (\bibinfo{year}{1976}).

\bibitem[{\citenamefont{Hughes et~al.}(2005)\citenamefont{Hughes, Drasco,
  Flanagan, and Franklin}}]{hughes2005}
\bibinfo{author}{\bibfnamefont{S.~A.} \bibnamefont{Hughes}},
  \bibinfo{author}{\bibfnamefont{S.}~\bibnamefont{Drasco}},
  \bibinfo{author}{\bibfnamefont{E.~E.} \bibnamefont{Flanagan}},
  \bibnamefont{and} \bibinfo{author}{\bibfnamefont{J.}~\bibnamefont{Franklin}},
  \bibinfo{journal}{Phys. Rev. Lett.} \textbf{\bibinfo{volume}{94}},
  \bibinfo{pages}{221101} (\bibinfo{year}{2005}), \eprint{gr-qc/0504015}.

\bibitem[{\citenamefont{Teukolsky}(1972)}]{teukolsky1972}
\bibinfo{author}{\bibfnamefont{S.~A.} \bibnamefont{Teukolsky}},
  \bibinfo{journal}{Phys.\ Rev.\ Lett.} \textbf{\bibinfo{volume}{29}},
  \bibinfo{pages}{1114} (\bibinfo{year}{1972}).

\bibitem[{\citenamefont{Teukolsky}(1973)}]{teukolsky1973}
\bibinfo{author}{\bibfnamefont{S.~A.} \bibnamefont{Teukolsky}},
  \bibinfo{journal}{Ap.\ J.} \textbf{\bibinfo{volume}{185}},
  \bibinfo{pages}{635} (\bibinfo{year}{1973}).

\bibitem[{\citenamefont{{Hughes}}(2000)}]{hughes2000}
\bibinfo{author}{\bibfnamefont{S.~A.} \bibnamefont{{Hughes}}},
  \bibinfo{journal}{\prd} \textbf{\bibinfo{volume}{62}},
  \bibinfo{pages}{044029} (\bibinfo{year}{2000}), \eprint{arXiv:gr-qc/0002043}.

\bibitem[{\citenamefont{{Press} and {Teukolsky}}(1973)}]{press1973}
\bibinfo{author}{\bibfnamefont{W.~H.} \bibnamefont{{Press}}} \bibnamefont{and}
  \bibinfo{author}{\bibfnamefont{S.~A.} \bibnamefont{{Teukolsky}}},
  \bibinfo{journal}{\apj} \textbf{\bibinfo{volume}{185}}, \bibinfo{pages}{649}
  (\bibinfo{year}{1973}).

\bibitem[{\citenamefont{{Teukolsky} and {Press}}(1974)}]{teukolsky1974}
\bibinfo{author}{\bibfnamefont{S.~A.} \bibnamefont{{Teukolsky}}}
  \bibnamefont{and} \bibinfo{author}{\bibfnamefont{W.~H.}
  \bibnamefont{{Press}}}, \bibinfo{journal}{\apj}
  \textbf{\bibinfo{volume}{193}}, \bibinfo{pages}{443} (\bibinfo{year}{1974}).

\bibitem[{\citenamefont{{Chandrasekhar}}(1975)}]{chandrasekhar1975}
\bibinfo{author}{\bibfnamefont{S.}~\bibnamefont{{Chandrasekhar}}},
  \bibinfo{journal}{Royal Society of London Proceedings Series A}
  \textbf{\bibinfo{volume}{343}}, \bibinfo{pages}{289} (\bibinfo{year}{1975}).

\bibitem[{\citenamefont{{Leaver}}(1985)}]{leaver1985}
\bibinfo{author}{\bibfnamefont{E.~W.} \bibnamefont{{Leaver}}},
  \bibinfo{journal}{Royal Society of London Proceedings Series A}
  \textbf{\bibinfo{volume}{402}}, \bibinfo{pages}{285} (\bibinfo{year}{1985}).

\bibitem[{\citenamefont{Haberman}(1998)}]{haberman}
\bibinfo{author}{\bibfnamefont{R.}~\bibnamefont{Haberman}},
  \emph{\bibinfo{title}{Elementary Applied Partial Differential Equations with
  Fourier Series and Boundary Value Problems}} (\bibinfo{publisher}{Prentice
  Hall}, \bibinfo{year}{1998}), \bibinfo{edition}{3rd} ed.

\bibitem[{\citenamefont{Glampedakis}(2005)}]{Glampedakis:2005hs}
\bibinfo{author}{\bibfnamefont{K.}~\bibnamefont{Glampedakis}},
  \bibinfo{journal}{Class. Quant. Grav.} \textbf{\bibinfo{volume}{22}},
  \bibinfo{pages}{S605} (\bibinfo{year}{2005}), \eprint{gr-qc/0509024}.

\bibitem[{\citenamefont{Sasaki and Nakamura}(1982)}]{sasaki1982}
\bibinfo{author}{\bibfnamefont{M.}~\bibnamefont{Sasaki}} \bibnamefont{and}
  \bibinfo{author}{\bibfnamefont{T.}~\bibnamefont{Nakamura}},
  \bibinfo{journal}{Phys. Lett. A} \textbf{\bibinfo{volume}{89A}},
  \bibinfo{pages}{68} (\bibinfo{year}{1982}).

\bibitem[{\citenamefont{Breuer}(1975)}]{breuer}
\bibinfo{author}{\bibfnamefont{R.~A.} \bibnamefont{Breuer}},
  \emph{\bibinfo{title}{Gravitational perturbation theory and synchrotron
  radiation}}, vol.~\bibinfo{volume}{44} of \emph{\bibinfo{series}{Lecture
  notes in physics}} (\bibinfo{publisher}{Springer-Verlag},
  \bibinfo{address}{the University of California}, \bibinfo{year}{1975}), ISBN
  \bibinfo{isbn}{0387075305, 9780387075303}.

\bibitem[{\citenamefont{Isaacson}(1968)}]{isaacson1968}
\bibinfo{author}{\bibfnamefont{R.~A.} \bibnamefont{Isaacson}},
  \bibinfo{journal}{Phys. Rev.} \textbf{\bibinfo{volume}{166}},
  \bibinfo{pages}{1272} (\bibinfo{year}{1968}).

\bibitem[{\citenamefont{Maggiore}(2008)}]{maggiore}
\bibinfo{author}{\bibfnamefont{M.}~\bibnamefont{Maggiore}},
  \emph{\bibinfo{title}{Gravitational Waves, Volume 1: Theory and Experiments}}
  (\bibinfo{publisher}{Oxford University Press}, \bibinfo{year}{2008}).

\bibitem[{\citenamefont{Hawking and Hartle}(1972)}]{hawking1972}
\bibinfo{author}{\bibfnamefont{S.~W.} \bibnamefont{Hawking}} \bibnamefont{and}
  \bibinfo{author}{\bibfnamefont{J.~B.} \bibnamefont{Hartle}},
  \bibinfo{journal}{Commun. Math. Phys.} \textbf{\bibinfo{volume}{27}},
  \bibinfo{pages}{283} (\bibinfo{year}{1972}).

\end{thebibliography}

\end{document}